\documentclass[
prd,nofootinbib,longbibliography,a4paper,12pt,eqsecnum]{revtex4-1}
\usepackage{amsmath} 
\usepackage{graphicx} 
\usepackage{amsthm}
\usepackage{amssymb} 
\usepackage{physics}
\usepackage{dsfont}
\usepackage{yfonts}
\usepackage{hyperref}
\usepackage{graphicx}
\usepackage{epigraph}
\usepackage{slashed,stmaryrd}
\usepackage{soul}
\usepackage{subcaption}
\usepackage{quiver}
\usepackage{enumerate}

\usepackage{array,xcolor,graphicx}
\usepackage{booktabs,multirow}
\usepackage[utf8]{inputenc}
\usepackage{mathtools}

\usepackage{tikz}
\usetikzlibrary{arrows.meta,decorations.markings,calc}

\usepackage{etoolbox}
\patchcmd{\section}
  {\centering}
  {\raggedright}
  {}
  {}
\patchcmd{\subsection}
  {\centering}
  {\raggedright}
  {}
  {}
\usepackage{diagbox}
\usepackage{adjustbox}
\usepackage{caption, multirow, makecell}
\setcellgapes{3pt}

\usepackage[all]{xy}
\usepackage{tikz}
\usetikzlibrary{arrows.meta}

\hypersetup{colorlinks=true,linkcolor=black,citecolor=red,urlcolor=black}

\usepackage{isomath}

\newcommand{\be}{\begin{equation}}
\newcommand{\ee}{\end{equation}}
\newcommand{\bea}{\begin{eqnarray}}
\newcommand{\eea}{\end{eqnarray}}

\newcommand{\nn}{\nonumber\\}

\def\la{\langle}
\def\ra{\rangle}
\def\Tr{{\mathrm{Tr}}}

\def\d{\partial}

\def\CA{\mathcal{A}}
\def\CB{\mathcal{B}}
\def\CC{\mathcal{C}}
\def\CD{\mathcal{D}}

\def\CG{\mathcal{G}}
\def\CH{\mathcal{H}}

\def\CL{\mathcal{L}}

\def\CN{\mathcal{N}}
\def\CS{\mathcal{S}}
\def\CT{\mathcal{T}}
\def\CO{\mathcal{O}}

\def\CW{\mathcal{W}}

\def\bx{\mathbf{x}}
\def\by{\mathbf{y}}

\def\ns{\text{NS}}

\def\rld{\mathrm{\Lambda}}

\def\nn{\nonumber}

\def\Tr{\text{~Tr~}}

\def\hom{\text{~Hom}}

\def\RPn (#1,#2){
  \fill (#1, #2) circle (3pt);
  \fill (#1, #2+1) circle (3pt);
}

\def\sqtwoL (#1,#2,#3){
  \draw[#3] (#1,#2) .. controls (#1-1,#2+1) .. (#1,#2+2);
}

\def\sqtwoR (#1,#2,#3){
  \draw[#3] (#1,#2) .. controls (#1+1,#2+1) .. (#1,#2+2);
}

\def \sqtwoCR (#1,#2,#3){
  \draw[#3] (#1,#2) .. controls (#1+1,#2+.5) and (#1+1.5,#2+2) .. (#1+2,#2+2);
}

\def \sqtwoCL (#1,#2,#3){
  \draw[#3] (#1,#2) .. controls (#1-1,#2+.5) and (#1-1.5,#2+2)  .. (#1-2,#2+2);
}

\def \sqone (#1,#2,#3){
  \draw[#3] (#1,#2) -- (#1,#2+1);
}

\def\Aone (#1,#2){
  \fill (#1, #2) circle (3pt);
  \fill (#1, #2+1) circle (3pt);
  \fill (#1, #2+2) circle (3pt);
  \fill (#1, #2+3) circle (3pt);
  \fill (#1+2, #2+3) circle (3pt);
  \fill (#1+2, #2+4) circle (3pt);
  \fill (#1+2, #2+5) circle (3pt);
  \fill (#1+2, #2+6) circle (3pt);
  \draw (#1, #2) -- (#1, #2+1);
  \draw (#1, #2+2) -- (#1, #2+3);
  \draw (#1+2, #2+3) -- (#1 + 2, #2+4);
  \draw (#1+2, #2+5) -- (#1+2, #2+6);
  \draw (#1, #2) .. controls (#1-1, #2+1) .. (#1, #2+2);
  \draw (#1+2, #2+4) .. controls (#1+3, #2+5) .. (#1+2, #2+6);
  \draw (#1, #2+1) .. controls (#1+1, #2+1.5) and  (#1+1.5 ,#2+3) .. (#1+2,#2+3);
  \draw (#1, #2+2) .. controls (#1+1, #2+2.5) and (#1+1.5, #2+4) .. (#1+2, #2+4);
  \draw (#1, #2+3) .. controls (#1+1, #2+3.5) and (#1+1.5, #2+5) .. (#1+2, #2+5);
}

\def\Aonecolor (#1,#2,#3){
  \fill[#3] (#1, #2) circle (3pt);
  \fill[#3] (#1, #2+1) circle (3pt);
  \fill[#3] (#1, #2+2) circle (3pt);
  \fill[#3] (#1, #2+3) circle (3pt);
  \fill[#3] (#1+2, #2+3) circle (3pt);
  \fill[#3] (#1+2, #2+4) circle (3pt);
  \fill[#3] (#1+2, #2+5) circle (3pt);
  \fill[#3] (#1+2, #2+6) circle (3pt);
  \draw[#3] (#1, #2) -- (#1, #2+1);
  \draw[#3] (#1, #2+2) -- (#1, #2+3);
  \draw[#3] (#1+2, #2+3) -- (#1 + 2, #2+4);
  \draw[#3] (#1+2, #2+5) -- (#1+2, #2+6);
  \draw[#3] (#1, #2) .. controls (#1-1, #2+1) .. (#1, #2+2);
  \draw[#3] (#1+2, #2+4) .. controls (#1+3, #2+5) .. (#1+2, #2+6);
  \draw[#3] (#1, #2+1) .. controls (#1+1, #2+1.5) and  (#1+1.5 ,#2+3) .. (#1+2,#2+3);
  \draw[#3] (#1, #2+2) .. controls (#1+1, #2+2.5) and (#1+1.5, #2+4) .. (#1+2, #2+4);
  \draw[#3] (#1, #2+3) .. controls (#1+1, #2+3.5) and (#1+1.5, #2+5) .. (#1+2, #2+5);
}

\def\rectangle (#1,#2,#3){   \draw[#3] (#1-0.15,#2-0.15) rectangle (#1+0.15,#2+0.15)}

\def\Eone (#1,#2){
  \fill (#1, #2) circle (3pt);
  \fill (#1, #2+1) circle (3pt);
  \fill (#1, #2+2) circle (3pt);
  \fill (#1, #2+3) circle (3pt);
  \draw (#1, #2) -- (#1, #2+1);
  \draw (#1, #2+2) -- (#1, #2+3);
  \draw (#1, #2) .. controls (#1-1, #2+1) .. (#1, #2+2);
  \draw (#1, #2+1) .. controls (#1+1, #2+2) .. (#1, #2+3);
}

\def\joker (#1,#2){
  \foreach \y in {#2, #2+1, #2+2, #2+3, #2+4}
  {\fill (#1,\y) circle (3pt);}
  \draw (#1,#2) -- (#1, #2+1);
  \draw (#1,#2+3) -- (#1, #2+4);
  \draw (#1,#2+0) .. controls (#1-1,#2+1) .. (#1, #2+2);
  \draw (#1,#2+2) .. controls (#1-1,#2+3) .. (#1, #2+4);
  \draw (#1,#2+1) .. controls (#1+1,#2+2) .. (#1, #2+3);
}

\def\jokercolor (#1,#2, #3){
  \foreach \y in {#2, #2+1, #2+2, #2+3, #2+4}
  {\fill[#3] (#1,\y) circle (3pt);}
  \draw[#3] (#1,#2) -- (#1, #2+1);
  \draw[#3] (#1,#2+3) -- (#1, #2+4);
  \draw[#3] (#1,#2+0) .. controls (#1-1,#2+1) .. (#1, #2+2);
  \draw[#3] (#1,#2+2) .. controls (#1-1,#2+3) .. (#1, #2+4);
  \draw[#3] (#1,#2+1) .. controls (#1+1,#2+2) .. (#1, #2+3);
}

\def\msopart (#1,#2,#3){
  \fill[#3] (#1,#2) circle (3pt); 
  \fill[#3] (#1, #2+1) circle (3pt);
  \fill[#3] (#1, #2+2) circle (3pt);
  \fill[#3] (#1+2, #2+2) circle (3pt);
  \fill[#3] (#1+2, #2+3) circle (3pt);
  \fill[#3] (#1+2, #2+4) circle (3pt);
  \fill[#3] (#1+2, #2+5) circle (3pt);
  \sqtwoCR(#1,#2, #3);
  \sqtwoCR (#1, #2+1, #3);
  \sqtwoCR (#1, #2+2, #3);
  \sqone (#1+2, #2+2, #3);
  \sqone (#1+2, #2+4, #3);
  \sqtwoR(#1+2, #2+3, #3);
  \sqone (#1, #2+1, #3); }

\def\amme (#1,#2,#3){
  \fill[#3] (#1,#2) circle (3pt) ;
  \sqtwoR(#1,#2,#3);
  \fill[#3] (#1,#2+2) circle (3pt) ;
  \sqone(#1,#2+2,#3);
  \fill[#3] (#1,#2+3) circle (3pt) ;
  \sqtwoR(#1,#2+3,#3);
  \fill[#3] (#1,#2+5) circle (3pt) ;}

\def\questionupsidedon (#1,#2,#3){
  \fill[#3] (#1,#2) circle (3pt) ;
  \sqtwoR(#1,#2,#3);
  \fill[#3] (#1,#2+2) circle (3pt) ;
  \sqone(#1,#2+2,#3);
  \fill[#3] (#1,#2+3) circle (3pt) ;}

\def\emm (#1,#2,#3,#4){
  \fill[#3] (#1,#2) circle (3pt) node[anchor=east] {$#4$};
  \fill[#3] (#1,#2+1) circle (3pt);
  \fill[#3] (#1,#2+2) circle (3pt);
  \fill[#3] (#1,#2+3) circle (3pt);
  \fill[#3] (#1,#2+4) circle (3pt);
  \fill[#3] (#1,#2+5) circle (3pt);
  \fill[#3] (#1,#2+6) circle (3pt);
  \fill[#3] (#1,#2+7) circle (3pt);
  \fill[#3] (#1,#2+8) circle (3pt);

  \sqone (#1,#2,#3);
  \sqtwoR (#1,#2+1,#3);
  \sqone (#1,#2+2,#3);
  \sqtwoL (#1,#2+2,#3);
  \sqone (#1,#2+4,#3);
  \sqtwoL (#1,#2+5,#3);
  \sqone (#1,#2+6,#3);
  \sqtwoR (#1,#2+6,#3);

  \draw[dashed,#3] (#1,#2+8) -- (#1,#2+9);
}

\def\Minf (#1,#2,#3){
  \fill[#3] (#1,#2) circle (3pt);
  \fill[#3] (#1,#2+2) circle (3pt);
  \fill[#3] (#1,#2+3) circle (3pt);
  \fill[#3] (#1,#2+4) circle (3pt);
  \fill[#3] (#1,#2+5) circle (3pt);
  \fill[#3] (#1,#2+6) circle (3pt);

  \sqtwoL (#1,#2,#3);
  \sqone (#1,#2+2,#3);
  \sqtwoL (#1,#2+3,#3);
  \sqone (#1,#2+4,#3);
  \sqtwoR (#1,#2+4,#3);

  \draw[dashed,#3] (#1,#2+6) -- (#1,#2+7);
}

\tikzset{
  phys/.style   = {gray!42,line width=.35pt},
  physdot/.style= {fill=gray!62},
  dual/.style   = {orange!35,dash pattern=on 2pt off 2pt,line width=.35pt},
  dualdot/.style= {draw=orange!60,line width=.5pt},
  Cst/.style    = {blue!70!black,line width=1.5pt,-{Stealth[length=2.6mm,width=2.2mm]}},
  Cfade/.style  = {blue!70!black,opacity=.20,line width=1.5pt},
  Dst/.style    = {black,line width=1.5pt,-{Stealth[length=2.6mm,width=2.2mm]}},
  Dfade/.style  = {black,opacity=.20,line width=1.5pt},
  Mst/.style    = {red!78!black,line width=1.5pt,-{Stealth[length=2.6mm,width=2.2mm]}},
  lbl/.style    = {font=\scriptsize},
}

\begin{document}


\bigskip
    
\title{Discrete anomaly inflow from SymTFT of $U(1)$ global symmetry: \\
the modified Villain resolution}



 \author{Napat Poovuttikul}
 \email{napat.po@chula.ac.th}
 \affiliation{High Energy Physics Research Unit, Department of Physics,
 Faculty of Science, Chulalongkorn University, Bangkok 10330, Thailand}

 \author{Kamonsap Sapmee}
 \email{kai.sapmee@gmail.com}
 \affiliation{Yukawa Institute for Theoretical Physics, Kyoto University, Kyoto 606-8502, Japan}

\begin{abstract}
\vspace{1cm}
We construct a lattice SymTFT for a $U(1)$ global symmetry with anomaly coefficient $k$ using the modified Villain formulation, in which the spin structure required for odd $k$ is realised explicitly by lattice Majorana fermions. We demonstrate that the same Majorana fermions sit at the junctions of the symmetry lines, exactly where the fusion of two $U(1)$ elements wraps around the circle. From the lattice algebra of lines and junctions we derive the $R$- and $F$-symbols in the bulk and on the gapped boundary, and show that for odd $k$ the boundary $F$-symbol obeys the Gu-Wen relation for the $U(1)$ symmetry defect, without passing through a bosonic shadow. Restricting to $\mathbb{Z}_2\subset U(1)$, we reproduce the mod-4 anomalous phase from a mapping torus and the fermionic stacking rule. As a simple application of this SymTFT, we also determine which subgroups $\mathbb{Z}_d\subset U(1)$ can be gauged and the symmetry of the resulting theories. This is a first step towards deriving, from the anomaly inflow of a continuous symmetry, the junction data of global anomalies beyond group cohomology.

\end{abstract}
\maketitle

\tableofcontents

\section{Introduction} \label{section:introduction}
It has long been realized that global anomalies of a group $G$ in a theory $\mathcal{T}_G$ can be represented using another theory $\mathcal{T}_{G'}$ exhibiting only local anomalies of a larger group \cite{Elitzur:1984kr,Kiritsis:1986mf}, i.e., $G\subset G'$. This approach has been made precise in terms of bordism classification \cite{Davighi:2020uab,Davighi:2020kok,Davighi:2020bvi} where it is shown that the APS $\eta-$invariant \cite{Atiyah:1975jf} characterising the global anomaly of $G$ can be computed from the local anomaly of $G'$ upon appropriately embedding $G$ inside $G'$.
This method enables us to easily calculate invertible phase that characterise the global anomaly of $\CT_G$, even in the case where the invertible phase cannot be expressed in terms of the differential form or cocycle of the background gauge fields. Nevertheless, information of the theories other than the invertible phase of $\CT_G$ remains inaccessible from $\CT_{G'}$. 

In this manuscript we develop a framework to answer a concrete question: \emph{Knowing the invertible phase of $\CT_{G'}$, is it possible to extract the categorical data of $\CT_G$ that characterise its global anomaly?} 
At a first glance the answer is not obvious, as the anomaly-induced observables $\mathcal{O}'$ in $\mathcal{T}_{G'}$ are well understood through the violation of conservation laws (e.g., the pion decay rate and anomalous conductivities), whereas the anomaly-induced observables $\mathcal{O}$ in $\mathcal{T}_G$ are often not. Consider, for example, the $(1+1)$-dimensional case where $G = \mathbb{Z}_2 \subset G' = U(1)$. It is known that $\mathcal{T}_{U(1)}$ is a theory of complex Weyl fermions with a $U(1)$ chiral anomaly characterized by the non-conservation of the $U(1)$ current. On the other hand, the $\mathcal{T}_{\mathbb{Z}_2}$ theory has no conserved current, and its anomalies are classified by the junctions of symmetry defects \cite{Chen:2011pg,Gu:2012ib,Kapustin:2014zva,Else:2014vma}.\footnote{
To be more mathematically precise in this specific case, one must consider the characterization of
these anomalies. The local anomaly in $\CT_{U(1)}$ is known to be characterized by $H^4(BU(1);\mathbb Z)$,
whereas the global anomaly of $T_{\mathbb Z_2}$ is characterized by the bordism group
$\Omega_3^{\rm Spin}(B\mathbb Z_2)\cong\mathbb Z_8$ \cite{Kapustin:2014dxa}, which is not a single cohomology group.
Rather, an element $\nu\in\mathbb Z_8$ is built from three layers of cohomology data
\cite{Delmastro:2021xox,Barkeshli:2021ypb},
\be
\nu=\nu_1+2\nu_2+4\nu_3 \bmod 8\,,\qquad \nu_1\in H^1(B\mathbb Z_2;\mathbb Z_2)\,,\quad
\nu_2\in H^2(B\mathbb Z_2;\mathbb Z_2)\,,\quad \nu_3\in H^3(B\mathbb Z_2;U(1))\,,
\ee
which are glued by non-trivial group extensions rather than forming a direct sum.
In particular, the group cohomology $H^\bullet(B\mathbb Z_2;A)$ has a nice interpretation in terms of
the fusion of the topological line operators. On the other hand, the Lie group cohomology is rather
subtle, see e.g. \cite{Jia:2025uun}, and it is not obvious how to translate these two types of information
into one another.
}
If one wants to chart the observable consequences of the latter without specifying the microscopic details of the theories, a systematic construction of a map between $(\CO',\CT_{G'})\to (\CO,\CT_{G})$ is desired.

There are two recent developments that enable us to address this question. The first is the symmetry
topological field theory (SymTFT) for continuous symmetries \cite{Brennan:2024fgj,Antinucci:2024zjp,Bonetti:2024cjk,Apruzzi:2024htg}, see also \cite{Gagliano:2024off,Arbalestrier:2025poq,Jia:2025vrj, Argurio:2024ewp}.
It translates the algebra of symmetry operators into a topological field theory in one higher dimension,
so that the embedding $G\hookrightarrow G'$ reduces to a manipulation of the TQFT and its boundary
conditions. Most SymTFT constructions, however, concern bosonic theories, whose TQFTs do not depend on
the spin structure, while the spin structure is essential in this work. For a $U(1)$ symmetry with odd
$k$, the continuum SymTFT is a spin Chern--Simons theory \cite{Antinucci:2024zjp}; see also
\cite{Belov:2005ze,Gaiotto:2015zta} for spin Chern--Simons theories. For finite symmetries, fermionic
SymTFTs are usually formulated with a bosonic bulk TQFT and a fermionic topological boundary, on which a
fermionic line is condensed \cite{Bhardwaj:2024ydc,Wen:2024udn,Kawabata:2025hfd}, building on the
bosonisation of fermionic theories and their anomalies
\cite{Delmastro:2021xox,Kapustin:2017jrc,Thorngren:2018bhj}.\footnote{See e.g. 
\cite{Bhardwaj:2016clt,Aasen:2017ubm} 
for the state-sum (fermionic Turaev--Viro) construction of spin TQFTs and for fermion condensation.}
This approach requires knowing the \emph{bosonic shadow}, namely the bosonic TQFT in which condensing the
fermionic line (gauging its $\mathbb Z_2$ one-form symmetry) yields the desired spin TQFT \cite{Gaiotto:2015zta,Bhardwaj:2016clt,Delmastro:2021xox}. However, it 
is not obvious what a shadow theory is if, for example, the only information on the invertible
phase is the APS $\eta$-invariant. This gap can be filled by the second ingredient, the modified Villain
construction \cite{Sulejmanpasic:2019ytl,Chen:2019mjw,Gorantla:2021svj}. It allows us to write the
SymTFT of a continuous symmetry $G'$ on a lattice, including the Chern--Simons term with and without
spin structure \cite{Jacobson:2024hov,Xu:2024hyo} and to realize topological defects of continuous symmetries, including those obtained by flat gauging directly on lattice \cite{Seifnashri:2023dpa, Argurio:2026txf},\footnote{See also \cite{delaFuente:2024aok} for
related continuous-variable stabiliser models.} so that the categorical data of the symmetry defects of
both the discrete $G$ and the continuous $G'$ can be extracted explicitly and on an equal footing. In
particular, for odd $k$ the spin structure is realised by lattice Majorana fermions, and these
fermions sit at the junctions of the symmetry lines, so that the fermionic junction data can be read off
directly from the lattice algebra.

While our ultimate goal is to employ this method to extract, a priori unknown, algebra of symmetry defects whose anomaly can only be captured by the APS $\eta-$invariant, the result presented in this manuscript will only contain the demonstration of our method where the algebra of symmetry defect in $\CT_G$ is already known. Here we consider $\CT_{\mathbb{Z}_2}$ in 1+1 dimensions whose anomaly encoded in the bordism group $\Omega^\text{Spin}_3(B\mathbb{Z}_2)$ is characterised by a triplet 
\be \label{eq:fermionic-anomaly-data}
(\mu, p,\alpha) \in Z^1(B\mathbb{Z}_2;\mathbb{Z}_2)\times Z^2(B\mathbb{Z}_2;\mathbb{Z}_2)\times C^3(B\mathbb{Z}_2;U(1))
\ee
satisfying the following relation \cite{Brumfiel:2016vpy, Gu:2012ib}
\be \label{eq:fermionic-anomaly-condition}
 \delta \alpha = \frac{1}{2}p\cup p
\ee
where $\delta$ denotes the coboundary operator in the group cohomology. The case where $
\mu = 0$ is known as the Gu-Wen anomaly which is reflected in the partition function whose phase ambiguity being the fourth root of unity. The decomposition \eqref{eq:fermionic-anomaly-data}-\eqref{eq:fermionic-anomaly-condition} has a physical interpretation that 
the fermionic point operator at the fusion junction of $\mathbb{Z}_2$ symmetry defect operator, represented by $[p]\in H^2(B\mathbb{Z}_2;\mathbb{Z}_2)\cong \mathbb{Z}_2$, violates the pentagon associativity $\delta\alpha \neq 0$, and extends the bosonic anomaly layer $[\alpha]\in H^3(B\mathbb{Z}_2;U(1))\cong \mathbb{Z}_2$ to $\mathbb{Z}_4$. We will show how all this information can be derived from a SymTFT of anomalous $U(1)$ symmetry whose anomaly coefficient is $k\in H^4(BU(1);\mathbb{Z})$. In other words, 
by formulating a fermionic-SymTFT on a lattice with the modified Villain construction, we are able to detect the Gu-Wen anomaly, bypassing the need for a bosonic shadow that was employed in the earlier SymTFT literature \cite{Wen:2024udn,Bhardwaj:2024ydc,Huang:2024ror}.

We summarise our main results in Section~\ref{subsection:summary-of-results}. The rest of the manuscript is organised as follows. In Section~\ref{section:lattice-symtft-theory}, we
formulate the lattice $U(1)$ SymTFT with the modified Villain construction and perform its canonical
quantisation. We show in Section~\ref{subsection:gauge-transformations-and-spin} that the theory depends on the spin structure when $k$
is odd, define the topological operators in Section~\ref{subsection:topological-operator}, and construct the Hilbert space on a
torus and the Lagrangian algebra in Section~\ref{subsection:u(1)-lagrangian-algebra}. In Section~\ref{section:u(1)-junction-computation}, we compute the
$R$- and $F$-symbols, derive the $F$-symbol on the gapped boundary, and compute the topological spin
from the modular transformations. In Section~\ref{section:mod-4-anomaly}, we restrict to $\mathbb Z_2\subset U(1)$,
compute the mod-4 anomalous phase, and derive the stacking rules in Section~\ref{subsection:stacking-rule}. The gaugeable
subgroups $\mathbb Z_d\subset U(1)$ and the symmetries of the gauged theories are discussed in
Section~\ref{section:gaugin-anomaly}. We list open problems in Section~\ref{section:discussions-and-outlooks}. Various appendices summarise the
technical notation and collect the longer derivations.
\subsection{Summary of results} \label{subsection:summary-of-results}
 
Let us now state our main results. We construct a lattice SymTFT for a $U(1)$ global symmetry with
anomaly coefficient $k$ by applying the modified Villain construction
\cite{Sulejmanpasic:2019ytl,Chen:2019mjw,Gorantla:2021svj} to the BF+CS theory of
\cite{Brennan:2024fgj,Antinucci:2024zjp}. For odd $k$, large-gauge invariance requires a spin
structure, which is realised explicitly by lattice Majorana fermions. From the lattice algebra of the
line operators, we compute the bulk $R$- and $F$-symbols. For odd $k$, we find that the same Majorana
fermions sit at the fusion junctions of the lines precisely when the fusion of two $U(1)$ elements wraps
around the circle, so that the fermion parity of a junction is the carry $p$.
Projecting the bulk
junctions onto the gapped boundary defined by the Lagrangian algebra \cite{Cong:2017ffh,Cheng:2020rpl,Zhang:2023wlu},
we then obtain the boundary $F$-symbol $\bar\alpha$, which for odd $k$ obeys the Gu--Wen relation
$\delta\bar\alpha=\tfrac12\,p\cup p$ for the whole $U(1)$. Following \cite{Jia:2025uun}, we interpret
$(\bar\alpha,p)$ as cochains on $U(1)$ with the discrete topology.
 
Such junction data have also been obtained directly in $1+1$ dimensions, by restricting the symmetry to an
interval \cite{Else:2014vma,Okada:2025kie} or from the fusion operators of symmetry defects
\cite{Kawagoe:2021gqi,Seifnashri:2023dpa}. Let us emphasise what is gained by obtaining them from the bulk
instead. On the boundary side, the junction data are extracted from a given realisation of the symmetry: a free
Dirac edge \cite{Else:2014vma}, a chiral boson edge \cite{Kawagoe:2021gqi}, the modified Villain compact boson
\cite{Seifnashri:2023dpa}, or a $U(1)$ current algebra of level $k$ \cite{Okada:2025kie}, and each new boundary
condition or gauging requires a separate computation. In the SymTFT, a single set of lattice operators
determines the $R$-symbols and topological spins in the bulk, the $F$-symbols on the gapped boundary, and the
gauging of any subgroup $\mathbb Z_d\subset U(1)$, which correspond to different Lagrangian algebras of the same
lattice theory. In particular, the fermionic boundary condition required for gauging $\mathbb Z_d$ at odd $k$,
which is not visible from braiding and topological spin alone \cite{Antinucci:2024zjp}, is constructed
explicitly in Section~\ref{section:gaugin-anomaly}. Moreover, the origin of the fermion parity at the junction
becomes explicit: it is carried by the Majorana fermions that realise the spin structure of the lattice
Chern--Simons term.
 
We then test this framework against known results. Restricting to $\mathbb Z_2\subset U(1)$, we
reproduce the mod-4 anomalous phase from the mapping torus of $\texttt{ST}^2\texttt{S}^{-1}$, following the anomaly
interplay of \cite{Davighi:2020uab}. Since the charged fermions of a $U(1)$ theory are complex, only the even
elements $\nu=2k\bmod 8$ of $\mathbb Z_8$
can be reached in this way. We also derive the stacking rules of bosonic and fermionic theories from the
$F$-move, under which four copies of the Gu--Wen theory are anomaly free, in agreement with
\cite{Else:2014vma,Okada:2025kie}. Finally, we determine which subgroups $\mathbb Z_d\subset U(1)$ can be gauged, namely
$d\,|\,k/2$ for even $k$ and $d\,|\,k$ for odd $k$. These conditions were suggested in \cite{Antinucci:2024zjp} from the braiding and topological spin of the condensed lines. For odd
$k$, however, the condensed lines include those with topological spin $-1$, whose condensation requires a fermionic boundary condition. We construct this boundary condition explicitly by dressing these lines and their junctions by the lattice Majorana fermions, and confirm the gaugeability from the $F$-move. The symmetry of the gauged theory is
$U(1)'\times\mathbb Z_g$, where $U(1)':= U(1)/\mathbb{Z}_g$, with $g=\gcd(d,k/2d)$ for even $k$ and $g=\gcd(d,k/d)$ for odd $k$ (together
with $\mathbb Z_2^F$), which reduces to the $U(1)'\times\mathbb Z_d$ of \cite{Antinucci:2024zjp} when $g=d$. The gauged $U(1)'$ has level $k':=k/g^2$. The mixed $U(1)'\times \mathbb{Z}_g$ anomaly cannot be removed by a change of basis if and only if $\gcd(g,k')>1$. For odd $k$ the gauged $U(1)'$ again has odd level and carries a Gu-Wen layer.

\section{Lattice $U(1)-\mathbb{R}$ TQFT and Quantization} \label{section:lattice-symtft-theory}

%

In this section, we explain how to place a $(2+1)$--dimensional TQFT with $U(1)$ gauge field $A$ and $\mathbb{R}$ gauge field $b$ on a lattice and perform a canonical quantisation. This theory was first formulated on a lattice in \cite{Chen:2019mjw} and later has been realised as a SymTFT for the theory with anomalous $U(1)$ global symmetry \cite{Brennan:2024fgj,Antinucci:2024zjp}. The key advantage of this approach is the appearance of the local large gauge transformation operator as well as the point dyonic operator and Majorana fermion which the line operators terminate. These operators are crucial in construction the junctions of symmetry defects in Section \ref{section:u(1)-junction-computation}. Furthermore, projecting the states into those invariant under the large gauge transformation necessitate the existence of the spin structure whose data can be encoded by the local Majorana fermionic operator. 

To start, we can consider the continuum description of the $U(1)-\mathbb{R}$ SymTFT theory in $(2+1)$ dimensions, namely
\be \label{eq: double R BF+CS action}
\CS_\text{cont} = \frac{i}{2\pi}\int_{X}b\wedge dA + \frac{k}{2}A\wedge dA
\ee
where $b$ is an $\mathbb{R}-$valued gauge field and $A$ is a $U(1)-$valued gauge field. For odd $k$, $X$ is required to be a spin manifold for the theory to be consistently defined, see e.g. \cite{Belov:2005ze}. As a first step toward obtaining the lattice action, we decompose the spacetime into a non-compact `temporal direction' and a slice of a square lattice $\rld$ equipped with the branching structure,\footnote{By a branching structure, we mean that, given any triangulated manifold, we equip each vertex with a distinct integer and draw arrows from the lower number to the higher number. In this approach, no face ($2$-simplex) contains an oriented cycle.} where we adopt the conventions of \cite{Jacobson:2024hov,Jacobson:2023cmr, Chen:2019mjw,Gorantla:2021svj,Hsin:2025jot}.
\be
X = \mathbb{R}\times \mathrm{\Lambda} \ni (x^0,x^1,x^2)
\ee
where $\mathbb{R}$ is the non-compact `temporal' direction in the modified Villain construction, and $\rld$ is the a square lattice,\footnote{There is a notion of \emph{dual lattice} used frequently in this work. The dual lattice is obtained by diagonally shifting the physical lattice $\tilde\rld = \rld +\frac{\hat{s}}{2}$, where $\hat{s} = \hat{1} + \hat{2}$. Both lattice and its dual are equipped with the same branching structure.} which is taken to be compact, $\rld\cong T^2$, as shown in Fig. \ref{fig:iota-map-explain}. The geometric objects on the lattice are the supports of the cochain associated to fields in the theory, i.e., links are supports of $1-$cochain fields. In the modified Villain construction, it is more convenient to start with a double $\mathbb{R}$ lattice action, namely
\begin{align} \label{eq: lattice double R BF+CS action}
\CS = \frac{i}{2\pi}\int_{\mathbb{R}\times\mathrm{\Lambda}}  b\cup da +\frac{ik}{4\pi}\int_{\mathbb{R}\times\mathrm{\Lambda}}a\cup d a
\end{align}
where we use lattice notation, e.g., $\wedge\rightarrow \cup$, $d\rightarrow d$, and replace fields by operators living on sites, links, and plaquettes. Here $a\in C^1(X;\mathbb{R})$ and $ b\in C^1(X;\mathbb{R})$. Throughout this paper, $d$ denotes the exterior/coboundary derivative on the lattice; however, we use the same notation as for the exterior derivative on differential forms, which appears only once here. More details about various derivatives on the lattice and cup products as well as how to translate them in standard lattice notation can be found in Appendix \ref{app:useful}. One can quantize this theory by first express the cup product expression in \eqref{eq: lattice double R BF+CS action} in terms the link variables $a_\ell,  b_\ell$ with $\ell =(\bx,i)$ denoting the position of the link on $\bx \in \Lambda$ and its direction $i=1,2$, then proceed to read off the symplectic structure from the terms containing time derivative of the link variables (see Appendix \ref{app: quantize-noncompact} for a more detailed derivation). This results in the commutation relation 
\be \label{eq:BF-CS-commutator}
\begin{aligned}
[a_{\bx,i},a_{\by,j}] = 0,\quad [a_{\bx,i}, b_{\by,j}] = -2\pi i \epsilon_{ij}\delta_{\bx,\by+\hat{j}},\
[ b_{\bx,i}, b_{\by,j}] = i\pi k\epsilon_{ij}(\delta_{\bx,\by+\hat{j}}+\delta_{\bx+\hat{i},\by})
\end{aligned}
\ee
To turn this into a compact $U(1)$ theory, we introduce additional information, namely the non-trivial homotopy group $\pi_1(U(1))\cong \mathbb{Z}$, corresponding to the winding number. This information is incorporated through the large gauge transformation $a\to a + 2\pi  m$, $ m\in C^{1}(X;\mathbb{Z})$. The action that is invariant under both small and large gauge transformation can be construct following the procedure in \cite{Jacobson:2023cmr} and results in 

\be
\begin{aligned}
\label{eq:action-BF-CS}
\CS = \frac{i}{2\pi} \int_X b\cup(d  a-2\pi  n) + i\int_X\varphi\cup d n + \frac{ik}{4\pi} \int_X \left( a\cup d  a - 2\pi( a\cup n +  n\cup  a +  a\cup_1 d  n)\right)
\end{aligned}
\ee
where $\cup_1$ is a higher cup product. 
Here $\varphi$ is a scalar living on a site acting as a Lagrange multiplier and $ n$ is the discrete flux. This strongly resembles the differential cohomology construction (see e.g. \cite{Hsieh:2020jpj}). However, unlike those cases where $n\in Z^2(X;\mathbb{Z})$, the implementation of $n$ in the modified Villain context becomes simpler if one writes $n$ as a cochain whose lattice realization is simply a plaquette. The closedness of $n$ then can be imposed by introducing the Lagrange multiplier, i.e., using the term $\varphi\cup dn$. On a closed spin manifold $X$, the action is completely invariant under the following gauge transformations:
\be \label{eq:complete-gauge-transformation-U1}
\begin{aligned}
 a&\to a + d\lambda_A + 2\pi  m;\quad \lambda_A \in C^0(X;\mathbb{R}),\quad  m\in C^1(X;\mathbb{Z}),\\
 n &\to  n + d m;\quad  n\in C^2(X;\mathbb{Z}),\\
 b &\to b + d \lambda_B;\quad \lambda_B\in C^0(X;\mathbb{R}),\\
\varphi &\to \varphi - k\lambda_A - \lambda_B + 2\pi r;\quad  r\in C^{0}(X;\mathbb{Z})
\end{aligned}
\ee
where $r$ imposes the compactness of the Lagrange multiplier. In this case, one can rewrite the action \eqref{eq:action-BF-CS} in the complete lattice expression by decomposing temporal and spatial components, such that $a=(a_{\bx,i},(a_0)_\bx)$, $b=(b_{\bx,i}, (b_0)_\bx)$, and $n=(n_{\bx,12}, n_{\bx,0i})$. The decomposed action reads
\be
\begin{aligned} \label{eq:action-BF-CS full lattice}
    -\CS = &i\int_\mathbb{R}dt\sum_\bx\left( \frac{1}{2\pi}\epsilon^{ij}b_{\bx,i}\dot{a}_{\bx+\hat{i},j}+\frac{k}{4\pi}\epsilon^{ij}a_{\bx,i}\dot{a}_{\bx+\hat{i},j} +\dot{\tilde{\varphi}}_{\bx}n_{\bx,12}\right) \\
    &-\frac{i}{2\pi}\int_\mathbb{R}dt\sum_\bx\left((db)_{\bx-\hat{s},12}(a_0)_{\bx}+\frac{k}{2}((a_0)_{\bx} + (a_0)_{\bx+\hat{s}})(da-2\pi n)_{\bx,12} \right) \\
    &- \frac{i}{2\pi}\int_\mathbb{R}dt\sum_\bx\left( (b_0)_\bx(da-2\pi n)_{\bx,12} \right) \\
    &-i\int_{\mathbb{R}}dt\sum_{\bx}n_{\bx,0j}\epsilon^{ij}\left( b_{\bx-\hat{i},i}+(d\tilde{\varphi})_{\bx-\hat{i},i}+\frac{k}{2}(a_{\bx-\hat{i},i}+a_{\bx+\hat{j},i}) \right) \\
    &+\frac{ik}{2}\int_{\mathbb{R}}dt\sum_{\bx}(a_0)_{\bx+\hat{s}}\,(dn)_{\bx,012}
\end{aligned}
\ee
where $\tilde\varphi_\bx := \varphi_\bx - \frac{k}{2}( a_{\bx,2} +  a_{\bx+\hat 2,1})$ is a dressed scalar. We will refer to $\varphi_\bx$ and $\tilde\varphi_\bx$ as the scalar and dressed scalar, respectively. Note that the last term comes from the higher cup term, but it does not contribute to anything since it is trivial by $\tilde\varphi$'s equation of motion. Equivalently, it can be removed by redefining the dressed scalar.
In particular, the dressed scalar operator transforms under both $\lambda_A$ and $\lambda_B$:
\be \label{eq:transformation-varphi}
\tilde\varphi_\bx \to \tilde\varphi_\bx - (\lambda_B)_\bx - \frac{k}{2}\left( (\lambda_A)_{\bx} + (\lambda_A)_{\bx+\hat{s}} \right)
\ee
This form of action is particularly useful since one can read off the symplectic structure from the first line, the second and the third line correspond to small gauge transformations, while the fourth line will be related to the large gauge transformation of the compact theory. The small and large gauge will be explained in more details in the following part, we first want to perform canonical quantization and read off the symplectic structure from the first line, namely
\begin{equation}
\label{eq: complete-commutators-compact}
\begin{aligned}
[ a_{\bx,i}, a_{\by,j}] &= 0,\quad
[ a_{\bx,i},b_{\by,j}] = -2\pi i \epsilon_{ij}\delta_{\bx,\by+\hat{j}},\\
[b_{\bx,i},b_{\by,j}] &= i\pi k\epsilon_{ij}(\delta_{\bx,\by+\hat j}+\delta_{\bx+\hat i,\by}),\quad
[\tilde\varphi_\bx,  n_{\by,12}] = i\delta_{\bx,\by}
\end{aligned}
\end{equation}
These commutation relations are almost identical to those of the non-compact theory, with the last one providing an additional relation. This last commutation relation will be essential for properly seeing the spin-structure dependence of the theory in the modified Villain construction. Then one can use these canonical commutation relations to construct lattice operators associated to gauge transformation in the following subsection. 
A user's manual on how to operate various differentials products on the lattice can be found summarised in Appendix \ref{app:useful}. 
%

\subsection{Projection to physical Hilbert space and spin structure} \label{subsection:gauge-transformations-and-spin}

\begin{figure}[tbh]
    \includegraphics[width=0.7\linewidth]{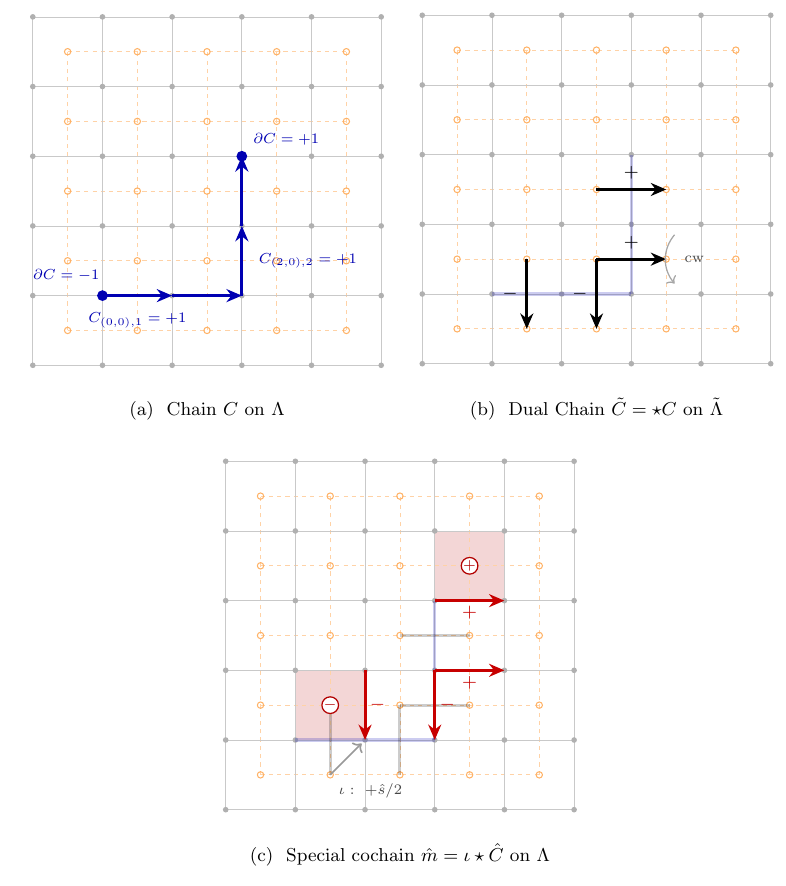}
    \caption{Illustration of the \textcolor{red!78!black}{\emph{stair}} configuration corresponding to the special cochain \textcolor{red!78!black}{$\hat{m}:=\iota\star\hat{C}$} on the physical lattice \textcolor{gray}{$\rld$}. \textbf{(a)} Given the support \textcolor{blue!70!black}{$C$}, defined on the physical lattice \textcolor{gray}{$\rld$}, we define an associated integer-valued cochain \textcolor{blue!70!black}{$\hat{C}\in C^{1}(\rld;\mathbb{Z})$}. \textbf{(b)} The lattice dual of the cochain $\hat{C}$ is given by $\star\hat{C}\in C^1(\tilde\rld;\mathbb{Z})$ whose supports are dual chain \textcolor{blue!70!black}{$\tilde C$} on the dual lattice \textcolor{orange}{$\tilde{\rld}$}. \textbf{(c)} By applying the translation map $\iota$, defined in \eqref{eq: iota-map}, we obtain the stair-configuration support of the cochain \textcolor{red!78!black}{$\hat{m}:=\iota\star\hat{C}$}. This cochain has special properties namely $\hat{m}_{\bx,1} = \hat{C}_{\bx-\hat{2},2}$ and $\hat{m}_{\bx,2} = -\hat{C}_{\bx-\hat{1},1}$, leading to $d\hat{m}_{\bx,12}=\widehat{(\partial C)}_\bx$. This means that endpoints of the chain \textcolor{blue!70!black}{$C$} coincide with the plaquettes of the stair configuration \textcolor{red!78!black}{$m$}.}
\label{fig:iota-map-explain}
\end{figure}

In this subsection, we introduce operators corresponding to small gauge transformations called the Gauss's law operators, while large gauge transformation is implemented by a large gauge operator. Then the physical states span the Hilbert space can be constructed as gauge invariant states.
Firstly, from the second and the third lines of \eqref{eq:action-BF-CS full lattice}, one can obtain the lattice Gauss's laws by integrating over the non-compact temporal component $ a_{\bx,0}$ and $ b_{\bx,0}$, one obtains
\be \label{eq:BF-CS-GuassLaw}
\begin{aligned}
(\CG_B)_{\bx,12} &= F_{\bx,12} := (d a)_{\bx,12} -2\pi  n_{\bx,12} = 0 \\
(\CG_A)_{\bx,12} &= (d b)_{\bx-\hat{s},12} + \frac{k}{2} \Big( F_{\bx,12} + F_{\bx-\hat{s},12} \Big)=0
\end{aligned}
\ee 
One can directly check by using the commutation relations \eqref{eq: complete-commutators-compact} that the Gauss's laws are associated to the small gauge transformations of the theory; thus, we defined Gauss's law operators as
\be \label{eq:gauss-law-operator-G}
G[\lambda_A,\lambda_B] := \exp\left[-\frac{i}{2\pi} \sum_\bx\left( (\lambda_A)_{\bx}(\CG_A)_{\bx,12} + (\lambda_B)_{\bx}(\CG_B)_{\bx,12}  \right) \right]
\ee
One can check that this operator actually generates the small gauge transformations of the theory, namely
\be \label{eq:action-of-gauss-law-operato}
\begin{aligned}
    G[\lambda_A,\lambda_B]^\dagger  a_{\by,i} G[\lambda_A,\lambda_B] &=  a_{\by,i} + (d\lambda_A)_{\by,i},\\
    G[\lambda_A,\lambda_B]^\dagger  b_{\by,i} G[\lambda_A,\lambda_B] &=  b_{\by,i} + (d\lambda_B)_{\by,i} \\
    G[\lambda_A,\lambda_B]^\dagger  n_{\by,12} G[\lambda_A,\lambda_B] &=  n_{\by,12},\\
    G[\lambda_A,\lambda_B]^\dagger \tilde\varphi_\by G[\lambda_A,\lambda_B] &= \tilde\varphi_{\by} - (\lambda_B)_{\by} -\frac{k}{2}[(\lambda_A)_\by + (\lambda_A)_{\by+\hat{s}}]
\end{aligned}
\ee
which are consistent from the action-level gauge transformations.
The first constraint on the physical Hilbert space comes from invariance under small gauge transformations, namely, the states spanning the Hilbert space must satisfy
\begin{align}\label{eq:Guasslaw-U1R}
    G[\lambda_A,\lambda_B]|\psi\ra = |\psi\ra
\end{align}
This physical Hilbert space is still unconstrained by the large gauge transformations. One is therefore required to construct operators that implement the large gauge transformations, which can be directly read off from the last line in \eqref{eq:action-BF-CS full lattice}, but a bit more tricky. One can check that upon $n_{\bx,0j}\to n_{\bx,0j}+\partial_0m_{\bx,j}$, where $m_{\bx,j}\in C^1(\rld;\mathbb{Z})$, the action is transformed under large gauge transformation \eqref{eq:complete-gauge-transformation-U1}, namely
\be \label{eq:large-gauge-on-the-action}
\CS\to \CS + i\sum_\bx \epsilon^{ij} m_{\bx+\hat{i},j}X_{\bx,i}\Big\vert_{t_i}^{t_f},\qquad X_{\bx,i}:= \left(b_{\bx,i}+(d\tilde{\varphi})_{\bx,i}+\frac{k}{2}(a_{\bx,i}+a_{\bx+\hat{s},i})\right)
\ee
where $t_i$ and $t_f$ are initial and final time, respectively. For $m=\hat{m}=\iota(\star\hat{C})$, one has $\epsilon^{ij}\hat{m}_{\bx+\hat{i},j}= - \hat{C}_{\bx,i}$, so that $\exp(i\sum_{(\bx,i)}\hat{C}_{\bx,i}X_{\bx,i})$ yields the large gauge transformation operator.\footnote{Note that the configuration of this cochain $m$'s support has to be the stair configuration, as shown in Fig. \ref{fig:iota-map-explain} in order to perform non-trivial commutation relations generating the large gauge transformations \eqref{eq:large-gauge-transformation-matrix}.}
\be
\begin{aligned} \label{eq:large-gauge-operator-U}
    U[C] &= \exp\left(i \sum_{(\bx,i)}(b_{\bx,i}+ \frac{k}{2}( a_{\bx,i}+  a_{\bx+\hat{s},i}))\hat{C}_{\bx,i} + i \sum_{\bx}\tilde{\varphi}_\bx\widehat{\partial C}_{\bx}\right) \\
    &= \exp(i\int_C b +i\int_\Lambda \varphi\cup d\hat{m} + \frac{ik}{2}\int_\Lambda(\hat{m}\cup a - a\cup \hat{m} + a\cup_1 d\hat{m}))
\end{aligned}
\ee
This operator is invariant under small gauge transformations, and generates the desired large gauge transformation, namely
\be \label{eq:large-gauge-transformation-matrix}
U[C]^\dagger  a_{\bx,i} U[C] =  a_{\bx,i} + 2\pi  m_{\bx,i},\;\quad U[C]^\dagger  n_{\bx,12} U[C] =  n_{\bx,12} + (d m)_{\bx,12}
\ee
Throughout this paper, we use $U[C]$ and $U[m]$ interchangeably to indicate that $\hat{m} = \iota(\star\hat{C}) \in C^1(\rld; \mathbb{Z})$. This operator puts a large gauge constraint on the physical Hilbert space, namely
\be \label{eq:large-gauge-proj-evenk}
U[C]\ket{\psi} = \ket{\psi}
\ee
However, for open chains ($\partial C \neq 0$), one can show that this operator acts projectively on the Hilbert space, namely
\be \label{eq:projective-U}
\begin{aligned}
U[C_1]U[C_2] &= U[C_2]U[C_1] \exp\left(i\pi k \int_\Lambda (\hat C_1 \cup \hat C_2 -\hat C_2\cup \hat C_1)\right) 
\end{aligned}
\ee
For closed chains this phase is trivial. This implies that the physical state is ill-defined whenever $k\in\mathbb{Z}_\text{odd}$.  To properly define the physical Hilbert space, we can introduce a composite operator with similar commutation relation to even out the phase following \cite{Jacobson:2024hov}. It turns out that such an operator can be constructed from a pair of Majorana fermions $\gamma_p,\gamma_p'$ defined on plaquettes \cite{Chen:2017fvr,Bhardwaj:2016clt} with the anticommutation relation 
\be 
\{\gamma_{\bx,12},\gamma_{\by,12}\} = \{\gamma'_{\bx,12},\gamma'_{\by,12}\} = 2\delta_{\bx,\by} \,,\qquad \{\gamma_{\bx,12},\gamma'_{\by,12}\} =0
\ee 
We can define the fermionic parity operator $(-1)^{f_p}$ on a plaquette $p$ as
\be 
(-1)^{f_p} = i\gamma'_p \gamma_p
\ee 
One then defines a link operator $S_{\ell}$ using two Majorana fermions residing on a pair of plaquettes located to the left and right of a given edge, with respect to the branching structure of the lattice.
\be \label{eq:link-operator-s}
S_{\bx,1} := i\gamma_{\bx,12}\gamma'_{\bx-\hat 2,12}\,,\qquad S_{\bx,2} := i\gamma_{\bx-\hat 1,12} \gamma'_{\bx,12}
\ee 
On a cochain $m$ dual to the collections of ordering links $\lambda =\{\ell: m_\ell =1 \text{ mod  }2\}$  
. One can define the transparent fermionic line $S$ (see  \cite{Aasen:2017ubm,Wen:2024udn,Bhardwaj:2024ydc}), namely
\be \label{eq:transparent-fermionic-line-S}
S[m] = \exp\left( i\pi \sum_{\ell < \ell'\in\lambda} m_\ell\cup m_{\ell'} \right) \vec{\prod}_{\ell\in\lambda} S_\ell,\;\quad (S[m])^2=1
\ee 
where $\vec{\prod}$ denotes ordering product of links in $\lambda$.  Here, the sign prefactor is assigned so that the operator $S[m]$ is independent of the order of the links in $\lambda$. The algebra of this operator is 
\be \label{eq:S-fusion}
\begin{aligned}
S[m]S[m'] &= S[m']S[m] \exp\left[i\pi \int_\Lambda (\hat{m}\cup \hat{m}' + \hat{m}'\cup \hat{m})\right]\,,\\
&= S[m+m'] \exp\left[i\pi \int_\Lambda \hat m\cup \hat m'\right]
\end{aligned}
\ee 
This extra projective factor cancels with the one in \eqref{eq:projective-U}, provided that $\hat{m} = \iota(\star\hat{C})$, yielding $\int \hat{m} \cup \hat{m}' = \int \hat{C} \cup \hat{C}'$. The combined operator $U[C]S[m]$ can now act faithfully on the physical Hilbert space, since
\be \label{eq:resolve-Um-commutativity-1}\begin{aligned}
(S[m]U[C])(S[m']U[C']) &=(S[m']U[C'])(S[m]U[C])\\
&\quad\times\exp\left(i\pi (k+1)\int_\rld \hat{m}'\cup \hat{m} - i\pi(k-1) \int_\rld \hat{m}\cup \hat{m}'\right)
\end{aligned}
\ee 
A comprehensive review and derivations of properties of this operator can be found in e.g. \cite{Chen:2017fvr,Jacobson:2024hov} but there are important properties that should also be stated here. 
First, this $S[m]$ operator encodes the information regarding spin structure in its eigenvalue. Notice that due to the $\mathbb{Z}_2$ nature of $\gamma_p,\gamma'_p$, one can define an operator that count the mod 2 number of fermion on each plaquette via $(-1)^{f_p} = i\gamma'_p \gamma_p$ and the fermionic loop $S[m]$ with $d\hat{m}=0$ can be written as 
\be \label{eq:Sm-on-cycle}
S[m] = -(-1)^{\int_\Lambda \hat{m}\cup_1 f}
\ee 
The eigenstate $|\rho\ra $of this operator is the configuration where even (or odd) number of fermion is placed on the plaquettes passed by the dual curve $\star m$ with eigenvalue $-1$ (or $+1$) respectively. These configurations corresponds to the periodic (or antiperiodic) boundary condition of a spectator fermion along the cycle $\star m$.
The eigenvalue problem for $|\rho\ra$ and $S[m]$, with $[\star m]\in H_1(\tilde\Lambda;\mathbb{Z}_2), $ can then be written as 
\be 
S[m]|\rho\ra = (-1)^{q_\rho[m]} |\rho\ra
\ee 
where the convention is chosen such that $q_\rho[m] = 0$ for an antiperiodic (NS) and $q_\rho[m]=1$ for a periodic boundary condition (R). 
The projectivity \eqref{eq:S-fusion} implies that the eigenvalue $q_\rho[m]$ satisfy 
\be \label{def:quadratic-refin}
q_\rho[m_1] + q_\rho[m_2] = q_\rho[m_1+m_2] + \int_\Lambda \hat{m}_1\cup \hat{m}_2 \text{ mod }2
\ee 
This is nothing but a defining property of the quadratic refinement that is one-to-one correspondence of the spin structure on $\Lambda$ (see \cite{Karch:2019lnn} for a nice physicist's review). In case where $[\star m_1], [\star m_2]$ are two generators of $H_1(T^2;\mathbb{Z})$, the relation between spin structure and values of $(q_\rho[m_1],q_\rho[m_2])$ are summarised in Table \ref{table:quadratic-refinement-dict}.

\renewcommand{\arraystretch}{1.3}
\begin{center}
\begin{tabular}{|  c  |c|c|}
\hline
 $(q_\rho[m_1],q_\rho[m_2])$  & $\mathrm{Arf}[\rho]$ & $\rho$  \\
\hline
 $(1,1)$  & $1$ & R--R \\
 $(1,0)$  & $0$ & R--NS \\
 $(0,1)$  & $0$ & NS--R \\
 $(0,0)$  & $0$ & NS--NS \\
\hline
\end{tabular}
\captionof{table}{Correspondence between the quadratic refinement $q_\rho$, the spin structure $\rho$ of a torus as well as the Arf invariant $\text{Arf}[\rho]:= q_\rho[m_1] q_\rho[m_2]$. }
\label{table:quadratic-refinement-dict}
\end{center}
It is then convenient to define a spin structure dependence operator $S_\rho[m] := (-1)^{q_\rho[m]}S[m]$ such that $S_\rho[m]|\rho\ra = |\rho\ra$. The physical Hilbert space $|\psi\ra_\rho$ for a given spin structure $\rho$ is defined as 
\be \label{eq:large-gauge-proj-oddk}
U[C]S_\rho[m] |\psi\ra_\rho = |\psi\ra_\rho\qquad \text{ for }k \in \mathbb{Z}_\text{odd}\,.
\ee 

Consequently, there is a \emph{spin-charge} relation which related discrete flux $n_p$ mod 2 to $F_p$ on each plaquette upon projecting on the physical Hilbert space \eqref{eq:large-gauge-proj-oddk}. To understand this, consider an open curve $C$, with $m=\iota(\star C)$, the operator $S[m]$ can be written as 
\be \label{eq:s-operator-with-endpoint}
S[m] = e^{i\pi\int \hat{m}\cup_1 F} \prod_p (\gamma_p)^{|(d\hat{m})_p|}
\ee 
For an semi-infinite line $\star m$ with only one end point, we use the $\mathbb{Z}_2$ nature of the fermionic line $S[m]$ to write
\be 
U[C]|\psi\ra_\rho = S_\rho[m]|\psi\ra_\rho = (-1)^{ q_\rho[m] + \int_\Lambda m\cup_1 f} \gamma_p |\psi\ra_\rho
\ee 
The r.h.s. contain a fermion on a plaquette $p$ which shift the fermion number $f_p$ by one. 
Thus, for this theory, we have 
\be 
(-1)^{f_p} S_\rho[m] (-1)^{f_p} = - S_\rho[m]
\ee 
On the other hand, $U[C]$ only shift $n_p \to n_p + (d\hat{m})_p$ with $(d\hat{m})_p=1$, we have 
\be 
(-1)^{n_p} U[C] (-1)^{n_p} = - U[C]
\ee 
Thus, for the projection to the physical Hilbert space to be consistent, one must have 
\be \label{eq:spin-charge-reln}
n_p = f_p \text{ mod }2\,.
\ee 
With the Hilbert space defined in this way, one shall proceed to consider nontrivial operators in this system. 
\subsection{Topological and Non-topological Operators} \label{subsection:topological-operator}

The previous section provides a concrete construction of the Hilbert space for an arbitrary value of $k$. However, there is no interesting information encoded in these states. To extract more information from these states, we can construct topological operators on the lattice and use the defect insertion picture in solid torus quantization to build more useful states. We thus list out all important topological operators in this section.
The gauge invariant operators (under both large and small) are the framed Wilson line $W_q^\text{fr}[m=\iota(\star C)]$ define along the ladder configuration $m$ and its conjugated $V_\phi[C]$, which will be called the 't Hooft line due to it being the conjugate of the Wilson line 
\be \label{eq: top-line-final}
\begin{aligned}
W^\text{fr}_q[m]
&= \exp\left(\frac{iq}{2} \sum_{\bx} \left[ m\cup  a - a \cup m + 2\pi m\cup_1 n\right]\right)\\
V_\phi[C] &:= \exp(i\phi\sum_{(\bx,i)}\hat{C}_{\bx,i}b_{\bx,i})
\end{aligned}
\ee 
for a close loop $C$ with $q\in\mathbb{Z}$ and $\phi\in\mathbb{R}$. See Appendix \ref{app: quantize-compact} for further discussions on these operators, especially why they are topological with explanation on the coefficients $\phi$ and $q$.
The framed Wilson line corresponds to diagonally splitting the normal Wilson line by shifting it into the $\hat{s}$ direction, with each resulting line carrying a charge of $q/2$. The term $m\cup_1 n$ is the integer flux inserted between the two half-integer Wilson line, added so that $W^\text{fr}[m]$ is invariant under the large gauge transformation.\footnote{
 To see how this works, notice that, under the large gauge transformation generated by $m'\in C^1(\Lambda;\mathbb{Z})$, the framed Wilson line gain a phase
\be 
\begin{aligned}
\pi q\sum \left[m\cup m' - m'\cup m + m\cup_1 dm'\right] &= \pi q \sum \left[ m\cup m'- m'\cup m -(m\cup m' + m'\cup m) + dm\cup_1 m' \right]\\
&= -2\pi  q \sum m'\cup m  + \pi q \sum dm\cup_1 m'
\end{aligned}
\ee 
 Here, the second equality is obtained using the definition of the $\cup_1$ product in e.g. \eqref{eq:cup-commutator} and we can see that such phase becomes nontrivial unless $q\in \mathbb{Z}$ and $dm = 0$.
 Without the term $m\cup_1 n$, the phase generate by the large gauge transformation will simply be $\pi q \sum (m\cup m' - m'\cup m) \in \pi q \mathbb{Z}$ that may produce sign ambiguity in $W^\text{fr}_q$ when $q\in \mathbb{Z}_\text{odd}$. 
}
One can also combine these two operator into a single \emph{dyonic} operator, namely
\be \label{eq:combined-dyonic-operator}
\CW_{(q,\phi)}[C] := W_q^\text{fr}[\iota(\star C)]V_\phi[C]
\ee
To derive the algebra of the dyonic operator \eqref{eq:combined-dyonic-operator}, we first derive the commutation relation between $\int_C a := \sum_{(\bx,i)}\hat{C}_{\bx,i}a_{\bx,i}$ and $\int_{C}b := \sum_{(\bx,i)}\hat{C}_{\bx,i}b_{\bx,i}$, namely
\be
[\int_C a ,\int_{C'}b]= -2\pi i \text{Int}(C,C'),\; \quad [\int_C b,\int_{C'}b] = 2\pi ik \mathrm{Int}(C,C')
\ee 
where the term in the sum is nothing but the intersection number when $C,C'$ are two closed 1-chain, i.e., $\partial C=\partial C' = 0$. The precise definition of the lattice intersection number is provided in Appendix \ref{app:useful}, i.e., see equation \eqref{eq:def-intersection-number}. These commutation relations imply that 
\be \label{eq:algebra-dyonic-operator-W}
\CW_{(q_1,\phi_1)}[C_1]\CW_{(q_2,\phi_2)}[C_2] = \CW_{(q_2,\phi_2)}[C_2]\CW_{(q_1,\phi_1)}[C_1] (\CB_{(q_1,\phi_1),(q_2,\phi_2)})^{\text{Int}(C_1,C_2)}
\ee
where the braiding phase $\CB_{(q_1,\phi_1),(q_2,\phi_2)}$ \textbf{}is 
\be \label{eq:braidng-CW}
\CB_{(q_1,\phi_1),(q_2,\phi_2)}:= \exp\left[2\pi i (q_1\phi_2+q_2\phi_1-k\phi_1\phi_2)\right]
\ee 
These operators reproduce the same algebraic structure of the SymTFT results found in the continuum derivation \cite{Antinucci:2024zjp}.

\begin{figure}[htbp]
    \centering
    \includegraphics[width=1\linewidth]{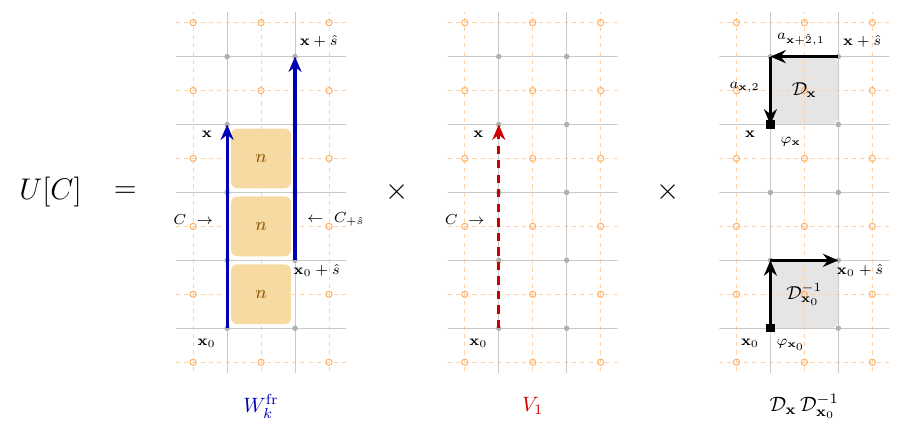}
    \caption{Large gauge operator $U[C]$ for $k$ even with endpoints $\bx_0$ and $\bx$ with local dyonic operators $\CD_{\bx_0}^{-1}$ and $\CD_\bx$, respectively. In this figure, we represent the framed Wilson line \textcolor{blue!70!black}{$W^\text{fr}_k$} on the lattice support $C$ and inside the framed Wilson line is filled with integer plaquettes \textcolor[HTML]{E8A317}{$n\in C^2(\rld;\mathbb{Z})$}.}
    \label{fig:bosonic-large-gauge-operator}
\end{figure}

\begin{figure}[htbp]
    \centering
    \includegraphics[width=1\linewidth]{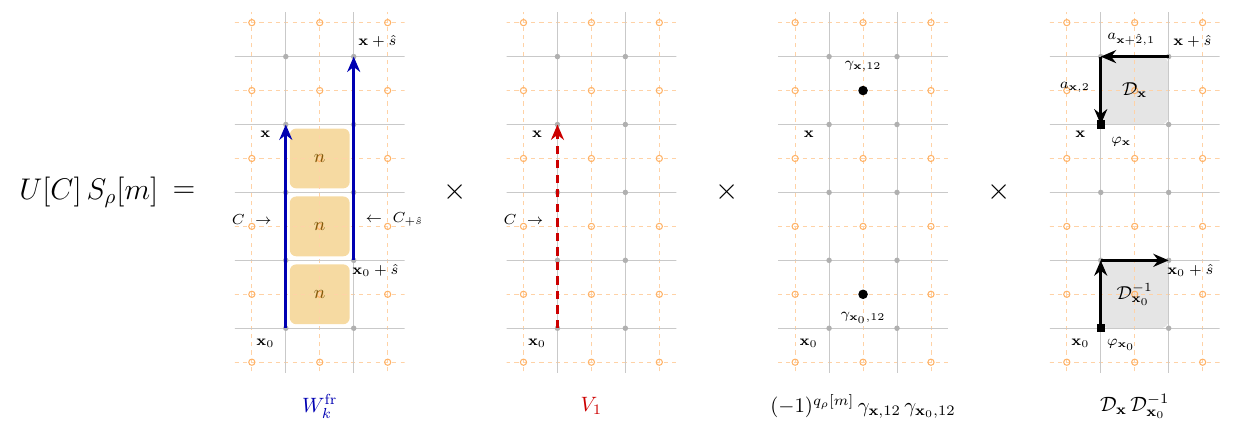}
    \caption{Large gauge operator $U[C]$ for odd $k$. It is almost identical to the bosonic one, but it has additional fermionic degree of freedom from odd Majorana fermions $\gamma_{\bx_0/\bx,12}$ living at the endpoints' plaquettes $p_{\bx_0/\bx}$, respectively.}
    \label{fig:fermionic-large-gauge-operator}
\end{figure}

There is another object in the theory, which is not transparent in the continuum, but it arises naturally in the lattice framework.\footnote{Localized degree of freedom at topological defects is a general feature of the modified Villain lattice construction, e.g., in the compact boson case, the interfaces obtained by flat gauging host non-compact edge modes at the defect sites \cite{Argurio:2026txf}.} It is important to remark that this operator is not topological as it is not gauge invariant. We will call this operator a \emph{local dyonic} operator $\CD$, given by
\be 
\CD_\bx :=\exp(i \tilde\varphi_\bx)
\ee 
which has a support on a plaquette due to the dress scalar $\tilde\varphi$ being a composition of point operator $\varphi$ and links $(a_{\bx,2} + a_{\bx+\hat 2,1})$. 
\be
\CD_\bx \to \CD_\bx e^{-i(\lambda_B)_\bx} e^{-\frac{ik}{2}((\lambda_A)_{\bx} + (\lambda_A)_{\bx+\hat s})}
\ee 
What makes this operator interesting is that it can live at the point from which the dyonic line $W^\text{fr}_k[C] V_1[C]$ emanates. For example, let's consider an open line $C_{\bx_0\to\bx}$ such that $\partial C =\bx-\bx_0$, where $\bx$ and $\bx_0$ are final and initial points, respectively. One can check that the dyonic operator $\CW_{(k,1)}[C_{\bx_0\to\bx}]$ transforms with additional factor, namely
\be
\begin{aligned}
    \exp(i(\lambda_B)_\bx + \frac{ik}{2}((\lambda_A)_\bx + (\lambda_A)_{\bx+\hat{s}}))\exp(-i(\lambda_B)_{\bx_0} - \frac{ik}{2}((\lambda_A)_{\bx_0} + (\lambda_A)_{\bx_0+\hat{s}})) \\
\end{aligned}
\ee
Therefore, the local dyonic operator $\CD$ can be attached to each endpoint of $\CW_{(k,1)}$ making the whole combination gauge invariant. Then it is obvious by construction that this operator can recast the definition of the large gauge operator \eqref{eq:large-gauge-operator-U} when $\partial C \neq 0$, as shown in Fig. \ref{fig:fermionic-large-gauge-operator}, namely\footnote{The prefactor $(-1)^{q_\rho}[m]$ is not defined on an open chain $m$. This does not affect the result since it can be absorbed into the state and cancels in every inner product. It can be defined properly by closing $m$ with a fixed reference path \cite{Jacobson:2024hov}.}
\be
\begin{aligned}
U[C]S_\rho[m] &= \left( W_k^\text{fr}[m] e^{i\pi k \int_\Lambda m\cup_1 n}\right) V_1[C]S_\rho[m]\mathcal{D}_\bx\mathcal{D}_{\bx_0}^{-1}\\
&= (-1)^{q_\rho[m]} W_k^\text{fr}[m] V_1[C] \gamma_{\bx,12}\mathcal{D}_\bx \gamma_{\bx_0,12}\mathcal{D}_{\bx_0}^{-1}
\end{aligned}
\ee
The second equality is obtained from expressing $S[m]$ as the Majorana end points and $(-1)^{\int_\Lambda m\cup_1 F}$ as in \eqref{eq:s-operator-with-endpoint} then eliminate the latter together with the phase $e^{i\pi k \int_\Lambda m\cup_1 n}$ from the framed Wilson line using the spin-charge relation \eqref{eq:spin-charge-reln}.
This equation will be a crucial ingredient for construction of the operators' fusion algebra in section \ref{section:u(1)-junction-computation}. The above operator corresponds to the odd $k$ case. For even $k$ case, the operator is much simpler since $S_\rho=1$, as shown in Fig. \ref{fig:bosonic-large-gauge-operator}. Using these topological and non-topological operators, we are now able to add more information to the Hilbert space in the next subsection.
%
\subsection{Hilbert space on a torus} \label{section:hilbert-space-construction}

\begin{figure}[thb]
\includegraphics{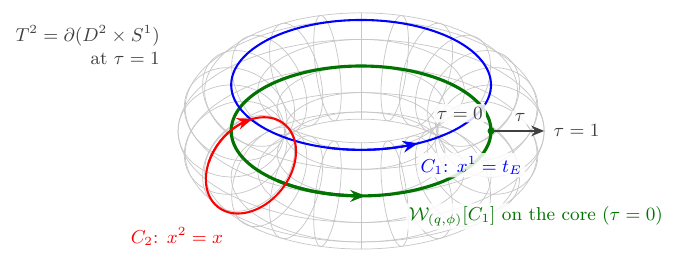}
\caption{The assignments of coordinate and line operator alignment on the solid torus $D^2\times S^1$.}
\label{fig:solid-torus}
\end{figure}

In this section, we are going to explicitly construct the physical Hilbert space of the theory with $U(1)$ global symmetry and show the spin structure dependency in the odd $k$ case. As shown previously that the physical Hilbert space are spanned by states that are invariant under both small and large gauge transformation. The definition of such states coincides with the Lagrangian algebra. Therefore, we can use the solid torus quantization picture to fill in the Hilbert space's details \cite{Witten:1988hf,Delmastro:2021xox,Chen:2024ulc}. To begin with, we consider a solid torus, $D^2\times S^1$, which is described by coordinate $(\tau,x^1,x^2)$
where $\tau\in [0,1]$ is the bulk radial direction with $\tau=1$ corresponding to the $T^2$ boundary where the theory \eqref{eq:action-BF-CS}  is placed and $\tau=0$ is the center of the solid torus. We define $C_1$ and $C_2$ to be the cycles along the longitudinal ($x^1$) and meridian
($x^2$) directions, respectively; $C_2$ bounds the disc $D^2$. The boundary theory is a
$(1+1)$-dimensional theory on $T^2$ whose Euclidean time runs along the longitude and
whose space is the meridian,
\begin{equation}
  \text{QFT:}\quad (t_E,\,x)\ \sim\ (C_1,\,C_2),
\end{equation}%
so that the Hilbert space of the boundary theory lives on the spatial circle $C_2=\partial
D^2$ and a line inserted along the centre of the solid torus, i.e.\ along $C_1$, labels a
sector of that Hilbert space. We orient $T^2$ by $(t_E,x)=(x^1,x^2)$, and the modular
parameter is $\tau=[C_1]/[C_2]$; the modular transformations $S:\tau\to-1/\tau$ and
$T:\tau\to\tau+1$ act on the cycles as in the standard convention. See Figure \ref{fig:solid-torus} for an assignment of the cycle and coordinate on the solid torus

To construct the Hilbert space, one is required to assume the existence of an \emph{empty} state $\ket{0}$, which corresponds to a solid torus state with no line insertion.
\be \label{eq:solid-torus-state}
\CW_{(q,\phi)}[C_1]\ket{0} = \ket{q,\phi}
\ee
The same can be applied for the $k\in\mathbb{Z}_\text{odd}$ but while the fermionic empty state has no line insertion, it has a non-trivial spin structure dependence which must be taken into account. The insertion of another line along the same $C_1$ direction, both topological lines can fuse into a single line, such that
\be
\CW_{\alpha'}[C_1]\CW_{\alpha}[C_1]= N_{\alpha',\alpha}^{\alpha'+\alpha}\CW_{\alpha+\alpha'}[C_1]
\ee
where $N_{\alpha',\alpha}^{\alpha'+\alpha}$ is the fusion matrix with a total charge of $\alpha'+\alpha$ where $\alpha$ denotes the pair $(q,\phi)$.
On the other hand, if one inserts a topological line along $C_2$, such a line can be contracted to identity. However, its contraction must pass the existing $\CW_\alpha[C_1]$ at the center, which corresponds to braiding phase, such that
\be
\CW_{\alpha'}[C_2]\ket{\alpha} = \CB_{\alpha'\alpha}\ket{\alpha}
\ee
with the braiding phase found in \eqref{eq:braidng-CW}. In the remaining section, we will start by explicitly construct the physical Hilbert space for even and odd $k$ using the solid torus quantization. Then we organise these states into a gapped boundary condition that corresponds to the $U(1)$ global symmetry in the SymTFT framework.
\subsubsection{Even $k$ Hilbert space} \label{subsection:even-k-hilbert-space}
Starting with the bosonic case, we can set $S_\rho[m]=1$ as spin structure is not required. One can check that the state \eqref{eq:solid-torus-state} is not invariant under large gauge transformation, namely
\be 
U[C_1]|q,\phi\ra = |q+k,\phi+1\ra 
\ee 
One can define a new basis where $U[C_1]$ acts trivially
\be 
|q,\phi\ra_\Omega : = \sum_{n\in\mathbb{Z}} | q+ kn,\phi+ n\ra,\;\quad U[C_1]|q,\phi\ra_\Omega =\sum_{n'\in\mathbb{Z}}| q+ kn',\phi+ n'\ra
\ee 
with the ground state $|0\ra_\Omega := |0,0\ra_\Omega$  invariant under both $U[C_1]$ and $U[C_2]$. The `ladder' operator acting on this state automatically inherit the periodicity $(q,\phi)\sim (q+k,\phi+1)$ 
\be 
\begin{aligned}
\CW_{(q',\phi')}[C_1]|q,\phi\ra_\Omega &= |q+q'-k\lfloor \phi+\phi'\rfloor,[\phi+\phi']\ra_\Omega\,,\\
\CW_{(q',\phi')}[C_2]  |q,\phi\ra_\Omega  &= \CB_{(q',\phi'),(q,\phi)} | q,\phi\ra_\Omega
\end{aligned}
\ee 
where $[a+b]_c$ is $a+b$ mod $c$ and $[a+b]:=[a+b]_1$. This basis $|q,\phi\ra_\Omega$ form a physical Hilbert space of this theory. 

\subsubsection{Odd $k$ Hilbert space} \label{subsection:odd-k-hilbert-space}
The situation is slightly different for odd $k$, where we are now required to construct the Hilbert space $\CH\ni |\psi\ra$ such that $U[C]S_\rho[m]|\psi\ra=|\psi\ra$. As we remarked previously, the empty state in the fermionic theory is not truly empty, since it initially inherits underlying fermionic configuration. The starting point is to consider the empty state $|0\ra_\rho$ which satisfies
\be 
U[C]S_\rho[m]|0\ra_\rho = |0\ra_\rho 
\ee 
One can try to employ the same strategy as we did in the bosonic case to construct this empty state $|0\ra$, on which there is no line operators inserted at the centre of the solid torus and that $(-1)^f=1$ on every plaquette. By \eqref{eq:Sm-on-cycle}, this is $S[m]|0\ra = -|0\ra$ for every closed $m$, indicating that it is a state with $R-R$ spin structure. To define a ground state for arbitrary spin structure, one  must consider replacing such state $|0\ra$ by a `fermionic empty state' $|\rho\ra$ with no line operator at the centre of the torus but with assignments of $\gamma_p,\gamma'_p$ of the plaquettes such that it satisfy
\be \label{eq:rho-empty-state-1}
 S_{\rho}[m]|\rho\ra = |\rho\ra \,,\qquad U[C_2]|\rho\ra = \CB_{(k,1),(0,0)}|\rho\ra = |\rho\ra
\ee 
%
The condition \eqref{eq:rho-empty-state-1} implies that the fermionic empty state is actually invariant under the large gauge transformation.
As a result, the physical Hilbert space can be constructed as
\be 
|0\ra_\rho := \sum_{n\in\mathbb{Z}}(U[C_1]S_\rho[m_1])^n|\rho\ra  \,,\qquad |q,\phi\ra_\rho = \CW_{(q,\phi)}[C_1]|0\ra_\rho
\ee 
and is invariant under any $U[C]S_\rho[m]$. Both states can only be defined once the spin structure $\rho$ is determined. A direct consequence of this state construction, under expansion of $|0\ra_\rho$, is that
\be 
\begin{aligned}
\CW_{(k,1)}[C_1]|0\ra_\rho = U[C_1]|0\ra_\rho =S_\rho[m_1]|0\ra_\rho
\end{aligned}
\ee 
realising the algebra $U[C]= S_\rho[m]$ upon acting on the state in the Hilbert space. The same conclusion can be made for any $|q,\phi\ra_\rho$ due to trivial braiding $\CB_{(q,\phi),(k,1)}=1$. This aligns well with the fact that one can just set the large gauge condition to be $U[C]S_\rho[m]=1$. This implies that on a closed chain $\d C=0$, the physical state labels $(q,\phi)$ are periodic under
\be \label{eq: periodic charge for U(1)}
(q,\phi) \sim \begin{cases}
(q+k,\phi+1) &\text{ for even }k\\
(q+2k,\phi+2) &\text{ for odd }k
\end{cases}
\ee 
Our lattice computation exhibits the same results as shown in \cite{Antinucci:2024zjp}, but with a properly defined spin structure.
\subsubsection{Lagrangian algebra of $U(1)$ theory and SymTFT interpretation} \label{subsection:u(1)-lagrangian-algebra}
From the previous section, we now have access to the physical Hilbert space of the $U(1)-\mathbb{R}$ SymTFT. To realize the Hilbert space of the theory with a $U(1)$ global symmetry, one has to wrap the center ($C_1$ direction) of the solid torus with $U(1)$ charges $q\in\mathbb{Z}$. This results in the $U(1)$ Lagrangian algebra corresponding to the gapped boundary state.
We are mainly interested in the Dirichlet boundary state \cite{Brennan:2024fgj}. In this section, we follow the procedure to construct the gapped boundary state $|\CL\ra$ from the Lagrangian algebra in \cite{Chen:2024ulc}. 
This state can be constructed by placing the line $\CL = \bigoplus_a \CN_a\CW_a$, with $\{ \CW_a\}$ be maximally commuting set of line operators $\CW_{(q,\phi)}$ and $\CN_a$ are multiplicities of the line $\CW_a$, at the centre of the solid torus. 
For a review of Lagrangian algebra and gapped boundary state with application to SymTFT see e.g. \cite{Bhardwaj:2023ayw}. A more recent generalisation to fermionic gapped boundary condition by dressing some operators with a transparent fermionic line, which turns out to be $S[m]$ in our case. A category theory treatment of this approach can be found in  \cite{Bhardwaj:2024ydc,pirsa_PIRSA:24030089}, see also \cite{Lou:2020gfq,You:2023gay,Wen:2024udn} for different approach to find gapped fermionic boundary conditions.
For even $k$ case, the corresponding Lagrangian algebra is given by
    \be \label{eq:even-lagrangian-state}
    |\CL\ra  =\sum_{q\in \mathbb{Z}} |q,0\ra_\Omega = \sum_{r\in \mathbb{Z}_k} \sum_{\phi \in \mathbb{Z}} |r,\phi\ra_\Omega
    \ee 
    with $|nk,0\ra_\Omega = |0,-n\ra_\Omega$ ensured by $U[C]|\psi\ra = |\psi\ra$ for all $|\psi\ra \in \CH$. Similarly, the gapped boundary state for the odd $k$ case must be given by
    \be \label{eq:odd-lagrangian-state}
    |\CL_\rho\ra = \sum_{q\in\mathbb{Z}} |q,0\ra_\rho = \sum_{r\in\mathbb{Z}_k} \left[ \sum_{\phi\in\mathbb{Z}_\text{even}}|r,\phi\ra_\rho + S_\rho[m_1]\sum_{\phi\in\mathbb{Z}_\text{odd}}|r,\phi\ra_\rho \right]
    \ee 
    which has an explicit dependence on the spin structure. Notice that the fermionic line operator $S_\rho$ is attached to the line operator $\CW_{(r,\phi)}$ with $\phi$ in $\mathbb{Z}_\text{odd}$ by construction of the state $|q,0\ra_\rho$. This is an explicit construction of the situation in \cite{Bhardwaj:2024ydc} where the transparent fermionic line, referred to as $\pi-$line in their notation, is attached to the line operator with topological spin $\theta = -1$ in order for the latter to condensed on the gapped boundary.

To see that this Lagrangian algebra actually realizes the usual $U(1)$ global symmetry, one can see that upon acting by $\CW_{(q,\phi)}[C]$ on $|\CL\ra$, we find that, for an integer $q$
\be 
\CW_{(q,\phi)}[C_2] |\CL\ra = V_\phi[C_2] |\CL\ra 
\ee 
namely the Wilson line $W_q[C]$ can be absorbed into $|\CL\ra$ due to $\CB_{(q,0),(q',0)} = \CB_{(q,0),(0,\phi\in \mathbb{Z})} = 1$ while $W_q[C_1]|\CL\ra = |\CL\ra$ by construction.  Combining this with the periodicity of the braiding phase \eqref{eq: periodic charge for U(1)}, we find that $V_\phi \sim V_{\phi+1}$ upon acting on $|\CL\ra$. Thus the parameter $\phi \in \mathbb{R}$ got projected down to $\mathbb{R}/\mathbb{Z}\cong U(1)$, indicating that $V_\phi$ acted as a topological line operator with $U(1)$ parameter $\phi$, just like a symmetry defect that generates the $U(1)$ global symmetry. Same conclusion can also be found for $|\CL_\rho\ra$ for odd $k$.

Once we defined the Lagrangian algebra, one can check that upon inserting $V_\phi$ with $\phi\in U(1)$ along another direction $C_2$, the resulting algebra is $U(1)$ by the braiding phase. Therefore, these gapped boundary states correspond to the global $U(1)$ symmetry on the boundary. This boundary condition will be used to construct the partition function and categorical data of the theory with anomalous $U(1)$ symmetry and its $\mathbb{Z}_2$ subgroup in Section \ref{section:mod-4-anomaly}. Alternative boundary conditions corresponds to theory with gauged discrete subgroup of anomalous $U(1)$ symmetry will be discussed in Section \ref{section:gaugin-anomaly}.

\section{$U(1)$ Theory Junctions computations} \label{section:u(1)-junction-computation}

From the previous section, we have access to the Hilbert space and operators defined on the lattice using the modified Villain construction. These operators allow us to perform certain manipulations on the lattice, which we refer to as \emph{move-and-split} operations, first introduced in \cite{Kawagoe:2019jbx,Kawagoe:2021gqi}. These operations can be used to extract categorical data directly from the microscopic setting. This information is essential for specifying a TQFT and can also be computed in the continuum formulation. The results of this section will provide understanding in lattice manipulation and will be used to study Gu-Wen system, which can be obtained as a sub-symmetry of the $U(1)$ global symmetry, in the following section.
 
\begin{figure}[htbp]
    \centering
    \includegraphics[width=0.8\linewidth]{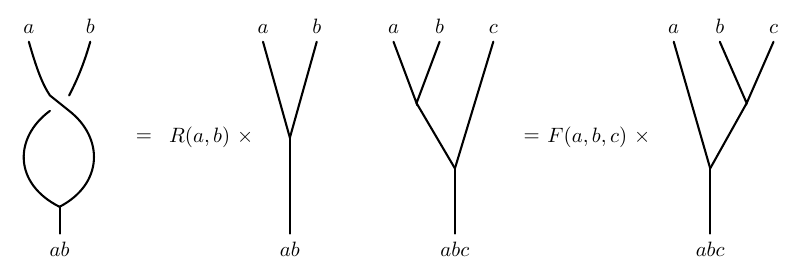}
    \caption{\textbf{Left:} $R-$move corresponds to exchanging two charges $a,b$. \textbf{Right:} $F-$move corresponds to fusing charged particles in different order. In our work, time evolution is taken to be upward.}
    \label{fig:general-r-f-move}
\end{figure}

The first operation, the $R-$move, amounts to exchanging two particles, resulting in a phase known as the `half-braiding' phase. The second operation, the $F-$move,\footnote{In continuum, the computation of $F-$move is highly non-trivial since it characterizes the associativity of fusion or splitting of topological operators; thus, the suitable choices of fusion bases are required. Such bases are less direct in the continuum compared to the lattice setting.} encodes the associativity of operators in the theory, which results from fusing charges in different orders, as shown in Fig. \ref{fig:general-r-f-move}. Therefore, in this section, we define a set of move-and-split operators for both even and odd $k$. The necessary ingredient is access to the Lagrangian algebra defining the Hilbert space, which will become clear in the following discussion. 
Moreover, in this section, we focus on the continuous $U(1)$ group. In the bosonic case, the
corresponding $F$-symbol has been obtained from the domain walls of a chiral boson edge
\cite{Kawagoe:2021gqi}, from the fusion operators of the modified Villain compact boson
\cite{Seifnashri:2023dpa}, and from the topological defect network \cite{Jia:2025uun}. For odd $k$ this result is obtained from the boundary side in \cite{Else:2014vma,Okada:2025kie}.

\begin{figure}
    \centering
    \includegraphics[width=0.7\linewidth]{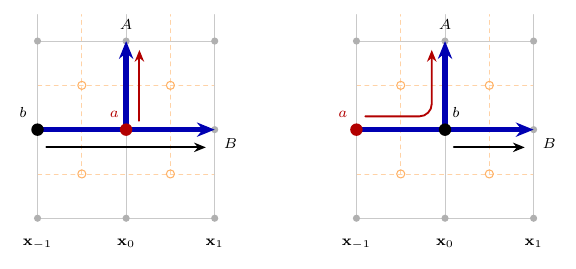}
    \caption{Both configurations start from the split $(N^{(b),(a)}_{ab})_{\bx_{-1}}$ (left) or $(N^{(a),(b)}_{ab})_{\bx_{-1}}$ (right) at $\bx_{-1}$, which leaves the first label at $\bx_{-1}$ and puts the second at $\bx_0$. \textbf{Left:} charge \textcolor{red!70!black}{$a$} moves vertically from $\bx_0$ to $A$, and then charge $b$ moves horizontally from $\bx_{-1}$ through $\bx_0$ to $B$. \textbf{Right:} charge \textcolor{red!70!black}{$a$} moves from $\bx_{-1}$ to $\bx_0$, charge $b$ moves from $\bx_0$ to $B$, and then \textcolor{red!70!black}{$a$} moves vertically to $A$. Both configurations end with \textcolor{red!70!black}{$a$} at $A$ and $b$ at $B$; only in the left one does the path of $b$ cross the leg of \textcolor{red!70!black}{$a$} at $\bx_0$. Their overlap is the exchange phase $R(\textcolor{red!70!black}{a},b)=\langle R_R|R_L\rangle$.}
    \label{fig:r-move}
\end{figure}

\begin{figure}
    \centering
    \includegraphics[width=0.6\linewidth]{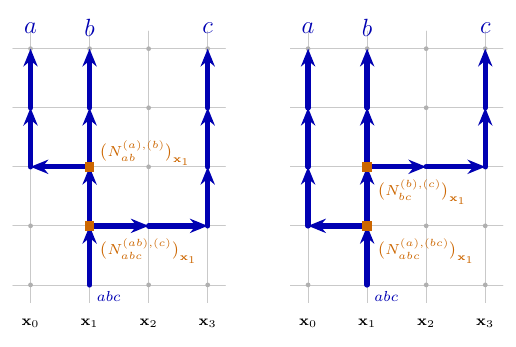}
    \caption{The $F-$move operation on lattice corresponds to splitting/fusing charged operators in different order. \textbf{Left:} The left configuration with (\textcolor{blue!70!black}{$abc$})$\to$(\textcolor{blue!70!black}{ab},\textcolor{blue!70!black}{c}) at the split junction \textcolor{orange!80!black}{$N_{abc}^{(ab),(c)}$} and the second split with (\textcolor{blue!70!black}{$ab$})$\to$(\textcolor{blue!70!black}{a},\textcolor{blue!70!black}{b}) at the split junction \textcolor{orange!80!black}{$N_{ab}^{(a),(b)}$}. \textbf{Right:} The right configuration is another possible splitting order where (\textcolor{blue!70!black}{$abc$})$\to$(\textcolor{blue!70!black}{a},\textcolor{blue!70!black}{bc}) at \textcolor{orange!80!black}{$N_{ab}^{(a),(b)}$} and then with (\textcolor{blue!70!black}{$bc$})$\to$(\textcolor{blue!70!black}{b},\textcolor{blue!70!black}{c}) at \textcolor{orange!80!black}{$N_{bc}^{(b),(c)}$}. Both split junctions, represented by \textcolor{orange!80!black}{$\blacksquare$} in each panel are located at $\bx_1$, which are chosen for convenience in computations.}
    \label{fig:f-move-lattice}
\end{figure}

To begin, we define the $R-$ and $F-$moves on the lattice, as shown in Figures \ref{fig:r-move} and \ref{fig:f-move-lattice}, respectively. It is clear that both pieces of data can be extracted once we know how charged objects can be moved, fused, and split on the lattice. We begin with the first ingredient: the move operator, which can be defined as $M^{a}_{\bx_{i+1},\bx_{i}}$, acting on the Hilbert space as
\be \label{eq:move-operator-def}
M^{a}_{\bx_{i+1},\bx_{i}}\ket{...,a_{\bx_i},...} \propto \ket{...,a_{\bx_{i+1}}}
\ee
which amounts to move operator of charge $a$ from sites $\bx_{i}$ to $\bx_{i+1}$ and $\bx_i$ denotes the horizontal position of the i-th site. Note that, in this work, charge $a$ actually represents a pair of charges, namely $a:=(q_a,\phi_a)$. The order of action is taken from right to left as normal. Another important related operation is $M^{a}_{\bx+\hat{2},\bx}$ that moves particle of charge $a$ vertically from site $\bx$ to its vertically nearest site. We will see that the move operators are the same for both cases of $k$. 

Another operation is the split operation, which is the inverse of the conventional fusion operation and depends on the parity of $k$, since the fusion junction can involve an $S$ operator for odd $k$, as shown in Fig.\ref{fig:fusion-split-even-odd-k}. In general, the split operator acts on the Hilbert space as
\begin{equation} \label{eq:split-operator-def}
    \begin{aligned}
        (N_{ab}^{a,b})_{\bx_i} \ket{...,(ab)_{\bx_i},...} \propto \ket{...,a_{\bx_i},b_{\bx_{i+1}},...} 
    \end{aligned}
\end{equation}
which amounts to split operator with a total charge $ab$ to two operators with charges $a$ and $b$, respectively. One can now realize the $R-$move by moving two charges, denoted as $a$ and $b$ in a different order, as shown in Fig. \ref{fig:r-move}.
\begin{equation}\label{def:state-R}
    \begin{aligned}
        \ket{R_L} &= M^{b}_{\bx_{1},\bx_{-1}}M^{a}_{\bx_0+\hat{2},\bx_0}(N^{b,a}_{ab})_{\bx_{-1}}\ket{ab_{\bx_-1}},\\
        \ket{R_R} &= M^{a}_{\bx_0+\hat{2},\bx_0}M^{b}_{\bx_1,\bx_0}M^{a}_{\bx_0,\bx_{-1}}( N^{a,b}_{ab})_{\bx_{-1}}\ket{ab_{\bx_-1}}
\end{aligned}
\end{equation}
where $\ket{ab_{\bx_0}}$ is the state corresponding to the fusion channel of $a$ and $b$. Then the $R-$move phase is realized as $R(a,b) = \bra{R_R}\ket{R_L}$. 
To realize $F-$move, one has to perform series of moves and splits as shown in Fig. \ref{fig:f-move-lattice}. The basic property of $F-$move is to detect non-associativity in fusion or splitting ordering as shown by the left and right configurations in Fig. \ref{fig:f-move-lattice}. Then $F-$move can be realized as a phase $F(a,b)=\bra{F_R}\ket{F_L}$, where each configuration reads
\begin{equation} \label{def:state-F}
    \begin{aligned}
        \ket{F_L} &= M^{b}_{\bx_1,\bx_2}M^a_{\bx_0,\bx_1}(N^{a,b}_{ab})_{\bx_1} M^{c}_{\bx_3,\bx_2}(N^{ab,c}_{abc})_{\bx_1}\ket{abc_{\bx_1}}, \\
        \ket{F_R} &= M^{c}_{\bx_3,\bx_2}(N^{b,c}_{bc})_{\bx_1}M^{bc}_{\bx_1,\bx_2}M^a_{\bx_0,\bx_1}(N^{a,bc}_{abc})_{\bx_1}\ket{abc_{\bx_1}}
     \end{aligned}
\end{equation}

%

Another important property of a continuous $U(1)$ theory that one has to keep in mind is that the charges can be fractional upon fusion or splitting. We introduce a \emph{carry} along with a modulo operation
\be \label{eq:carry-modulo-operation}
\begin{aligned}
    p_c(a,b) := \lfloor \frac{a+b}{c} \rfloor,\;\quad  a+b := [a+b]_c + cp_c(a,b)
\end{aligned}
\ee
with $[\cdot]\coloneqq[\cdot]_1$ is a modulo 1 operation and $c\in\mathbb{Z}$. This operation is important because it demonstrates the splitting of a `charge' during fusion into a fractional part $[a+b]_c$ and an integer `carry' (or spill-over) part $cp_c(a,b)$ (see \cite{Okada:2025kie} for more details).
For example, if $a,b\in {[0,1)}$ then the fractional part is $[a+b]$ and the integer part occurring at the fusion junction is given by
\be
\begin{aligned}
    p(a,b) = \begin{cases}
                0\;\text{if}\; a+b< 1, \\
                1\;\text{if}\; 1\leq a+b < 2
             \end{cases}
\end{aligned}
\ee
\subsection{$R-$ and $F-$ move for general $k$ without Lagrangian algebra} \label{subsection:r-and-f-move-no-lagrangian}

In this section, we explain how to extract the move and split operator and how to use them to compute $R-$ and $F-$ move for a general $\CW_{(q,\phi)}$. The computation is very straightforward with no complications beyond commuting the line operators among themselves and with the local dyonic operator $\CD$. The next section will explain how to turn these categorical data to those on the gapped boundary that corresponds to $U(1)$ symmetry. 

\subsubsection{Even $k$ case}
\label{subsection:move-and-split-even-k}
\begin{figure}[htbp]
    \centering
    \includegraphics[width=0.6\linewidth]{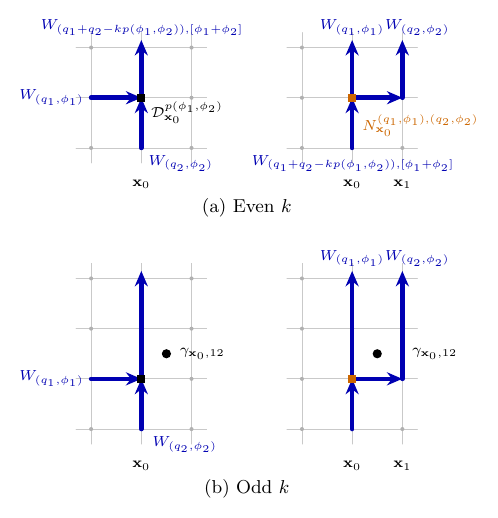}
    \caption{Fusion (left) and splitting (right) operations on the lattice for even $k$ \textbf{(a)} and odd $k$ \textbf{(b)}. In each panel, the physical charge lines \textcolor{blue!70!black}{$W_{(q_1,\phi_1)}$} and \textcolor{blue!70!black}{$W_{(q_2,\phi_2)}$} meet at a fusion vertex \textcolor{black}{$\mathcal{D}^{p(\phi_1,\phi_2)}_{\bx_0}$}, represented by a \textcolor{black}{$\blacksquare$}, and combine into the composite line \textcolor{blue!70!black}{$W_{(q_1+q_2-kp(\phi_1,\phi_2)),[\phi_1+\phi_2]}$}. Alternatively, read in reverse, the composite line splits back into its two constituents, where \textcolor{orange!80!black}{$\blacksquare$} represents the \textcolor{blue!70!black}{$N$} operator (right panels). For odd $k$ \textbf{(b)}, the composite line is additionally dressed by the plaquette operator \textcolor{red!70!black}{$S_\rho$} with the solid black circle $\bullet$ as $\gamma_p$, the Majorana fermion on the plaquette.}
    \label{fig:fusion-split-even-odd-k}
\end{figure}
In even $k$ case, fusion of operators whose charges have periodicity with respect to \eqref{eq: periodic charge for U(1)} is given by
\begin{equation} \label{eq:fusion-junction-even-k}
\CW_{q_1,\phi_1}\CW_{q_2,\phi_2} = W^\text{fr}_{q_1+q_2}V_{[\phi_1+\phi_2]}(V_1)^{p(\phi_1,\phi_2)}
\end{equation}
To properly work on fusion operator, we consider the large gauge constraint with $\bx_0$ and $\bx$ being initial and final endpoints, respectively. This constraint implies that
\be
(V_1)^{p(\phi_1,\phi_2)} = (W^\text{fr}_ke^{\pi i k\int m\cup_1 n})^{-p(\phi_1,\phi_2)}[C]\CD^{p(\phi_1,\phi_2)}_{\bx_0}\CD^{-p(\phi_1,\phi_2)}_{\bx}
\ee
To derive fusion and split, one considers the following situation. When lines $a$ and $b$ merge at $\bx_0$ into $ab$, the junction carries a local dyonic operator $\CD^{p(\phi_1,\phi_2)}$, and vice versa. The configuration of local dyonic operator is not a matter of choice, but it is dictated by orientation of large gauge operator and charge conservation at the fusion and splitting junctions. Thus the fusion algebra at the fusion junction $\bx_0$ is described by
\be 
\begin{aligned}
\CW_{q_1,\phi_1} \CW_{q_2,\phi_2} = \CW_{q_1+q_2-k p(\phi_1,\phi_2), [\phi_1+\phi_2]} \CD^{p(\phi_1,\phi_2)}_{\bx_0}
\end{aligned}
\ee 
One can then define the move operator $M_{\bx_2,\bx_1}^{(q,\phi)}$, which moves the composite charge $(q,\phi)$ from site $\bx_1$ to site $\bx_2$ as
\be 
\begin{aligned}
M^{(q,\phi)}_{\bx_2,\bx_1} &:= \CW_{(q,\phi)}[C] \qquad \text{where }\CC \text{ is a horizontal line from } \bx_1 \to \bx_2\\
\end{aligned}
\ee 
The split operation, defined as \eqref{eq:split-operator-def}, also contains a point-like dyonic operator at the split junction, namely
\be
(N^{(q_1,\phi_1)(q_2,\phi_2)}_{q_1+q_2-kp(\phi_1,\phi_2),[\phi_1+\phi_2]} )_\bx =\CD^{-p(\phi_1,\phi_2)}_\bx M^{(q_2,\phi_2)}_{\bx+\hat 1,\bx}
\ee 
where $\CD$ and $M$ here commute with each other as they are defined from the same site.\footnote{Note that ordering in $\CD^{-p(\phi_1,\phi_2)}_\bx M^{(q_2,\phi_2)}_{\bx+\hat 1,\bx}$ does not matter since for the path $\bx\to\bx+\hat 1$, the only non-zero link is related to $\hat{C}_{\bx,1}=1$. This results in $s_\bx=\hat{C}_{\bx,2}-\hat{C}_{\bx-\hat{1},1}=0$. Therefore, they always commute.}
The next important ingredient is the algebra between the move operators that can be regarded as a direct consequence of moving two dyonic lines $\CW$ passing each other.
\be \label{eq:general-move-dynon-algebra}
\begin{aligned}
M^{(q_1,\phi_1)}_{\bx'_1,\bx_1} M^{(q_2,\phi_2)}_{\bx'_2,\bx_2} &=M^{(q_2,\phi_2)}_{\bx'_2,\bx_2} M^{(q_1,\phi_1)}_{\bx'_1,\bx_1} \exp\Big( 2\pi i (q_1\phi_2 + q_2\phi_1-k\phi_1\phi_2) \text{Int}(C_{\bx_1\to \bx_1'},C_{\bx_2\to \bx_2'})\Big) \\
M^{(q_1,\phi_1)}_{\bx',\bx}\CD_\by^{-p(\phi_1,\phi_2)} &= \CD_\by^{-p(\phi_1,\phi_2)}M^{(q_1,\phi_1)}_{\bx',\bx}\exp(-\frac{2\pi i q_1p(\phi_1,\phi_2)}{2}s(\bx\to\bx',\by))
\end{aligned}
\ee 
where we define, for path $C$ from site $\bx$ to site $\bx'$,
\be
\begin{aligned}
    \text{Int}:= \text{Int}(C_{\bx_1\to \bx_1'},C_{\bx_2\to \bx_2'}),\quad s(\bx\to\bx',\by) := \hat{C}_{\by,2} - \hat{C}_{\by-\hat{1},1}
\end{aligned}
\ee
where the first object $\text{Int}$ denote number of intersections between the supports of two moves operators, i.e., see Fig. \ref{fig:r-move}.
The first relation is obtained directly from the braiding phase while the second relation is obtained from the canonical commutation relation between $n$ and $\tilde\varphi$.

Using the above ingredient, one can proceed to compute the $R-$symbol from $\la R_R|R_L\ra$. The detailed derivations can be found in Appendix \ref{app:r-and-f-detailed} and one finds 
\be \label{eq:general-r-move-phase}
R(\CW_{(q_1,\phi_1)},\CW_{(q_2,\phi_2)}) = \exp\left( i\pi  \left( q_1\phi_2 + q_2\phi_1 - k\phi_1\phi_2 \right) \right)
\ee 
One can then produce the topological spin by setting $\phi_1=\phi_2 =\phi$ and $q_1=q_2=q$, namely
\be \label{eq:U(1)-topological-spin-phase}
\theta(\CW_{(q,\phi)}):= \exp(2\pi i(q\phi-\frac{k\phi^2}{2}))
\ee
This phase is a result of exchanging two identical charges, and has a nice interpretation from Dehn twist as a topological charge.
Furthermore, the product of two `half-braid' yield a `complete-braiding phase, namely
\be \label{eq:U(1)-full-braiding-phase}
\begin{aligned}
\CB(\CW_{(q_1,\phi_1)},\CW_{(q_2,\phi_2)}) &= R(\CW_{(q_1,\phi_1)},\CW_{(q_2,\phi_2)}) R(\CW_{(q_2,\phi_2)},\CW_{(q_1,\phi_1)}) \\
&= \exp\left( 2\pi i (q_2 \phi_1 + q_1\phi_2 - k \phi_1\phi_2) \right)
\end{aligned}
\ee 
which is the same braiding phase obtained earlier in the manuscript (as well as in the literature \cite{Antinucci:2024zjp}).
Similarly, the $F-$move can be directly obtained as
\be \label{eq:general-f-move-phase}
F(\CW_{(q_1,\phi_1)}\CW_{(q_2,\phi_2)}\CW_{(q_3,\phi_3)}) = \exp\left(\pi i q_1p(\phi_2,\phi_3) \right) 
\ee 
 It is then convenient to introduce a shorthand notation $\CW_a$ for the dyonic line where $a$ represent the pair $(q,\phi)$ and the composite line $ab$ carries charge $(q_{ab} := q_a +q_b - kp(\phi_a,\phi_b), [\phi_a+\phi_b])$. One can check that it satisfies the pentagon equation:
\be \label{eq:ptentagon-keven}
F(a,b,c) F(a,bc,d) F(b,c,d) = F(ab,c,d) F(a,b,cd)
\ee 
By writing $F(a,b,c) = \exp(2\pi i \alpha(a,b,c))$, the pentagon identity can be recast in terms of group cohomological relation such that $\delta \alpha = 0$ mod $1$, since
\be \label{eq:fake-Gu-Wen}
\begin{aligned}
\delta \alpha &:= \alpha(b,c,d) - \alpha(ab,c,d) + \alpha(a,bc,d) \\
&\quad- \alpha(a,b,cd) + \alpha(a,b,c) \\
&= \frac{k}{2} \lfloor \phi_a+\phi_b\rfloor \lfloor \phi_c+\phi_d\rfloor
\end{aligned}
\ee 
vanishes modulo 1 for $k\in\mathbb{Z}_\text{even}$ where $\delta$ is the coboundary operator in the group cohomology. 

Note however that the form of $F-$ and $R-$ symbols are not unique as one could redefine a junction through expressing $N^{a,b}_{ab}$ with different basis. This amounts to assigning an arbitrary phase $e^{2\pi i \beta(a,b)}$ to the split operator $N^{a,b}_{ab}$, see e.g. \cite{Kitaev:2005hzj,Barkeshli:2014cna}. This amounts to a change in $\alpha(a,b,c)$ by $\alpha\to \alpha+\delta \beta$ where the shift is the coboundary, in the group cohomology, of $\beta$, defined as 
\be \label{eq:coboundary-group}
\delta\beta(a,b,c)=\beta(b,c)+\beta(a,bc)-\beta(a,b)-\beta(ab,c)
\ee 
This property of $N^{a,b}_{ab}$ and therefore $\alpha$, is crucial in determining the anomaly of the system. In plain words, the phases $\alpha \ne 0 $ mod 1 measure the failure of the associativity of operator $\CW_a$ and if redefinition through $\beta$ cannot remove it, the global symmetry generated by $\CW_a$ is anomalous. This property is crucial in determining the gaugeable subgroup of $U(1)$ symmetry in Section \ref{section:gaugin-anomaly}.

This implies that the non-associativity is classified by element in $H^3(BG^\delta,U(1))$ as expected in the bosonic theory \cite{Chen:2011pg}. See e.g. \cite{Dijkgraaf:1989pz,Chen:2011pg} for a practical introductions to group cohomology. 

\subsubsection{Move-and-Split with odd $k$} \label{subsection:move-and-split-odd-k}

The structure of fusion algebra becomes richer due to the presence of $S_\rho[m]$ operator when $k$ is odd. The large gauge constraint requires
\be
U[C]S_\rho[m] = (-1)^{q_\rho[m]}W^\text{fr}_k[m]V_1[C]\gamma_{\bx,12}\CD_\bx\gamma_{\bx_0,12}\CD^{-1}_{\bx_0}\,,
\ee
which implies that
\be
(V_1)^{p(\phi_1,\phi_2)} = (-1)^{p(\phi_1,\phi_2)q_\rho[m]}(W^\text{fr}_{k})^{-p(\phi_1,\phi_2)}\CD_{\bx_0}^{p(\phi_1,\phi_2)}\gamma_{\bx_0}^{p(\phi_1,\phi_2)}\CD_\bx^{-p(\phi_1,\phi_2)}\gamma_{\bx,12}^{p(\phi_1,\phi_2)}
\ee
where $(\gamma_{\bx,12})^{-p(\phi_1,\phi_2)}=(\gamma_{\bx,12})^{p(\phi_1,\phi_2)}$.
Thus, the complete fusion algebra occurring at the fusion junction is given by
\begin{equation} \label{eq:fusion-junction-odd-k}
\begin{aligned}
\CW_{(q_1,\phi_1)}\CW_{(q_2,\phi_2)} &= W^\text{fr}_{q_1+q_2} V_{[\phi_1+\phi_2]} V_1^{p(\phi_1,\phi_2)}\\
&=\CW_{(q_1+q_2-kp(\phi_1,\phi_2),[\phi_1+\phi_2])}\CD_{\bx_0}^{p(\phi_1,\phi_2)}\gamma_{\bx_{0}}^{p(\phi_1,\phi_2)}
\end{aligned} 
\end{equation}
where the configuration of this Majorana fermion is shown in Fig. \ref{fig:fusion-split-even-odd-k}. The spin-dependent sign $(-1)^{q_\rho[m]}$ depends on the prescribed spin structure, and cancels in the inner product. Therefore, we will write fusion and splitting operators up to this spin-structure sign. 

The move operation is the same as the even $k$ case; however, the split junction has a subtler structure, namely
\be \label{eq:split-operator-odd-k}
(N^{(q_1,\phi_1),{(q_2,\phi_2})}_{q_1+q_2-kp(\phi_1,\phi_2),[\phi_1+\phi_2]})_\bx = \CD^{-p(\phi_1,\phi_2)}_{\bx} (\gamma_{\bx,12})^{p(\phi_1,\phi_2)} M^{(q_2,\phi_2)}_{\bx+\hat 1,\bx}
\ee 
It turns out that the computation of $R-$ and $F-$ move is identical to those in even $k$ since it commutes with $M^a_{\bx',\bx},\CD_\bx$ and, as $\gamma$ inside $|L\ra,|R\ra$ sit on the same plaquette, they are cancelled out in the overlap. So that one only needs to replace $\CD_\bx^{-p}$ to $\CD_\bx^{-p} \gamma_{\bx,12}^{p}$ and carrying out identical computations. It turns out that the form of $R-$ and $F-$symbols are identical to those presented in \eqref{eq:general-r-move-phase} and \eqref{eq:general-f-move-phase} due to the fact that $\gamma^2=1$ dropout of the overlap $\la R_R|R_L\ra$ and $\la F_R|F_L\ra$ in the computation of both symbols. 

The key distinction from the $k\in \mathbb{Z}_\text{even}$ case is the pentagon equation. As one can see from \eqref{eq:fake-Gu-Wen}, that it is not trivial whenever $k$ is odd. This is a concrete sign of a contribution from fermionic anomaly. We first notice that the exponentiated carry behaves as a map to $\mathbb{Z}_2$, namely
\be
\exp(i\pi p(\phi_a,\phi_b)): U(1)\times U(1) \to \mathbb{Z}_2
\ee
This indicates that $p$ has a property similar to cocycle. In particular, it is straightforward to see that $p$ is actually a cocycle as it vanishes upon the coboundary operation
\be 
\begin{aligned}
(\delta p)(\phi_1,\phi_2,\phi_3)&:= p(\phi_2,\phi_3)- p([\phi_1+\phi_2],\phi_3) + p(\phi_1,[\phi_2+\phi_3]) - p(\phi_1,\phi_2)\\
&=\lfloor \phi_2+\phi_3\rfloor - \lfloor [\phi_1+\phi_2]+\phi_3\rfloor + \lfloor \phi_1+[\phi_2+\phi_3]\rfloor - \lfloor\phi_1+\phi_2\rfloor\\
&=0
\end{aligned}
\ee 
where $\delta$ is the coboundary operator in the group cohomology. One can then check that the violation of the cocycle condition, $\delta \alpha=0$, for $k\in\mathbb{Z}_\text{odd}$ can be written as 
\be \label{eq:gu-wen-anomaly}
(\delta \alpha)(a,b,c,d) = \frac{k}{2} p(\phi_a,\phi_b) p(\phi_c,\phi_d) = \frac{1}{2}p(\phi_a,\phi_b) p(\phi_c,\phi_d) \text{ mod } 1
\ee 
This result indicates that $\alpha$ is not valued in cohomology class, but it is a more subtle object related to Gu-Wen anomaly \cite{Brumfiel:2016vpy,Gu:2012ib} introduced in Section \ref{section:introduction}. We will explore this theory in more detail in section \ref{subsection:stacking-rule} and discuss its relationship to the mod-4 anomalous phase of the partition function in section \ref{section:mod-4-anomaly}.

\subsection{Symmetry algebra inflow to the gapped boundary} \label{subsec:symalg-inflow}
\begin{figure}[tbh]
\includegraphics{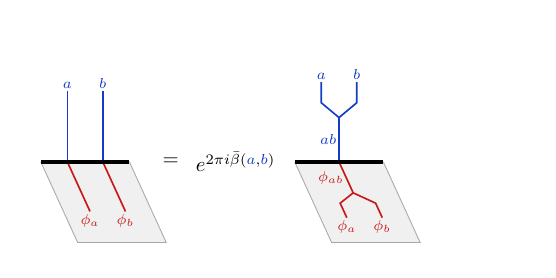}
\caption{Pictorial definition of the phase obtained upon combing the bulk split to the boundary fusion function in \eqref{def:M6jsymbol}}
\label{fig:M6jsymbol}
\end{figure}

\begin{figure}[tbh]
\includegraphics{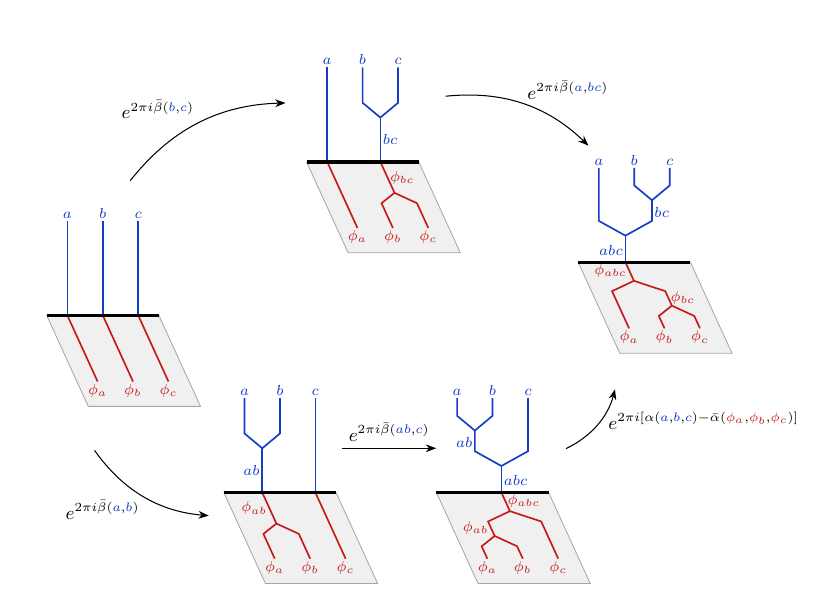}
\caption{Pictorial definition of the pentagon consistency condition where the phase on the arrow denotes the phase gain after a move at each step. For the phase gained on the upper and lower paths to agree, this implies the relation \eqref{eq:M6jpentagon}.}
\label{fig:M6jpentagon}
\end{figure}

To understand the symmetry of the boundary theory, one needs to understand how the junction is projected on the boundary. It turns out that a restricted class of junctions are compatible with the Lagrangian algebra \eqref{eq:even-lagrangian-state}, see e.g. \cite{Cong:2017ffh,Cheng:2020rpl,Zhang:2023wlu}. The essential idea is to distinguish the bulk split operator $N^{a,b}_{ab}$ to the boundary split operator $\bar N^{\phi_a,\phi_b}_{\phi_{ab}}$ related to one another upon the projection
\be \label{def:M6jsymbol}
N^{a,b}_c \to \bar N^{\phi_a,\phi_b}_{\phi_{ab}} := e^{2\pi i \bar\beta(a,b)} N^{a,b}_c
\ee 
for a specific phase $\bar\beta(a,b)$ that is compatible with the boundary condition\footnote{
This phase $\bar\beta(a,b)$ is a special case of the $M-6j$ symbol in \cite{Cong:2017ffh} where the fusion of the operators are group-like and that the multiplicity running over the dimensions of $\text{Hom}(\CW_a,V_{\phi_a})$ are all 1.
}.
Pictorially, this means that the bulk split junction can be projected down to the boundary and `annihilate' the fusion junction at the cost of the additional phase $e^{2\pi i \bar\beta(a,b)}$ as shown in Figure \ref{fig:M6jsymbol} 
This phase $\bar\beta(a,b)$ must satisfy the pentagon identities, which relates to F-symbol of the bulk and one that is projected on the boundary, shown in Figure \ref{fig:M6jpentagon}, 
\be 
\label{eq:M6jpentagon}
\bar\alpha(\phi_a,\phi_b,\phi_c) = \alpha(a,b,c) + \delta\bar\beta(a,b,c):= \Omega(a,b,c)\,.
\ee 
where $\bar\alpha$ is related to the boundary F-symbol via $\bar F(\phi_a,\phi_b,\phi_c) = e^{2\pi i \bar\alpha(\phi_a,\phi_b,\phi_c)}$.

\begin{figure}[thb]
\includegraphics{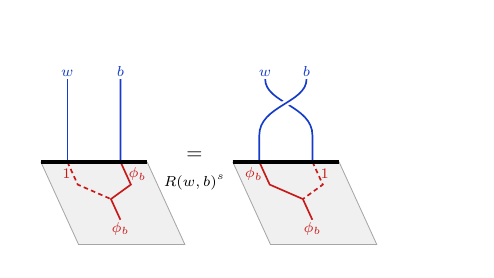}
\caption{Pictorial definition of \eqref{eq:M6j-braid}}
\label{fig:M6j-braid}
\end{figure}

There are two ingredients to determine the form of $\bar\beta(a,b)$. First, the boundary F-move is trivial if the framed Wilson line $W^\text{fr}_q = \CW_{(q,0)}$ can be attached to the bulk line $\CW_a,\CW_b,\CW_c$ in Figure \ref{fig:M6jpentagon} as it becomes trivial operator upon projection on the boundary $|\CL\ra_\Omega$. In the language of \eqref{eq:M6jpentagon} this means that 
\be \label{eq:condensate-Omega}
\Omega(w,b,c) = \Omega(a,w,c) = \Omega(a,b,w) = 0  \quad \text{ mod }1
\ee 
where $w$ represents the line operator $\CW_{(q,0)}$. With the result for $\alpha(a,b,c)$ in \eqref{eq:result-Fsymbol-bulk} and the ansatz for $\bar\beta(a,b)$ to be 
\be 
\bar\beta(a,b) = \frac{1}{2}\left(\epsilon_0 q_aq_b + \epsilon_1 q_a \phi_b + \epsilon_2 q_b\phi_a + \epsilon_3 \phi_a\phi_b\right) + g(\phi_a,\phi_b) \text{ mod }1
\ee 
The coefficient $\epsilon_0 = 0$ mod 2 since the boundary junction is trivial when the bulk line $\CW_{(q_a,0)},\CW_{(q_b,0)}$ becomes trivial upon projection on the boundary. Shuffling $w$ into each slot in $\Omega(a,b,c)$, we use the definition of $\Omega$ in \eqref{eq:M6jpentagon} to find 
\be 
\begin{aligned}
\Omega(w,a,b) = \frac{1+\epsilon_1}{2} q p(\phi_a,\phi_b)\,,\quad
\Omega(a,w,b) = 0 \,,\quad 
\Omega(a,b,w) = -\frac{\epsilon_2}{2} q p(\phi_a,\phi_b)\,.
\end{aligned}
\ee 
The consistency condition \eqref{eq:condensate-Omega} implies that $\epsilon_1 \in \mathbb{Z}_\text{odd}$ and $\epsilon_2 \in \mathbb{Z}_\text{even}$. Now plug in this solution to $\bar\alpha$, we find\footnote{
For a lack of better notation, despite being written as $\delta[\phi_a\phi_b]$, this object depends on $\phi_a,\phi_b,\phi_c$ as in $\bar\alpha$. Here we use this shorthanded notation to represent 
\be 
\delta[\phi_a\phi_b] = -\phi_a\phi_b - \phi_{ab} \phi_c + \phi_b \phi_c + \phi_a \phi_{bc} = -\phi_a p(\phi_b,\phi_c) +p(\phi_a,\phi_b)\phi_c
\ee 
with $\phi_{ab}:= \phi_a +\phi_b - p(\phi_a,\phi_b)$. The same goes for $\delta[g(\phi_a,\phi_b)] = \delta g(\phi_a,\phi_b,\phi_c)$ defined in \eqref{eq:coboundary-group}.
}  
\be 
\bar\alpha(\phi_a,\phi_b,\phi_c) = \frac{\epsilon_1-\epsilon_2}{2} k \phi_a p(\phi_b,\phi_c) + \delta \left[ \frac{k\epsilon_1+\epsilon_3}{2}\phi_a\phi_b + g(\phi_a,\phi_b)\right]
\ee 
for some arbitrary odd number $\epsilon_1-\epsilon_2$.
The second constraint is the boundary version of the braiding relation of \cite{Cong:2017ffh} i.e. when the bulk line $\CW_{q,0}$ condensed on $|\CL\ra$  its endpoint
is trivial and can be slid freely along the boundary past the endpoint of a second line. Pictorially, this is represented by the diagram in Figure \ref{fig:M6j-braid} or equivalently 
\be \label{eq:M6j-braid}
e^{2\pi i\bar\beta(w,b)} = e^{2\pi i \bar\beta(b,w)} R(w,b)^{s} \,,\qquad s= \pm 1
\ee 
where the sign $s=\pm 1$ represents whether the line $\CW_{(q,0)}$ and $\CW_b$ are braided clockwise or counterclockwise before being projected down on the state $|\CL\ra_\Omega$.
With the solution of $R(a,b)$ in \eqref{eq:R-symbol-bulk}, we can solve \eqref{eq:M6j-braid} to find that $\epsilon_2-\epsilon_1 = \pm 1$. Thus, upon combining the two requirements of $\bar\beta(a,b)$, we find that the F-symbol $\bar F(\phi_a,\phi_b,\phi_c) = e^{2\pi i \bar\alpha(\phi_a,\phi_b,\phi_c)}$ on the gapped boundary $|\CL\ra_\Omega$ to be 
\be \label{eq:F-move-projected}
\bar\alpha(\phi_a,\phi_b,\phi_c) = \pm \frac{k}{2} \phi_a p(\phi_b,\phi_c) + \delta \left[ \frac{k\epsilon_1+\epsilon_3}{2}\phi_a\phi_b + g(\phi_a,\phi_b)\right] \quad \text{ mod }1\,.
\ee 
The term $\delta[...]$ can be removed upon redefining the junction $V_{\phi_a} V_{\phi_b} = e^{2\pi i[...]} V_{\phi_{ab}}$ on the boundary operator. Removing it results in the F-symbol for $U(1)$ symmetry defect operator reported in \cite{Jia:2025uun,Okada:2025kie}. Notice also that for $k\in\mathbb{Z}_\text{even}$, we find $\delta\bar \alpha =0$ signifying the group cohomological nature of this anomaly. The sign $\pm k$ does not affect the cohomological nature of this anomaly but turns out to have physical meaning for odd $k$ as we shall demonstrate in Section \ref{section:mod-4-anomaly}. 

There are two additional important remarks. 
\begin{itemize}
\item 
First, there is no notion of $R-$symbol on the boundary simply because there is no way to braid two lines in two dimensions. One can extract the topological spin of the line operator by simply consider operator $V_{\phi}$ outside the Lagrangian algebra and employ the modular transformation as explained in Section \ref{subsection:modular-and-topological-spin}. 
\item Second, the above results and calculations also hold for odd $k$. This is simply due to the fact that the fusion/split junction only contains the Majorana fermion when $p(\phi_a,\phi_b) =1 $ mod 2 regardless of whether it is $\CW_a \times \CW_b \to \CW_{ab}$ or $V_{\phi_a}\times V_{\phi_b} \to V_{\phi_{ab}}$ junction. Thus, if they exist, both bulk and boundary junction $N^{a,b,}_c$ and $\bar N^{\phi_a,\phi_b}_{\phi_{ab}}$ carry the same number of Majorana making the phase $e^{2\pi i \bar\beta(a,b)}$ fermionic even. As a result, the key relation \eqref{eq:M6jpentagon} between the bulk and boundary $F-$symbols $\bar\alpha = \alpha + \delta \bar \beta$ is unaffected. Either by using $\delta \alpha = \delta\bar\alpha$ or directly compute from \eqref{eq:F-move-projected}, one finds the Gu-Wen type relation generalised to parameter $\phi\in \mathbb{R}/\mathbb{Z}\cong U(1)$ namely
\be \label{eq:GuWen-U1}
\delta \bar\alpha(\phi_a,\phi_b,\phi_c,\phi_d) = \pm \frac{k}{2} p(\phi_a,\phi_b) p(\phi_c,\phi_d)
\ee 
\end{itemize}

Let us also briefly comment on the group-cohomological nature of $\bar\alpha$ and $p$,
following the interpretation of \cite{Jia:2025vrj,Jia:2025uun}. First of all,
$\bar\alpha(\phi_a,\phi_b,\phi_c)$ cannot be a class in $H^3(BU(1);\mathbb R/\mathbb Z)$: the
classifying space $BU(1)\cong\mathbb{CP}^\infty$ has homology only in even degrees, all of it
free, so this group vanishes. Nor is $\bar\alpha$ a continuous $\mathbb R/\mathbb Z$-valued
function of the labels, as it jumps across $\phi_b+\phi_c=1$. Likewise, the carry
$p(\phi_a,\phi_b)$ cannot be a continuous $\mathbb Z_2$-valued cochain, since such a function on
the connected group $U(1)\times U(1)$ is constant, while $p$ jumps from $0$ to $1$. The
appropriate interpretation, pointed out in \cite{Jia:2025vrj,Jia:2025uun}, is to regard them as
arbitrary functions $U(1)^n\to A$, i.e.\ as cochains in $C^n(BU(1)^\delta;A)$ of the group $U(1)$
equipped with the discrete topology \cite{Milnor1983}. The space $BU(1)^\delta$ is the
classifying space of flat $U(1)$ bundles \cite{KamberTondeur1967,Jia:2025vrj}. This
interpretation is natural in our setup, as the $U(1)-\mathbb{R}$ SymTFT enforces the flatness of
the background gauge field through the Gauss law \eqref{eq:Guasslaw-U1R}.

In this language, the carry is the coboundary of the real-valued $1$-cochain
$[\phi]\colon U(1)\to[0,1)$, namely
$p(\phi_a,\phi_b)=[\phi_a]+[\phi_b]-[\phi_a+\phi_b]=(\delta[\,\cdot\,])(\phi_a,\phi_b)$, so that
$\delta p=0$. It is, however, not exact, even mod~$2$. Evaluating on the subgroup $\mathbb Z_2=\{0,\tfrac12\}\subset U(1)$, the point $(0,0)$ gives
$c(0)=p(0,0)=0$, whereas $(\tfrac12,\tfrac12)$ gives $1=p(\tfrac12,\tfrac12)=2c(\tfrac12)-c(0)
\equiv c(0)\bmod 2$, i.e.\ $c(0)=1$, a contradiction. Hence $p\bmod2$ represents
a non-trivial class in $H^2(BU(1)^\delta;\mathbb Z_2)$. This is in line with the result of
Milnor \cite{Milnor1983} that, for Lie groups whose identity component is solvable, such as
$U(1)$, the map $BG^\delta\to BG$ induces an isomorphism in cohomology with finite coefficients;
under this isomorphism $[p\bmod2]$ corresponds to $c_1\bmod2\in H^2(BU(1);\mathbb Z_2)$. For even
$k$, $\bar\alpha$ is a cocycle, and its class lies in the image of $\pm\tfrac k2\,c_1^2$ under the
injection $H^4(BU(1);\mathbb Z)\hookrightarrow H^3(BU(1)^\delta;\mathbb R/\mathbb Z)$
\cite{Jia:2025vrj,Jia:2025uun}. For odd $k$, \eqref{eq:GuWen-U1} suggests that the anomaly is
characterised by the pair
\begin{equation}
  \big(\bar\alpha,\,p\big)\in C^3(BU(1)^\delta;\mathbb R/\mathbb Z)\times
  Z^2(BU(1)^\delta;\mathbb Z_2)\,,\qquad \delta\bar\alpha=\tfrac12\,p\cup p\,,
\end{equation}
with $[p]\neq0$ and $\bar\alpha$ defined up to $\delta\bar\beta$. This is the Gu--Wen structure of
\eqref{eq:fermionic-anomaly-condition} with $\mathbb Z_2$ replaced by $U(1)^\delta$, and it
restricts to that data on $\mathbb Z_2\subset U(1)$. We do not attempt a classification here.

\subsection{Modular Transformation and Topological Spin} \label{subsection:modular-and-topological-spin}
An alternative way to extract the topological spin is to directly perform the Dehn's twist or modular-$\texttt{T} \in SL(2,\mathbb{Z})$ transformation. The main purpose of this section is not just to develop the modular transformation on the SymTFT Lagrangian algebra, but to later allows us to perform mapping torus computation for extracting bordism information in section \ref{section:mod-4-anomaly}. Since there are some subtleties in implementing the modular-$\mathtt{T}$ transformation in the SymTFT's method we are about to introduce, it is important to recall some basic facts. 
Recall that the modular transformations on the torus, whose non-contractible cycles are $([C_1],[C_2])$, where $[C_{1,2}]$ being the two generators of $H_1(T^2;\mathbb{Z})$, corresponding to longitudinal and meridian cycle of the torus respectively, are given by
\begin{equation} \label{eq:modular-transformation-on-c}
    \begin{aligned}
        \mathtt{S}: ([C_1],[C_2]) \to (-[C_2],[C_1]),\quad \mathtt{T}: ([C_1],[C_2]) \to ([C_1]+[C_2],[C_2])
    \end{aligned}
\end{equation}
%
We have another notion of cycles resulted from having a special cochain $\hat{m}$ whose supports are defined by \eqref{eq:support-of-m-cochain}. The induced modular transformations are then satisfy
\begin{align} \label{eq:modular-transformation-on-m}
\mathtt{S}: (m_1,m_2) \to (-m_2,m_1), \quad \mathtt{T}: (m_1,m_2) \to (m_1+m_2,m_2)
\end{align}
Suppose that the representations of modular transformations on the Hilbert space are denoted by $U_{\mathtt{S},\mathtt{T}}$ for modular-$\mathtt{S}$ and modular-$\mathtt{T}$, respectively. In this section, we focus only on the modular-$\mathtt{T}$ transformation, since the modular-$\mathtt{S}$ transformation is more straightforward. To see how modular transformations act on the Lagrangian algebra, one can examine their action on the dyonic operators $\CW$. For example, the action of $U_\mathtt{T}$ on $\CW[C]$, reads
\begin{equation} \label{eq:action-of-modular-T}
\begin{aligned}
    U_\texttt{T} \CW_{(q,\phi)}[C_1] (U_\texttt{T})^{-1} &= \CW_{(q,\phi)}[C_1+C_2] = \theta_{(q,\phi)} \CW_{(q,\phi)}[C_1] \CW_{(q,\phi)}[C_2] \\
    U_\texttt{T} \CW_{(q,\phi)}[C_2] (U_\texttt{T})^{-1} &= \CW_{(q,\phi)}[C_2]
\end{aligned}
\end{equation}
where $\theta{(q,\phi)}$ is half the braiding phase between the same dyonic line that is coincide with the $R-$symbol $R(\CW_{(q,\phi)},\CW_{(q,\phi)})$ namely
\be 
\theta_{(q,\phi)} = \exp\left(i\pi (2q\phi-k\phi^2)\right)
\ee 
These operators can then act on the Lagrangian algebra under the assumption that an arbitrary empty state defined together with the Lagrangian algebra is an eigenstate of the Dehn twist operator $U_\texttt{T}$. Since the results depend on the parity of $k$, we split the derivation into two parts.
\subsubsection{Modular Transformation on even $k$}
For even $k$, one can simply apply the Dehn's twist on the bosonic states defined in the earlier section and finds that
\be 
\begin{aligned}
U_\texttt{T}|q,\phi\ra = \theta_0 \theta_{q,\phi} |q,\phi\ra,\;\quad U_\texttt{T}|q,\phi\ra_\Omega = \theta_0 \theta_{q,\phi}| q,\phi\ra_\Omega
\end{aligned}
\ee 
where we used the fact that $\theta_{k,1} =1$ for even $k$. Therefore, the topological spin can be expressed as an expectation value of $U_\texttt{T}$ of the state $|q,\phi\ra_\Omega$ 
\be 
\theta_{(q,\phi)} = \frac{_\Omega\la q,\phi| U_\texttt{T}|q,\phi\ra_\Omega}{_\Omega \la 0| U_\texttt{T}|0\ra_\Omega}
\ee 
Here, we denote topological spin of the ground state by $\theta_0$ which is unfixed by this procedure and could potentially related to chiral central charge as shown in \cite{Xu:2024hyo,Xu:2026ygx}.

\subsubsection{Modular Transformation on odd $k$}
For odd $k$, one first recast the definition \eqref{eq:rho-empty-state-1} by considering $S[m]$ as a measurement of quadratic refinement, namely
\be 
S_{\rho}[m]|\rho\ra = |\rho\ra \quad\Longrightarrow\quad S[m]|\rho\ra = (-1)^{q_\rho[m]}|\rho\ra
\ee 
Then the modular transformation acts on these operators as 
\be 
\begin{aligned}
U_\texttt{T}^{-1}S[m_2]U_\texttt{T}= S[m_2],\;\quad U_\texttt{T}^{-1}S[m_1]U_\texttt{T} = - S[m_1]S[m_2]
\end{aligned}
\ee 
where the minus sign follows from \eqref{eq:S-fusion} with $\int_\Lambda \hat{m}_1\cup \hat{m}_2 = 1$. Then one can use $S[m]$ to measure the spin structure in of the state $U_\mathtt{T}\ket{\rho}$, namely
\be 
\begin{aligned}
S[m_2] U_\texttt{T}|\rho\ra = (-1)^{q_\rho[m_2]}U_\texttt{T}|\rho\ra,\quad S[m_1] U_\texttt{T}|\rho\ra =  (-1)^{q_\rho[m_1] + q_\rho[m_2] +1} U_\texttt{T}|\rho\ra 
\end{aligned}
\ee 
Therefore, the modular transformation $\mathtt{T}$ changes the prescribed spin structure as $\mathtt{T}:(q_\rho[m_1],q_\rho[m_2])\to (q_\rho[m_1]+q_\rho[m_2]+1,  q_\rho[m_2])$ per convention in e.g. \cite{Karch:2019lnn}. Comparing this with the dictionary in Table \ref{table:quadratic-refinement-dict}, we find that this indeed generate the orbit of the $Mp_1(\mathbb{Z})$, which not the conventional mapping class group. Because of this direct observation, one recovers the crucial fact that the \texttt{T} transformation is not well-defined in spin TQFT due to the fact that $|\rho\ra$ is not an eigenstate of $U_\texttt{T}$. One can, nevertheless. compute the eigenvalue of $U_\texttt{T}^2$ which can be extracted via 
\be 
\theta^2[\CW_{(q,\phi)}]= \frac{_\rho \la q,\phi| U_\texttt{T}^2|q,\phi\ra_\rho  }{_\rho \la 0| U_\texttt{T}^2|0\ra_\rho}
\ee 
Again, we choose this definition to subtract the vacuum contribution. 
\section{The Gu-Wen anomaly from $U(1)-\mathbb{R}$ SymTFT} \label{section:mod-4-anomaly}

In the previous sections, we introduced the $U(1)$ Lagrangian algebra capable of describing theories in $(1+1)$ dimensions, including how one can perform modular transformations on them. In this section, we focus on the $\mathbb{Z}_2$ subgroup of the fermionic $U(1)$ theory obtained by explicitly break into the subgroup. The corresponding global symmetry is then  
$\mathbb{Z}_2^f\times\mathbb{Z}_2$, where $\mathbb{Z}_2^f$ is a fermionic parity symmetry generated by $(-1)^f$.
It has been known that this theory has an anomaly that is best captured by the cobordism computation \cite{Kapustin:2014dxa} (see also \cite{Kaidi:2019tyf,Tachikawa:2018njr} for the APS $\eta-$invariant perspective and \cite{Fidkowski:2009dba}). Namely, the anomaly is captured by\footnote{The generator $\nu=1$ of $\mathbb{Z}_8$ can be realized by a single non-chiral Majorana fermion $(\chi_L,\chi_R)$ satisfying \eqref{eq:total-partition-function-cft}, whose partition function picks up phase $\exp(-2\pi i /8)$  under $\mathtt{S}\mathtt{T}^2\mathtt{S}^{-1}$  \cite{Kawabata:2024gek}. The  Gu-Wen system corresponds to $\nu=2\mod 4$, and $\nu=4$ is purely bosonic, as we show in section \ref{subsection:stacking-rule}. Embedding $\mathbb{Z}_2\hookrightarrow U(1)$ requires the charged fermions to be complex, which can be constructed using pairs of $(\chi_L,\chi_R)$. This doubles $\nu$, resulting in squaring the phase; thus, one can at best realize anomalous mod--4 phase from the $U(1)$ SymTFT.}
\be
\mathrm{\Omega}^\text{Spin}_3(B\mathbb{Z}_2) \cong \mathbb{Z}_8
\ee
The main goal of this section is to compute this information using the SymTFT framework developed in the previous section. The first step is to realize such a system and show correspondence between the SymTFT results and the usual CFT results of such system. Our strategy is to employ the \emph{anomaly interplay} mechanism introduced in \cite{Davighi:2020uab}, which uses mapping tori \cite{Witten:2019bou} and the action of modular transformations on the Lagrangian algebra established in the previous section. Since a single Weyl-Majorana fermion cannot be embedded into the $U(1)$ with spin structure theory, at least a pair of Weyl-Majorana fermions is required to be embedded into a single complex Weyl fermion. Thus, the best we can extract is the mod-4 anomalous phase. In particular, such an anomalous theory is referred to as the Gu-Wen anomaly ($\nu=2\mod{4}$) \cite{Gu:2012ib}, which is captured by supercohomology. After realizing the Gu-Wen anomalous phase, with a hint from the $F-$move derived in section \ref{subsection:move-and-split-odd-k}, it is possible to trivialize the $F-$move by stacking identical Gu-Wen theories. We thus derive the stacking rules for SPT phases in $(1+1)$ dimensions using the modified Villain approach and show that all anomalies can be trivialized by stacking four copies of the Gu-Wen theory. The results of this section agree with those obtained by restricting the symmetry to an interval in
\cite{Else:2014vma}, where the endpoint operators of $\mathbb Z_2$ acting on the right-moving
fermions of a Dirac edge were shown to be fermionic, with four copies being anomaly free, and in
\cite{Okada:2025kie} for general $\mathbb Z_n\subset U(1)$.

To understand the mapping torus construction, one can observe the fermionic CFT description of the $\nu=1$ theory which consists of one left-handed and right-handed Weyl-Majorana fermions $(\chi_L,\chi_R)$ on a torus with the central charges $c_L = c_R = 1/2$. The $\mathbb{Z}_2$ and $\mathbb{Z}_2^F$ acted on these variables as 
\be \label{eq:total-partition-function-cft}
\mathbb{Z}_2:\quad (\chi_L,\chi_R)\longrightarrow(\chi_L,-\chi_R)\,,\quad \mathbb{Z}_2^f:\quad (\chi_L,\chi_R)\longrightarrow(-\chi_L,-\chi_R)\,.
\ee 
 The partition function of the theory in this $\mathbb{Z}_2$ twisted background can be written in terms of left- and right-handed free fermion partition function as\footnote{
Following the standard convention in \cite{Ginsparg:1988ui,Lin:2019hks}, the partition function of a Weyl-Majorana fermion on a torus with complex parameter $\tau$ are 
\be
Z^L_{a-b}(\tau) = \text{Tr}_{a}\!\left[U_b\, q^{L_0-\frac{1}{48}}\right] ,\qquad
Z^R_{a-b}(\bar\tau) = \text{Tr}_{a}\!\left[U_b\, \bar q^{\bar L_0-\frac{1}{48}}\right] ,
\qquad q=e^{2\pi i\tau},\quad \bar q=e^{-2\pi i\bar\tau},
\ee
for left- and right-handed fermion respectively and the shift corresponds to $c=\bar{c}=1/2$. The indices $\{ a,b\}= \text{NS,R}$ label the spin sector with $\Tr_a$ label the sector of the Hilbert space and $U_b$ is the operator insertion where $U_\text{NS}=1$, $U_\text{R}=(-1)^f$.
}
\be 
Z_\text{total}(\tau,\bar\tau) = Z^L_\text{NS-NS} Z^R_\text{NS-R}(\tau,\bar\tau)
\ee 
Then the mod-4 anomaly can be derived using the facts that under the modular transformation $\texttt{T}: \tau \to \tau+1$ and $\texttt{S}: \tau \to -1/\tau$. For the modular $\texttt{T}-$transformation, we have
\begin{subequations}
\be
\begin{aligned}
Z^L_\text{NS-NS}(\tau+1) &= e^{-\frac{i\pi}{24}} Z^L_\text{NS-R}(\tau)\,,\\
Z^L_\text{NS-R}(\tau+1)&= e^{-\frac{i\pi}{24}} Z^L_\text{NS-NS}(\tau)\,,\\
Z^R_\text{R-NS}(\tau+1)&= e^{-\frac{i\pi}{12}} Z^R_\text{R-NS}(\tau)    \,,
\end{aligned}
\ee 
and 
\be 
Z_\text{NS-NS}^L(-\frac{1}{\tau}) = Z_\text{NS-NS}^L(\tau) \,,\quad Z_\text{NS-R}^L(-\frac{1}{\tau}) = Z^L_\text{R-NS}(\tau) \,,\quad Z^L_\text{R-NS}(-\frac{1}{\tau}) = Z^L_\text{NS-R}(\tau)
\ee 
\end{subequations}
With $\sigma=-1/\tau$ and $\rho=\sigma+2$, so that $\texttt{ST}^2\texttt{S}^{-1}: \tau \to -1/\rho$ yield
\be 
\begin{aligned}
  Z^L_{\rm NS\text{-}NS}(-\tfrac1\rho)&=Z^L_{\rm NS\text{-}NS}(\sigma+2)
  =e^{-\frac{i\pi}{12}}Z^L_{\rm NS\text{-}NS}(-\tfrac1\tau)=e^{-\frac{i\pi}{12}}Z^L_{\rm NS\text{-}NS}(\tau),\notag\\
  Z^L_{\rm NS\text{-}R}(-\tfrac1\rho)&=Z^L_{\rm R\text{-}NS}(\sigma+2)
  =e^{\frac{i\pi}{6}}Z^L_{\rm R\text{-}NS}(-\tfrac1\tau)=e^{\frac{i\pi}{6}}Z^L_{\rm NS\text{-}R}(\tau),\notag
\end{aligned}
\ee 
and the phase from the right-moving block is obtained via the complex conjugate $Z^R_{\rm NS\text{-}R}(\bar\tau)=\overline{Z^L_{\rm NS\text{-}R}(\tau)}$, which gives
$e^{-i\pi/6}$ under $\texttt{ST}^2\texttt{S}^{-1}$. 
Therefore the total partition function transforms as
\be 
Z_\text{total}(-1/\rho,-1/\bar\rho) = e^{-\frac{2\pi i}{8}} Z_\text{total}(\tau,\bar\tau)\;\;\text{per}\;\;(\chi_L,\chi_R)\;\;\text{pair}\;\;(\nu=1)\,.
\ee 
For Gu-Wen theory, we need two pair of $(\chi_L,\chi_R)$; thus, the phase total partition function transforms with a phase $\exp(-\frac{2\pi i }{4})$.
As pointed out in \cite{Davighi:2020uab}, this phase can be realized using the mapping torus construction. 
There, the torus is parametrised by two angles $\theta\sim\theta+2\pi$,
$\chi\sim\chi+2\pi$, places the $\mathbb Z_2$ background with holonomy around the
cycles that wrap $\theta$ or equivalently a $\mathbb Z_2$ defect along the cycles that
wrap $\chi$, with antiperiodic spin structure along $\theta$ and glues the mapping torus by $(\theta,\chi,1)\sim(\theta-2\chi,\chi,0)$. Our $\mathbb Z_2$ operator $V_{1/2}[C_2]$ is inserted along space, so the
holonomy is around the time cycle: $\theta\leftrightarrow t_E\leftrightarrow C_1$,
$\chi\leftrightarrow x\leftrightarrow C_2$, and their gluing map acts on homology as
\begin{equation}
  C_1\mapsto C_1,\qquad C_2\mapsto C_2-2C_1 ,
\end{equation}
which indicating that the mapping torus of \cite{Davighi:2020uab} is
the modular transformation $\texttt{ST}^2\texttt{S}^{-1}$ acting on the torus with the $\mathbb Z_2$ operator
inserted, as also used in \cite{Kawabata:2024gek}. It is a double Dehn twist
along the time cycle, the cycle that carries the holonomy; it preserves the inserted
operator and reads off the anomalous phase of its sector.

\subsection{SymTFT Construction of the $\mathbb{Z}_2\times\mathbb{Z}_2^f$ Theory} \label{subsection:symtft-for-gu-wen}

To describe the $\mathbb{Z}_2$ symmetry of QFT, we can simply restrict the bulk field to a $\mathbb{Z}_2$ valued one. In continuum \cite{Antinucci:2024zjp}, this is done by simply changing $a\to \pi a$ and $da/2\pi \to \mathrm{Bock}(a)$ with $\mathrm{Bock}(\cdot)$ being an appropriate Bockstein homomorphism. At the level of line operator, this amounts to the following. A $\mathbb{Z}_2$ gauge field $a$ reduces the charge $q$ to its mod 2 valued i.e. $\CW_{(q,0)} = W^\text{fr}_q \sim W^\text{fr}_{q+2}$ with $\CW_{(2,0)}$ set to the identity. The line that braided non-trivially with $W^\text{fr}_2$ must also be removed, which, via \eqref{eq:U(1)-full-braiding-phase}, means $\CB_{(2,0),(q,\phi)} = e^{4\pi i \phi}\ne 1$ or $\phi \notin \frac{1}{2}\mathbb{Z}$. The remaining lines are therefore those with label $q\in \mathbb{Z}_2$ and $\phi \in \{0,\frac{1}{2},1,\frac{3}{2}\}$. Among these, the maximum set of mutually transparent lines that also contain $\CW_{(1,0)}$ is $\{(0,0),(1,0),(0,1),(1,1)\}$ where the last two are the first two dressed by $S_\rho$. In the notation of \eqref{eq:odd-lagrangian-state} where $|q,0\ra_\rho$ already contains its image under $U[C_1]S_\rho[m_1]$. This is 
\be 
|\CL\ra_\rho^{\mathbb{Z}_2} = \sum_{q\in\mathbb{Z}_2} |q,0\ra_\rho
\ee
representing the Lagrangian algebra of $\mathbb{Z}_2\times \mathbb{Z}_2^f$ SymTFT, which we will be working with in this section. 

The global $\mathbb{Z}_2$ symmetry is generated by inserting $V_{1/2}[C_2]$ in the $C_2$ direction, which is the $\mathbb{Z}_2$ global symmetry generator/defect. To confirm this construction, one can check that these symmetry generators actually behave as we discussed previously. We first claim that
\be
\partial(S[m_1])\sim \chi_L,\;\quad \partial(S[m_1]W_1[C_1])\sim \chi_R
\ee
where $\partial(\bullet)$ represents endpoint of the operator $\bullet$ and $\hat{m}:=\iota(\star \hat{C})$. Then one can observe that these operators transform as
\begin{equation}
    \begin{aligned}
        V_{1/2}[C_2] &: \partial(S[m_1]) \to \partial(S[m_1]),\;\quad V_{1/2}[C_2] : \partial(S[m_1]W_1[C_1]) \to -\partial(S[m_1]W_1[C_1])\,,\\
        (-1)^{f}&: \partial(S[m_1]) \to -\partial(S[m_1]),\;\quad (-1)^{f}: \partial(S[m_1]W_1[C_1]) \to -\partial(S[m_1]W_1[C_1])
    \end{aligned}
\end{equation}
Since the endpoint of $S[m]$ contains a single Majorana fermion which anti-commutes with $(-1)^f$ and $V_{1/2}[C_2]$ has a non-trivial braiding with $W_1[C_1]$. Therefore, we can realize boundary theory with the same algebra as the CFT construction, confirming that the obtained theory is the Gu-Wen system. In the following part, we will use this system to extract both the fourth root of unity phase of the ratio of the partition function as well as the fact that the symmetry can be gauged upon stacking four copies of the theory together, as predicted by the cobordism framework.

\subsection{SymTFT Derivation of Mod-4 Anomaly} 

From the previous subsection, we have realized the Gu-Wen system using SymTFT construction with properly defined Lagrangian algebra. The next step is to compute the bordism data. In the SymTFT framework, to have a properly well-defined partition function, one needs to define the corresponding dynamical QFT boundary state $\ket{\text{QFT}}$ spanned by the Lagrangian algebra \cite{Brennan:2024fgj}.
\be
_\rho\bra{\text{QFT}} := \sum_{q\in\mathbb{Z}_2}Z_\text{QFT}[q,\rho]\,{_\rho\bra{q,0}}
\ee
From the previous section, we have shown that $V_{1/2}[C_2]$ acts as a $\mathbb{Z}_2$ global symmetry generator. To realize the mod-4 anomaly, one has to compute a partition function in the $\mathbb{Z}_2$-twist with prescribed spin structure $\ns-\ns$. Concretely, with $V_{1/2}[C_2]$ insertion, the left-moving fermion gains a phase $+1$ and the right-moving fermion gains a phase $-1$ after moving around the $C_1$ cycle. This implies that starting with the $\ns-\ns$ spin structure, with $V_{1/2}[C_2]$ insertion, the SymTFT partition function will split into left- and right-moving modes as in \eqref{eq:total-partition-function-cft}.
Then the desired total partition function should be given by
\be
Z_\text{total} = _{\ns-\ns}\bra{\text{QFT}}V_{1/2}[C_2]\ket{\CL}_{\ns-\ns}^{\mathbb{Z}_2}
\ee
To retrieve the mod-4 anomaly in this computation one has to perform the mapping class $\mathtt{S}\mathtt{T}^2\mathtt{S}^{-1}$ due to the difference in the geometric/topology setting.
Now, to extract the anomalous phase one computes as a topological spin of  $V_{1/2}$ and finds\footnote{Note that here we use
\be
\mathtt{S}V_{1/2}[C_2]\mathtt{S}^{-1} = V_{1/2}[C_1]
\ee
So that the topological spin corresponding to $\mathtt{S}\mathtt{T}^2\mathtt{S}^{-1}V_{1/2}[C_2] = \mathtt{T}^2V_{1/2}[C_1]$, which can be easily computed using the previous results.
}
\be
\frac{_{\ns-\ns}\bra{\text{QFT}}\mathtt{S}\mathtt{T}^2\mathtt{S}^{-1}V_{1/2}[C_2]\ket{\mathcal{L}}_{\ns-\ns}^{\mathbb{Z}_2}}{_{\ns-\ns}\bra{\text{QFT}}V_{1/2}[C_2]\ket{\CL}_{\ns-\ns}^{\mathbb{Z}_2}} = \exp(-\frac{2\pi i k}{4})
\ee
For $k=1$, this phase is $e^{-2\pi i/4}$, the phase of one Dirac fermion under the same $\texttt{ST}^2\texttt{S}^{-1}$ in the CFT picture with $\mathbb{Z}_2$ acting on $\chi_R$ as in \eqref{eq:total-partition-function-cft}. Note that $k=-1$ corresponds to the boundary condition where the boundary $\mathbb{Z}_2$ symmetry with $S V_{1/2}$ operator which yield topological spin with opposite sign. This is the choice of identifying the $\mathbb{Z}_2$ boundary with the bulk is the physical meaning of the sign of boundary F-symbol $\bar\alpha$ alluded in Section \ref{subsec:symalg-inflow}.

Therefore, we can compute the mod-4 anomalous phase using the SymTFT partition function calculation. As one can see that in the SymTFT construction, it is important to firstly define the corresponding Lagrangian algebra with properly defined spin structure if one were to compute a fermionic structure. While the spin structure is not transparently defined in the continuum description of SymTFT, the lattice SymTFT can explicitly define them.
\subsection{Stacking Rules and Trivializing Anomalous Theories} \label{subsection:stacking-rule}
\begin{figure}
    \centering
    \includegraphics[width=0.5\linewidth]{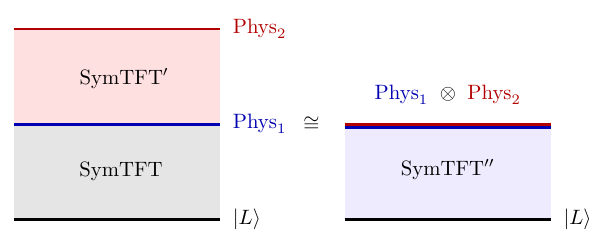}
    \caption{The SymTFT sandwich construction. \textbf{Left:} the physical theory \textcolor{blue!70!black}{$\mathrm{Phys}_1$} interface is separated from a second physical boundary theory \textcolor{red!70!black}{$\mathrm{Phys}_2$} by slabs of SymTFTs, with the same topological boundary condition $\ket{L}$ at the bottom. \textbf{Right:} The stacked SymTFT can be understood as finding the diagonal Lagrangian algebra, i.e., see section VI.B of \cite{Wen:2024udn}. The stacked theory is described by the SymTFT at level $k_A+k_B$ with the diagonal Lagrangian algebra. The stacked physical boundary state is basically a tensor product of the original boundary states, whose Lagrangian algebra is being fixed.}
    \label{fig:stacking-symtft}
\end{figure}

An interesting observation on $F-$move in Gu-Wen theory can be made that it is trivializable with fourth power, namely $F(\phi_1,\phi_2,\phi_3)^4\sim 1$. This rough observation suggests that one might be able to trivialize the anomalous theory by \emph{stacking} them. Stacking mechanism is not completely new, and there are existing results provided in, for SPT phases, \cite{Else:2014vma,Turzillo:2023yyr, Ren:2023kik, Lan:2018fqw, Aksoy:2022hua,Okada:2025kie}, and in more SymTFT language \cite{Wen:2024udn}. However, the notion of stacking here is yet unclear. In this section, we introduce the notion of stacking in the modified Villain picture for the SPT phases described in this work, providing generic treatment for both bosonic and fermionic theory. Moreover, we can show that stacking four identical copies of Gu-Wen anomalous theories actually yields a trivial theory.

Before proceeding, we can give a generic picture of stacking mechanism. Given theories $\mathcal{T}_{A,B}$ with the same (internal) symmetry structure, stacking, denoted as $\mathcal{T}_A\bigotimes \mathcal{T}_B$, which means we physically stack them on top of each other, as shown in Fig. \ref{fig:stacking-symtft}. The stacking mechanism in the Hilbert space depends on whether you are stacking bosonic or fermionic theories. For stacking bosonic theories, the Hilbert space is a conventional tensor product, namely
\be \label{eq:bosonic-stacking-hilbert-space}
\mathcal{H}_{AB} = \mathcal{H}_A\otimes_\CB \mathcal{H}_B
\ee
where $\otimes_\CB := \otimes$ denotes conventional tensor product, representing the bosonic stacking mechanism. Since the bosonic anomalous theory is characterized by $[\alpha] \in H^3(BG;U(1))$, we can derive their stacking rule by considering the stacked $F-$move as
\begin{equation}
    \ket{F_{L,R}^{AB}} = \ket{F_{L,R}^A}\otimes_\CB \ket{F_{L,R}^B}
\end{equation}
The derivation of the $F-$move phase is then straightforward as each sector acts independently without any graded structure, namely
\be
\bra{F_{R}^{AB}}\ket{F_{L}^{AB}} = \exp(\frac{2\pi i k \phi_1p_A(\phi_2,\phi_3)}{2}+ \frac{2\pi i k \phi_1p_B(\phi_2,\phi_3)}{2})
\ee
In terms of cocycle equation, the bosonic stacking rule reads
\be \label{eq:bosonic-stacking-rule}
\alpha_1\otimes_\mathcal{B}\alpha_2 = \alpha_1+\alpha _2\mod{1}
\ee
This result aligns well with the usual understanding that the bosonic anomaly data is linear in the group cohomology sense; thus, they can be stacked up using conventional group law.

While stacking fermionic theories can be tricky since the fermionic Hilbert space is a super-vector space that splits into even and odd spaces, $\mathcal{H}^f_{A,B} = \mathcal{H}_{A,B}^{0}\oplus \mathcal{H}_{A,B}^{1}$, since the Hilbert space must endow with the $\mathbb{Z}_2$-graded structure of fermions. The stacked Hilbert space is then given by a fermionic tensor product which preserves this $\mathbb{Z}_2$-grading structure, namely
\be \label{eq:fermionic-stacking-hilbert-space}
\mathcal{H}^f_{AB} = \mathcal{H}^f_A\otimes_\mathcal{F}\mathcal{H}^f_B
\ee
where $\otimes_\mathcal{F}$ denotes tensor product representing fermionic stacking that preserves $\mathbb{Z}_2-$graded structure of both Hilbert spaces.
In this $\mathbb{Z}_2-$graded vector space, any operator in the stacked Hilbert space is defined as $\mathcal{O}_{AB}:=\mathcal{O}_A\otimes_\mathcal{F}\mathcal{O}_B$. This operator behaves as
\be \label{eq:stacked-operator}
(\mathcal{O}_A\otimes_\mathcal{F} \CO_B ) (\mathcal{Q}_A\otimes_\mathcal{F} \mathcal{Q}_B ) = (-1)^{\vert \CO_B\vert \vert \mathcal{Q}_A\vert}(\mathcal{O}_A \mathcal{Q}_A)\otimes_\mathcal{F}(\mathcal{O}_B \mathcal{Q}_B)
\ee
where $\vert \bullet \vert = \{0,1 \}$ denotes even or odd fermionic parity, respectively. Then one can derive that the split operator in each layer caries a possibly non-trivial fermionic parity through its Majorana fermion at the fusion junction, namely $\vert N^{\phi_1,\phi_2}_{[\phi_1+\phi_2]} \vert = \vert \gamma_{p_\bx}^{p(\phi_1,\phi_2} \vert = p(\phi_1,\phi_2)$ given that $\vert\gamma_{p_\bx} \vert =1$ and the rest of the contribution to $N$ has even parity. The stacked splitting operator can be simply defined as
\be
N^{\phi_1,\phi_2}_{AB} := N^{\phi_1,\phi_2}_A \otimes_\mathcal{F} N^{\phi_1,\phi_2}_B
\ee
Since Majorana fermions between these layers do not commute $\{\gamma_A,\gamma_B \}=0$, the split operators on different layers $A$ and $B$ satisfy the following $\mathbb{Z}_2$-graded structure
\be
N^{\phi_1,\phi_2}_{A}N^{\phi_1',\phi_2'}_{B} = (-1)^{p_A(\phi_1,\phi_2)p_B(\phi_1',\phi_2')}N^{\phi_1',\phi_2'}_{B}N^{\phi_1,\phi_2}_{A}
\ee
Using the stacked splitting operator, one can derive a stacked $F-$move as we did with the bosonic case. In particular, the computation can be simplified since the move operators do not change upon stacking, by acting independently on their own corresponding layer. Thus, the only different contribution comes from the fermionic structure encoded in the splitting operators, namely the left and the right configurations read
\begin{align}
    \ket{F_{L}^{AB}} &= \ket{F_L^A} \otimes_{\mathcal{F}} \ket{F_L^B} \sim N^{\phi_1,\phi_2}_{AB} N^{[\phi_1+\phi_2],\phi_3}_{AB} \nonumber \\
    &\sim (-1)^{p_A([\phi_1+\phi_2],\phi_3) p_B(\phi_1,\phi_2)} \left( N^{\phi_1,\phi_2}_A N^{[\phi_1+\phi_2],\phi_3}_A \right) \otimes_{\mathcal{F}} \left( N^{\phi_1,\phi_2}_B N^{[\phi_1+\phi_2],\phi_3}_B \right), \\
    \ket{F_{R}^{AB}} &= \ket{F_R^A} \otimes_{\mathcal{F}} \ket{F_R^B} \sim N^{\phi_2,\phi_3}_{AB} N^{\phi_1,[\phi_2+\phi_3]}_{AB} \nonumber \\
    &\sim (-1)^{p_A(\phi_1,[\phi_2+\phi_3]) p_B(\phi_2,\phi_3)} \left( N^{\phi_2,\phi_3}_A N^{\phi_1,[\phi_2+\phi_3]}_A \right) \otimes_{\mathcal{F}} \left( N^{\phi_2,\phi_3}_B N^{\phi_1,[\phi_2+\phi_3]}_B \right)
\end{align}
Finally, the stacked $F$--move in $k$ odd case is given by
\begin{align}
\bra{F_R^{AB}}\ket{F_L^{AB}} = &(-1)^{p_A([\phi_1+\phi_2],\phi_3) p_B(\phi_1,\phi_2) + p_A(\phi_1,[\phi_2+\phi_3]) p_B(\phi_2,\phi_3)}\nonumber \\\times &F_A(\phi_1,\phi_2,\phi_3)F_B(\phi_1,\phi_2,\phi_3)
\end{align}
We can again express this in terms of cocycle equation, which yields the fermionic stacking rule
\be \label{eq:fermionic-stacking-rule}
(p_1,\alpha_1)\otimes_\mathcal{F}(p_2,\alpha_2) = (p_1+p_2 \mod{2}, \alpha_1+\alpha_2 + \frac{1}{2}p_1\cup_1 p_2 \mod{1})
\ee
where $(p,\alpha)\in Z^2(BG;\mathbb{Z}_2)\times C^3(BG;U(1))$. This pair of data is known to actually characterize the Gu-Wen anomaly. One can further check that for Gu-Wen system, $\phi_i=n_j/2$, where $n_i=\{0,1 \}$, the higher cup product term vanishes $(p_A\cup_1 p_B)(n_1,n_2,n_3) = 0$ since $p(n_i,n_j) = 1$ only if $n_i=n_j=1$. This result coincides with the operator-algebra computation in \cite{Okada:2025kie}. Fermionic stacking rule can be understood as bosonic stacking with a fermionic correction from the term $p_1\cup_1 p_2$. Using these stacking rule, we can show that stacking two identical Gu-Wen system characterized by $(p,\alpha)$ and $\phi_i = n_i/2$ with $n_i\in\{0,1 \}$ is given by
\be
(p,\alpha)\otimes_\mathcal{F}(p,\alpha) = (p+p \mod{2}, 2\alpha + \frac{1}{2}p\cup_1 p \mod{1}) = (0, 2\alpha)
\ee
where direct computation can show that $(p\cup_1 p)(\phi_1,\phi_2,\phi_3)= 0$.
The resulting stacked theory no longer has fermionic anomaly, but the bosonic anomaly still exist. Then we can proceed to perform three and four stacking, namely
\begin{equation}
    \begin{aligned}
        &(p,\alpha)\otimes_\mathcal{F}(p,\alpha)\otimes_\mathcal{F}(p,\alpha) = (p,3\alpha), \\
        &(p,\alpha)\otimes_\mathcal{F}(p,\alpha)\otimes_\mathcal{F}(p,\alpha)\otimes_\mathcal{F}(p,\alpha)= (0,0)
    \end{aligned}
\end{equation}
Therefore, stacking exactly four copies of Gu-Wen anomalous system trivializes both fermionic and bosonic anomalies.
\section{Gaugeable Subgroups of the $U(1)$ Theory} \label{section:gaugin-anomaly}

In the previous sections, we have constructed a SymTFT for theories with global $U(1)$ symmetry. One can then ask which discrete subgroups $\mathbb{Z}_d$ of the anomalous $U(1)$ theory can be gauged. We use the standard SymTFT dictionary on gauging \cite{kaidi2021higher, Kaidi:2022cpf, Antinucci:2024zjp, Brennan:2024fgj}. A topological boundary is specified by a Lagrangian algebra $\CL$, whose lines ends on the boundary and furnish the charges. The boundary symmetry is $\CG=\CA/\CL$, where $\CA$ denotes bulk lines. Gauging a subgroup $\CH\subset \CG$ replaces $\CL$ by a new Lagrangian algebra $\CL'$ obtained by condensing bulk lifts $\widetilde{\CH}$ of the $\CH$ symmetry lines, together with the charges of $\CL$ that are neutral under them.\footnote{In this section, we use the standard SymTFT language, but the construction is the same as in the solid torus quantization picture used so far. Namely, the condensed lines are those inserted along the $C_1$ direction, while the symmetry defect lines are those in the $C_2$ direction. \label{fn:symtft-and-torus}} The subgroup $\CH$ is gaugeable if and only if the lifts $x\in\widetilde{\CH}$ can be chosen and $\CL'$ satisfies:\footnote{From the definition of lifts, it satisfies a simple requirement that it is invariant upon fusion with Lagrangian algebra line, $x\sim x\otimes l$, where $l\in\CL'$}

\begin{itemize}
    \item[(I)] $\CL'$ is closed under fusion and maximal.
    \item[(II)] Each line in $\CL'$ is a boson, $\theta(l)=1$ for $\forall l\in\CL'$. In the fermionic case $\theta(l)=\pm 1$ is allowed and lines with $\theta(l)=-1$ can end on the boundary (form the Lagrangian algebra) only after being dressed by lines $S$ and become a composite boson.
    \item[(III)] Elements in the Lagrangian algebra must braid trivially with one another, i.e., if $l,l'\in \CL'$, it is required that $\CB(l,l')=1$. For abelian theory, this also implies that $\theta(l)^2=1$ for $l\in\CL'$ \cite{Kapustin:2010hk, Antinucci:2024zjp}.
    \item[(IV)] The $F$--moves of elements in $\CL'$ are cohomologically trivial \cite{Seifnashri:2023dpa, Tu:2025bqf, Seiberg:2016rsg}.\footnote{Restricting the bulk lines $a:=(q_a,\phi_a)$ to the $\mathbb{Z}_d$ subgroup amounts to setting $\phi_i:=j_i/d$, where $j_i\in\mathbb{Z}$, $0\le j_i<d$. The carry then becomes
    \be
    \begin{aligned}
        p(j_1,j_2)&:= \lfloor \frac{j_1+j_2}{d} \rfloor =
            \begin{cases}
                0\;\text{if}\; j_1+j_2 < d \\
                1\;\text{if}\; d\leq j_1+j_2 < 2d
            \end{cases},\; j_1+j_2 &:= [j_1+j_2]_d+dp(j_1,j_2)
    \end{aligned}
    \ee
    Therefore, the carry is an integer 2--cocycle representing the generator of $[p]\in H^2(B\mathbb{Z}_d,\mathbb{Z})\cong\mathbb{Z}_d$. This encodes the embedding $\mathbb{Z}_d\subset U(1)$. When $p$ enters the $F$--move only through $(-1)^p$, the relevant datum is defined modulo 2 such that
    \be
    [\bar p]\in H^2(B\mathbb{Z}_d,\mathbb{Z}_2)\cong
    \begin{cases}
    0 & d \text{ odd},\\
    \mathbb{Z}_2 & d \text{ even},
    \end{cases}
    \ee
    } In the fermionic case, they are trivial only up to Gu-Wen data, which we show can be removed by the $S$--dressed lines.
\end{itemize}
where braiding, topological spin, and $F$--move for both bosonic and fermionic theories are given by
\begin{gather}
    \theta(a) := \exp(2\pi i\left( q_a\phi_a-\frac{k\phi_a^2}{2}\right)),\quad \CB(a,b) := \exp\left(2\pi i (q_a\phi_b+q_b\phi_a-k\phi_a\phi_b)\right) \\
    F(a,b,c) := \exp\left( \pi i \phi_a kp(\phi_b,\phi_c) \right).
\end{gather}
Conditions (I)-(III) are standard in the SymTFT construction proposed in \cite{Brennan:2024fgj, Antinucci:2024zjp}. While condition for trivial $F$--move may be viewed as another consistency check, it helps one to precisely determine the anomalies in the theory more systematically. In this case, we use one choice of the boundary $F$--move (see Section \ref{subsec:symalg-inflow} for the general form) that is convenient to work with in gauging the theory.

Another non-trivial question is to find the resulting symmetry after gauging, $\CG'=\CA/\CL'$. In general, determining the gauged symmetry $\CG'$ is non-trivial, and may involve non-group symmetry, e.g., \cite{Tachikawa:2017gyf}. In our case, it follows directly from the Pontryagin duality \cite{Tachikawa:2017gyf,Gaiotto:2014kfa, kaidi2021higher, Kaidi:2022cpf} derivations. Equivalently, it can be read off from the braiding between the symmetry defect lines and $\CL'$. The symmetry defects lines are the non-trivial classes $x\in \CA/\CL'$; by maximality of the Lagrangian algebra, each satisfies $\CB(x,l)\neq 1$ for some $l\in\CL'$, and they form a group isomorphic, though not canonically to $U(1)\times \mathbb{Z}_g$ where $g$ depends on whether the theory is bosonic or fermionic. 

\begin{figure}[htbp]
    \centering
    \includegraphics[width=1\linewidth]{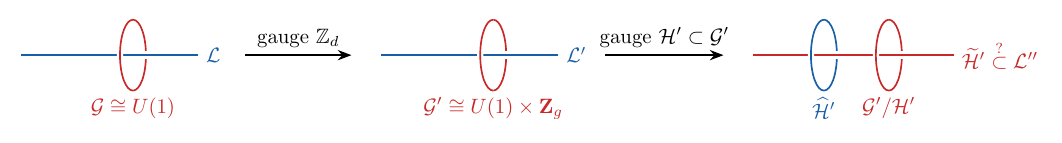}
    \caption{Illustration of the logic in this section. The horizontal line represents the $C_1$ torus direction while the loop around it represents the $C_2$ direction. We start from the $U(1)$ boundary theory with Lagrangian algebra $\CL=\{ \CW_{(q,0)}\}\cong \mathbb{Z}$ and the symmetry lines $\CG\cong U(1)$, i.e., $\CW_{(0,\alpha)}$. Gauging $\mathbb{Z}_d\subset U(1)$ amounts to finding a new Lagrangian algebra $\CL'$ with symmetry lines $\CG'$. We show below that $\CG'\cong U(1)'\times\mathbb{Z}_g$ (times $\mathbb{Z}_2^F$ for the fermionic case), where the value of $g$ depends on whether $\CL$ is bosonic or fermionic. A subgroup $\CH'\subseteq \CG'$ can be gauged if its lifts $\widetilde{\CH}'\subseteq \CL''$, together with the charges of $\CL'$ neutral under $\CH'$, generate a new Lagrangian algebra $\CL''$. 
    In that case, the resulting symmetry $\CG''$ is an extension of $\CG'/\CH'$ by the dual $\widehat{\CH}':= \hom(\CH',U(1))$ \cite{Tachikawa:2017gyf}. This extension can be non-split, which is the point of Section \ref{subsection:bosonic-gauging}.}
    \label{fig:gauging_summarized}
\end{figure}

Furthermore, it is also possible for the gauged symmetry $\CG'$, or its subgroup, to be anomalous. Here, one can use the same procedure (I)-(IV) described above to determine whether the gauged symmetry is anomalous or not. Suppose that we want to gauge some subgroup $\CH'\subseteq\CG'$ of the gauged theory. What we have to do is to check whether the lifts $\widetilde{\CH}'$ can form another Lagrangian algebra $\CL''$ or not. Classes of the lifts are determined from $x\in\widetilde{\CH}'=\CA/\CL'$ as they are originally the symmetry lines of the gauged theory. Primarily, we determine the anomalies of the gauged theory from topological spin and braiding. As a complementary check, we also compute the boundary $F$--move $F(x,y,z)$ of the symmetry defect lines. As illustrated above, Lagrangian algebra of bosonic and fermionic are similar, we will illustrate detailed derivations for the bosonic case with additional discussion in the fermionic theory. We summarize the flow of work in this section in Fig. \ref{fig:gauging_summarized}.

\subsection{Gaugeable Subgroup in Bosonic Theory} \label{subsection:bosonic-gauging}

In this case, SymTFT conditions (I)-(III) and the $F$--move conditions provide the same gaugeable condition, namely $k/2d\in\mathbb{Z}$, i.e., $k/2$ must be divisible by $d$. The order in which they are solved does not matter, so we begin with checking the $F$--move, which classifies the possible anomaly of the $\mathbb{Z}_d$ subgroup. In general, one can determine whether $\mathbb{Z}_d$ is anomalous or not by examining the third cohomology group $H^3(\mathbb{Z}_d;U(1))\cong \mathbb{Z}_d$, whose generator is given by 
\be \label{eq:z-d-cohomology-generator}
\omega(j_1,j_2,j_3) = \exp(\frac{2\pi i j_1}{d}p(j_2,j_3)),\qquad j_i\in\{ 0,1,..., d-1 \}
\ee
Writing $a=(n_a,j_a/d)$, our $F$--move comparing to $\omega$ is given  by
\be \label{eq:f-and-omega-compare}
F(a,b,c)=\exp(2\pi i \alpha(a,b,c)) = \omega(j_a,j_b,j_c)^{k/2}\,.
\ee
Hence $[F]\equiv\frac{k}{2}\text{ mod }d$, which vanishes if and only if $k/2d\in\mathbb{Z}$, yields the gaugeable constraint that $k/2d\in\mathbb{Z}$. Since a vertex redefinition shifts $F$ only by some coboundary, this condition cannot be relaxed.\footnote{As discussed in Section \ref{section:u(1)-junction-computation}, the fusion vertices can be redefined by a 2--cochain $\beta$, which shifts $F$ by some $\delta\beta$ and leaves $[F]$ unchanged.}

The remaining constraints follow from the topological spin and braiding, which fix the form of $\CL'$. Lines in $\CL'$ have $\phi\in\frac{1}{d}\mathbb{Z}$, so we write them as $(n,j/d)$ with $n,j\in\mathbb{Z}$. Conditions (II)-(III) require
\be \label{eq:solving-relation-bosonic}
\frac{nj}{d} -\frac{kj^2}{2d^2} \in \mathbb{Z},\qquad\frac{n_1j_2}{d} + \frac{n_2j_1}{d} -\frac{kj_1j_2}{d^2}\in \mathbb{Z}
\ee
To gauge all of $\mathbb{Z}_d$, $\CL'$ must also contain a lift $(n_1,1/d)$ of the generator. The spin condition gives $n_1 = \frac{k}{2d}+dl_1$, where $l_1\in\mathbb{Z}$. This has an integer solution only if $\kappa\equiv\frac{k}{2d}\in\mathbb{Z}$, reproducing the gaugeability obtained from $F$--move. The charge $(q,0)$ is neutral under $(\kappa,1/d)$ satisfying $e^{2\pi i q/d}=1$, i.e., $q\in d\mathbb{Z}$. The closure condition (i) further fixes
\be \label{eq:gauged-lagrangian-bosonic}
\begin{aligned}
\CL' = \Big\{\left(\kappa j+dl, \frac{j}{d}\right):j,l\in\mathbb{Z} \Big\},\quad \ket{\CL'} = \sum_{j,l\in\mathbb{Z}}\ket{\frac{kj}{2d}+dl,\frac{j}{d}},\qquad \kappa\equiv\frac{k}{2d}
\end{aligned}
\ee
where $\CL'$ in our solid torus quantization is given by the second equality in the equation above. One can check that topological spin and braiding between elements in $\CL'$ read $\exp(2\pi i jl)$ and $\exp(2\pi i (l_1j_2+l_2j_1))$, implying that $\CL'$ is mutually transparent. Moreover, any line braiding trivially with $(d,0)$ and $(\kappa,1/d)$ lies in $\CL'$, so $\CL'$ is maximal. Hence, we conclude that $\CL'$ is bosonic, maximal and mutually transparent, i.e., Lagrangian. 

From this Lagrangian algebra, the only independent parameters are $j,l\in\mathbb{Z}$, which implies that the charges corresponding to $\CL'$ are best labeled by $(j,l)\in\mathbb{Z}^2$. Since the theory is abelian, the actual charges must be defined modulo certain identifications. This is known as \emph{discriminant} group in general abelian TQFT theory \cite{Kapustin:2010hk, Cano:2013ooa}. To determine it, we first note that the bulk charge identification descends to the boundary, namely
\be
\overbrace{(q,\phi)\sim (q+k,\phi+1)}^{\text{Bulk charge identification}}
\Rightarrow
\overbrace{(j,l)\sim (j+d, l + \kappa)}^{\text{Boundary charge identification}}
\ee
Using Smith normal form \cite{Cano:2013ooa}, one can compute the discriminant charges of this theory. Namely, what one want to compute is the quotient of $\mathbb{Z}^2/\langle(d,\kappa) \rangle$, where $\langle(d,\kappa) \rangle$ is the boundary charge identification rule. We briefly explain the derivation using Smith normal form in Appendix \ref{app:for-sec-v}. Therefore, the actual group of the boundary charges read
\be
\CL' = \frac{\mathbb{Z}^2}{\langle (d, \kappa) \rangle} \cong \mathbb{Z}_{g_b}\oplus \mathbb{Z},\qquad g_b\equiv\text{gcd}(d,\kappa)
\ee
Using this one can then derive the symmetry defect lines of the gauged boundary. As we discuss above the symmetry lines are bulk lines modulo the Lagrangian algebra, $\CG'=\CA/\CL'$, where $\CA=\mathbb{Z}\times\mathbb{R}$. Since the bulk identification line $(k,1)$ already belongs to the gauged Lagrangian algebra $\CL'$, working with $\CA=\mathbb{Z}\times\mathbb{R}$ automatically accounts for the bulk identification. A symmetry line is characterized by how it braids with the condensed lines. For an arbitrary probe $(n,\alpha)\in\CA$ and a condensed line $l(j,l):= (\kappa j + dl,j/d)\in\CL'$, the braiding reads
\be
\CB_{(n,\alpha),l(j,l)} = e^{2\pi i\left(j(\frac{n}{d}-\kappa\alpha) + ld\alpha\right)} = e^{2\pi i (jx+yl)},\qquad (x,y):=(\frac{n}{d}-\kappa \alpha,\alpha d)
\ee
The braiding is linear in both arguments and obviously trivial on $\CL'\times\CL'$ since the condensed lines are mutually transparent. It defines a pairing $\CB:\CA/\CL' \times \CL' \to U(1)$. Equivalently, each probe acts on the condensed lines, and hence on the boundary charges. Therefore, braiding defines the homomorphism\footnote{For more detailed discussion on this structure, see, e.g., footnote 9 and 10 of \cite{kaidi2021higher}.}
\be
    \rho:\ \mathbb{Z}\times\mathbb{R}\ \to\ T^2,\qquad 
          (n,\alpha)\ \mapsto\ (x,y)\ \text{mod}\ \mathbb{Z}^2.
\ee
The kernel follows directly, $\rho(n,\alpha)=0$ requires $j:= \alpha d\in\mathbb{Z}$ and $n/d-\kappa \alpha \in \mathbb{Z}$, i.e., $n\equiv\kappa j\pmod{d}$ so $(n,\alpha)=( \kappa j+ ld, j/d )\in\CL'$. Hence we conclude that $\ker\rho = \CL'$, which means a probe that braids trivially with every condensed line is itself condensed, as required by the maximality. Every probe satisfies $xd+\kappa y=n \in\mathbb{Z}$, and conversely $(x,y)$ is the image of the probe $(n,\alpha)=(xd+\kappa y, y/d)$; hence $\text{im}\rho = \{(x,y)\in T^2:xd+\kappa y\in\mathbb{Z}\}$, and by the first isomorphism theorem,
\be \label{eq:gauged-symmetry-group-bosonic}
\CG' = \frac{\mathbb{Z}\times \mathbb{R}}{\text{ker}(\rho)}= \left\{ (x,y)\in T^2:xd+\kappa y \in\mathbb{Z} \right\}\cong U(1)'\times\mathbb{Z}_{g_b}, \quad g_b:=\text{gcd}(d,\kappa),
\ee
where the last isomorphism is shown in Appendix \ref{app:for-sec-v} and we use the parameterization such that $U(1)':= U(1)/\mathbb{Z}_{g_b}$.\footnote{As we are going to show in this section that the $U(1)'$ symmetry is generated by $u_\sigma:=\CW_{(0,\sigma/{g_b})}$, which closes at $\sigma=1$. Similar parameterization is also used in the fermionic case.}
Alternatively, we can understand $\CG'$ as the group of characters (braiding phase) of the boundary charge group, i.e.,
\be
\CG' \cong \hom\left( \mathbb{Z}^2/\langle (d,\kappa) \rangle, U(1) \right) \cong \mathbb{Z}_{g_b}\times U(1)'
\ee
where the second isomorphism is obtained by using Pontryagin dual and Smith normal form. This is the usual structure in the generalized symmetry \cite{Gaiotto:2014kfa, Tachikawa:2017gyf}. Therefore, we conclude that the gauged symmetry in the bosonic case is\footnote{ In \cite{Antinucci:2024zjp}, the symmetry lines on the gauged boundary are labeled by $(n,\alpha)$ with $n\in\mathbb{Z}_d$ and $\alpha\in\mathbb{R}/\frac{1}{d}\mathbb{Z}$, and the symmetry is identified as $U(1)'^{(0)}\times \mathbb{Z}_d^{(0)}$. As group elements, however, these lines are not independent. The symmetry lines are bulk lines modulo the Lagrangian algebra $\CL'$. Since $(\kappa,1/d)\in\CL'$, the lines $(n,\alpha)$ and $(n-\kappa,\alpha-1/d)$ act identically on the boundary, namely reducing $\alpha$ modulo $1/d$ shifts $n$ by $\kappa$. For example, for $k=4$ and $d=2$ one has $\kappa=1$ and $(1,1/2)\in\CL'$, so the generator $(1,0)$ of the would-be $\mathbb{Z}_2$ acts as the element $(0,-1/2)$ of the $U(1)'$. Therefore, the would-be $\mathbb{Z}_2$ lies in $U(1)'$. In general, the symmetry is $(U(1)/\mathbb{Z}_{g_b})\times \mathbb{Z}_{g_b}$, where $g_b:=\gcd(d,\kappa)$, which reduces to the direct product of \cite{Antinucci:2024zjp} when $g_b=d$, equivalently $2d^2\vert k$.} 
\be \label{eq:gauged-bosonic-symmetry}
\CG'_b = U(1)'\times \mathbb{Z}_{g_b}
\ee

The next question is whether the gauged symmetry $\CG'=\CA/\CL'$ is anomalous or not. Equivalently, we are asking whether a subgroup $H'\subseteq \CG'$ can be gauged or not. In our case, the gauged symmetry $\eqref{eq:gauged-bosonic-symmetry}$ is a product, so we can either gauge $\CH'=U(1)'$, or $\mathbb{Z}_{g_b}$, or both simultaneously. Therefore, there are three types of anomaly, with coefficients $(k',\nu',\mu') \in 2\mathbb{Z}\times \mathbb{Z}_{g_b}\times\mathbb{Z}_{g_b}$, as shown in the following discussion. We show the derivations for symmetry defects (lifts in the bulk) corresponding to this theory in Appendix \ref{app:for-sec-v}. Namely, the generators of $U(1)'$ and $\mathbb{Z}_{g_b}$ symmetries are

\be \label{eq:gauged-symmetry-lines}
\begin{aligned}
    &u_\sigma := \CW_{(0,\sigma/g_b)},\qquad \sigma\in[0,1),\\
    &w_r := \CW_{(r,-rt/g_b d)},\qquad r\in\mathbb{Z}_{g_b},
\end{aligned}
\ee
where $t$ is an integer satisfying $vd'-\kappa' t=1$ for some $v\in\mathbb{Z}$, i.e., $-\kappa' t \equiv 1 \pmod{d'}$. We can show that the anomalies are classified by topological spin and braiding statistics,
\begin{itemize}
    \item[(i)] Pure $U(1)'$ anomaly classified by some integer $k'$. This can be obtained from topological spin of $u_\sigma$
    \be
    \theta(u_\sigma) = \exp(-\frac{2\pi i k'\sigma^2}{2}),\qquad k':= \frac{k}{g_b ^2}\in 2\mathbb{Z},
    \ee
    %
    The $U(1)'$ anomaly is determined by $k'=k/g_b^2 = 2d'\kappa'\in 2\mathbb{Z}$. Since $k'\neq 0$, the gauged $U(1)'$ is always anomalous.
    \item[(ii)] Mixed $U(1)'\times \mathbb{Z}_{g_b}$ classified by some coefficient $\nu'$. This is detected by the braiding of the two symmetry lines.
    \be
    \CB(u_\sigma,w_r) = \exp(\frac{2\pi i r\sigma}{g_b}\nu'), \qquad \nu':= 1+2\kappa't,
    \ee
    where we define another integer coefficient $\kappa':= \kappa/g_b \in\mathbb{Z}$. 
    \item[(iii)] Pure $\mathbb{Z}_{g_b}$ anomaly classified by some coefficient $\mu'$, which is again obtained from the topological spin
    \be
    \theta(w_r)=\theta(\CW_{(r,-\frac{rt}{g_bd})})= \exp(-\frac{2\pi i}{g_b^2}r^2\mu'),\qquad \mu' := \frac{t}{d'}\left(1+\kappa' t\right)\in\mathbb{Z}.
    \ee
    The coefficient $\mu'$ is also an integer.\footnote{By Bézout's identity \eqref{eq:bezout-identity}, there are integers $t,v$ with $d'v-\kappa't=1$, so $v = \frac{1+\kappa't}{d'}\in\mathbb{Z}$ and $\mu'=tv\in\mathbb{Z}$.} The pure $\mathbb{Z}_{g_b}$ $\mu'$ anomaly coefficient is determined modulo $g_b$.
\end{itemize}

We remark that the $\mathbb{Z}_{g_b}$ generator is not unique. Since $t$ is fixed only modulo $d'$ by
$-\kappa't\equiv1\pmod{d'}$, we may replace $t\to t+d's$ with $s\in\mathbb{Z}$, which gives another
generator of order $g_b$,
\be
w_r\ \to\ \CW_{\left(r,\,-\frac{r(t+d's)}{g_bd}\right)}=w_r\,u_{-rs/g_b}\,.
\ee
Since $2\kappa'd'=k'$, this shifts the mixed coefficient as $\nu'\to\nu'+k's\pmod{g_b}$, so the mixed
anomaly can be removed if and only if $\nu'\equiv0\pmod{\gcd(g_b,k')}$. Moreover, the identity
$\nu'^2-2k'\mu'=1$ holds for every admissible $t$, hence $\nu'$ is a unit modulo $\gcd(g_b,k')$.
Therefore the anomaly is necessarily mixed if and only if $\gcd\big(g_b,\tfrac{k}{g_b^2}\big)>1$, in
particular whenever $g_b$ is even. Otherwise, choosing $s$ with $\nu'+k's\equiv0\pmod{g_b}$ leaves a purely $\mathbb{Z}_{g_b}$ anomaly with $\mu'\equiv-(2k')^{-1}\pmod{g_b}$.
    
Equivalently, we show that these anomalies can be completely identified from the computation of boundary $F$--move constructed in Section \ref{section:u(1)-junction-computation}. The detailed derivation of the following boundary $F$--move is presented in Appendix \ref{app:for-sec-v}. The boundary $F$--move captures non-associativity in fusion of the symmetry lines $x\in\CG'_b$, which reads
\be
\bar{\alpha}(x_1,x_2,x_3) = p(\sigma_1,\sigma_2)\left[-\frac{k'}{2}\sigma_3 + \frac{\kappa't}{g_b}r_3 \right] + p_g(r_1,r_2)(1+\kappa' t) \left[ \sigma_3 - \frac{t}{d}r_3\right]\,.
\ee
The first term $-p(\sigma_1,\sigma_2)\frac{k'}{2}\sigma_3$ is the level $k'$ $U(1)$ $F$--move. The terms $\frac{\kappa' t}{g_b}r_3p(\sigma_1,\sigma_2)$ and $(1+\kappa' t)\sigma_3p_g(r_1,r_2)$ are the two mixed terms; their coefficients add up to $\nu'$. Finally, $-(1+\kappa' t)tr_3 p_g/d$ is the pure $\mathbb{Z}_{g_b}$ anomaly term.

\subsection{Gaugeable Subgroup in Fermionic Theory} \label{subsection:fermionic-gauging}

Similar to the bosonic case, we can carry out analysis on gauging discrete subgroup and studying the resulted gauged theory. There are several distinctions in both cases, but the logic is almost identical. The first distinction is that there are some lines in the Lagrangian algebra that have fermionic statistics, and require to be dressed by the transparent fermionic line $S$. The dressed lines become composite boson, and form a Lagrangian algebra. This is the odd Lagrangian structure discussed in Section \ref{section:hilbert-space-construction}. This demonstrates the first necessity of $S$ lines in the theory. 

Apart from that, we can show that the existence of $S$ lines in the Lagrangian algebra also allow us to trivialize the $F$--move and gauge the $\mathbb{Z}_d$ subgroup. The second distinction is that the gauged condition also depends on the parity of $d$. Since, upon embedding $\mathbb{Z}_d$ into $U(1)$, the $F$--move determined by the relation $\delta\bar{\alpha} = \frac{k}{2}p\cup p$ becomes
\be
\delta\bar{\alpha} = \frac{k}{2}\bar{p}\cup\bar{p}, \qquad [\bar{p}]\in H^2(B\mathbb{Z}_d;\mathbb{Z}_2)\,.
\ee
Here we chose to work with the positive version of $\delta\bar{\alpha}$. This class is non-trivial when $d$ is even and trivial when $d$ is odd. Thus, the gaugeability crucially depends on parity of $d$ which we summarize in the following.
\begin{itemize}
    \item \textbf{Even d:} In this case, the class $[\bar{p}]\cong \mathbb{Z}_2$. Thus $F$--move is not trivial, implying that $\mathbb{Z}_d$ subgroup cannot be gauged. This can also be seen through the simple observation that, for even $d$, no lift $\CW_{(q, 1/d)}$ of the generator braids trivially with itself,
    \be
    \CB=\exp\left(2\pi i (2q/d - k/d^2)\right) = 1\Leftrightarrow k\equiv 2qd \pmod{d^2}\,,
    \ee
    which forces $k$ even and contradicts the initial setup. Even if one attempts to make specific gauge choice to somehow trivialize the $F$-move, it remains impossible to trivialize the braiding statistics. Therefore, we conclude that we cannot gauge $\mathbb{Z}_d$ subgroup when $d$ is even.
    \item \textbf{Odd d:} In this case, $[\bar p] =0$, but the Gu-Wen relation still exists. So naively, it seems that one cannot perform discrete gauging. We will soon show that the Gu-Wen relation obstructing gauging can be removed in this case by using fact that $[\bar p] =0$. This trivializes the $F$--move. Moreover, the SymTFT gauging conditions also align with this method, which reads $k$ must be divisible by $d$.
\end{itemize}
    
Similar to the bosonic case, we can generally determine whether $\mathbb{Z}_d$ is anomalous or not by examining \eqref{eq:z-d-cohomology-generator}, and compare it with the expression of $F$--move in the fermionic case which reads \eqref{eq:f-and-omega-compare}. We state the condition here again for convenience 
%
\be
\alpha(a,b,c) = \frac{k}{2}\frac{j_a p(j_b,j_c)}{d}
\ee
This implies that $[\alpha] \equiv \frac{k}{2} \pmod d$. By basic modular arithmetic, this class can be trivialized if and only if $k/d \in \mathbb{Z}$. Therefore, this imposes a strong constraint on the possible gaugeable subgroups.\footnote{The factor of $1/2$ might seem a bit subtle, but it is well-defined in modular arithmetic. Specifically, since $d$ is odd, there exists a modular inverse of $2$ such that $2^{-1} \cdot 2 \equiv 1 \pmod d$.} However, if one checks the expression of $F$ move, it seems like there are some possible non-trivial configurations since $p(\phi_a,\phi_b)$ might not be trivial (only its class is). 

Next, we can show that $F$--move can be trivialized using the fact that $[\bar{p}]$ is trivial. For odd $d$ case, $[\bar p]=0$ so it can be expressed as a coboundary of some $\mathbb{Z}_2$ $1$-cochain $h$, i.e., $h:\mathbb{Z}_d\to \mathbb{Z}_2$. Concretely, we can rewrite $\bar p$ as
\be
\bar{p}(j,j') := h(j)+h(j') - h([j+j']_d) =(\delta h)(j,j')\pmod{2},\; h(j)\equiv j \pmod2\,,
\ee
for $j\in\{0,...,d-1 \}$. Then, we can redefine the Gu-Wen phase as follows
\be \label{eq:removing-gu-wen-phase}
\alpha \to \alpha' = \alpha - \frac{1}{2}h\cup \bar{p} \Rightarrow \delta \alpha ' = \delta\alpha - \frac{1}{2}\delta h \cup \bar{p} \pmod{1} = 0
\ee
Therefore, the Gu-Wen relation can be removed and trivialization of $F$--move is possible. We will come back to the explicit lattice mechanism of this trivialization after we realize the Lagrangian algebra.

From the SymTFT perspective, we can solve for the explicit form of the Lagrangian algebra, which will also imply that k must be divisible by $d$ for such Lagrangian algebra to exist. Suppose that lines in $\CL_\rho'$ have $\phi\in\frac{1}{d}\mathbb{Z}$, so we write them as $(n,j/d)$ with $n,j\in\mathbb{Z}$. Again by solving \eqref{eq:solving-relation-bosonic}, we obtain that lines $\CW_{(\frac{k}{d}\frac{d+1}{2}j + ld, \frac{j}{d})}$ braid trivially among themselves if and only if $k/d\in\mathbb{Z}$. Thus, we can use this to solve for the Lagrangian algebra. However, the topological spin condition reads
\be
    \theta(\CW_{(\frac{k}{d}\frac{d+1}{2}j+ld, \frac{j}{d})}) = (-1)^{j^2} = (-1)^j\,,
\ee
where we used the fact that $k/d\in\mathbb{Z}$ and $(-1)^{j^2}=(-1)^j$. Therefore, odd $j$ lines cannot form a Lagrangian algebra since they violate condition (II). Similar to the construction in Section \ref{section:hilbert-space-construction}, we can dress the odd $j$ lines with the transparent fermionic line $S$. Furthermore, we will show later that this is the mechanism used to trivialize the $F$--move for the fermionic theory. From the previous discussion, we have $h$ that assigns $h(a)\in\mathbb{Z}_2$, where $a$ is some line label. Then, for every line $a$ such that $h(a)=1$, we dress $a$ by the fermionic line $S$ along its support, whose endpoint carries a Majorana fermion, the redefined operator reads
\be \label{eq:redefined-topological-lin-with-s}
\widetilde{\CW}_a := \CW_a S[m_a]^{h(a)}
\ee
We will refer to this structure $S$-dressing in this section. Using $S$-dressing mechanism on the odd $j$ lines, they become bosons, namely
\be
\theta(\CW_{(\zeta j+ld,j/d)}[m]S^{h(j)}) =1\,.
\ee
This understanding aligns well with the odd Lagrangian algebra \eqref{eq:odd-lagrangian-state}. Therefore, the correct Lagrangian algebra for this case is given by
\be
\begin{aligned}
\CL_\rho' &:= \left(\zeta j + ld, \frac{j}{d} \right),\quad \zeta:=  \frac{k}{d}\frac{d+1}{2},\qquad j,l\in\mathbb{Z} \\
\ket{\mathcal{L}'}_\rho &= \sum_{j,l\in\mathbb{Z}}\ket{\zeta j+ld, \frac{j}{d}}_\rho  \\
&= \ket{0,0}_\rho + \sum_{\substack{l\in\mathbb{Z} \\ j\in\mathbb{Z}_\text{even}}} \ket{\zeta j+ld, \frac{j}{d}}_\rho + \sum_{\substack{l\in\mathbb{Z} \\ j\in\mathbb{Z}_\text{odd}}} S_\rho[m]\ket{\zeta j+ld, \frac{j}{d}}_\rho\,,
\end{aligned}
\ee
%
Before proceeding to compute the corresponding gauged symmetry, we would like to remark that dressing $S$ lines also help in $F$--move trivialization. Namely, we can derive the dressed $F$--move phase that cancels the Gu-Wen relation. Using \eqref{eq:redefined-topological-lin-with-s}, we first define the dressed move-and-split operators as follows
\be
\widetilde{M}_{\bx',\bx}^a := M^a_{\bx',\bx}T^{h(a)}_{\bx',\bx}, \quad (\widetilde{N}_{ab}^{(a),(b)})_\bx := T^{h(b)}_{\bx+\hat{1},\bx}\gamma_\bx^{\bar{p}(a,b)}\CD^{-p(\phi_a,\phi_b)}_\bx M^{b}_{\bx+\hat{1},\bx}
\ee
where $T$ is responsible for moving the Majorana fermion at the original fusion junction $\bx$ to the final fusion junction $\bx'$, which can be given as
\be
T^{h(a)} \propto (\gamma_{\bx'}\gamma_{\bx})^{h(a)}
\ee
The precise form does not matter as they will be canceled out in taking inner product in the final step. In this split operator, there is an original Majorana fermion $\gamma_\bx$ that is inherited from the fermionic structure in the theory. One can then carry out the computation of $F$--move similar to what we have done in section \ref{section:u(1)-junction-computation}. Finally, the modified $F$--move reads\footnote{The expression $-\frac{1}{2}h\cup \bar{p}$ in \eqref{eq:removing-gu-wen-phase} and $+\frac{1}{2}\bar{p}\cup h$ in \eqref{eq:modified-f-move-odd-k} have the same coboundary $\frac{1}{2}\bar{p}\cup\bar{p}$, so their difference is a 3--cocycle. Since $H^3(\mathbb{Z}_d;\mathbb{Z}_2)=0$ for odd $d$, they only differ by a coboundary.}
\be \label{eq:modified-f-move-odd-k}
\widetilde{F}(a,b,c) = (-1)^{\bar{p}(a,b)h(c)}F(a,b,c)
\ee
Detailed derivation is shown in Appendix \ref{app:for-sec-v}. By using \eqref{eq:removing-gu-wen-phase}, one can show that Gu-Wen phase can be removed using this modified $F$--move. Therefore, we conclude that for odd $k$, at the level of $F$--move it is possible to gauge the $\mathbb{Z}_d\subset U(1)$ subgroup with the condition that $d\vert k$.

Next, we discuss the group structure of the gauged theory. All derivations in the fermionic case are similar to the bosonic case discussed in detailed previously. The theory is again abelian so the Lagrangian algebra must be defined modulo identification. In this case, the bulk identification also descends to the boundary identification, namely
\be
\overbrace{(q,\phi)\sim (q+2k,\phi+2)}^{\text{Bulk charge identification}}
\Rightarrow
\overbrace{(j,l)\sim ( j+2d, l - 2\zeta' )}^{\text{Boundary charge identification}},\quad \zeta':= \frac{k}{d}\frac{d-1}{2}
\ee
Thus, the group of the boundary charges is defined modulo this charge identification as
\be \label{eq:fermionic-gauged-charge}
\CL_\rho' = \frac{\mathbb{Z}^2}{\langle (2d,-2\zeta') \rangle} \cong \mathbb{Z}\oplus \mathbb{Z}_{2g_f} \cong \mathbb{Z}\oplus \mathbb{Z}_{g_f}\oplus \mathbb{Z}_2,\quad g_f:=\gcd(d,\zeta') = \gcd(d,k/d)
\ee
where the split $\mathbb{Z}_{2g_f}\cong \mathbb{Z}_2\times\mathbb{Z}_{g_f}$ comes from the fact that $g_f$ is odd and $\gcd(2,g_f)=1$. This implies that the fermionic theory has an additional $\mathbb{Z}_2$ charge in the Lagrangian algebra compared to the bosonic case. We show in the following analysis that this, indeed, corresponds to the fermionic charge presented in the theory. To see this, we may first try to observe the symmetry structure using the braiding probe. Consider arbitrary probes $\CW_{(n,\alpha)}$ with $n\in\mathbb{Z}$ and $\alpha\in\mathbb{R}$ whose braiding with line in the Lagrangian algebra reads
\be
\CB_{(n,\alpha),l(j,l)} = \exp(2\pi i \left( \frac{jn}{d} +\zeta'\alpha j +ld\alpha \right)) := \exp(2\pi i \left(  xj+yl\right))\,,
\ee
where $x:= \frac{n}{d}+\zeta'\alpha$ and $y:=\alpha d$ satisfying the constraint $dx-\zeta'y \in \mathbb{Z}$. We then use the method illustrated in Appendix \ref{app:for-sec-v} to derive the group structure, we obtain that
\be
\CG_f' = \left\{ (x,y)\in T^2:xd-\zeta'y\in\mathbb{Z} \right\} \cong U(1)'\times \mathbb{Z}_{g_f}\,,
\ee
where again we use $U(1)':= U(1)/\mathbb{Z}_{g_f}$. From the braiding level, the $\mathbb{Z}_2$ factor corresponding to the $\mathbb{Z}_2$ charge obtained in \eqref{eq:fermionic-gauged-charge} does not contribute. On the other hand, if one uses the standard Pontryagin dual method to compute the symmetry structure, one obtains
\be
\CG'_f = U(1)'\times \mathbb{Z}_{g_f}\times \mathbb{Z}_2^F\,,
\ee
with an additional $\mathbb{Z}_2$ factor which we intentionally label as $\mathbb{Z}_2^F$. Thus, it seems like the braiding method is blind to see this additional factor. And from the Pontryagin computation alone, it is not obvious how the additional $\mathbb{Z}_2$ is related to the fermionic symmetry. But we can demonstrate from the structure Lagrangian algebra that it is related to the fermionic symmetry. The first observation is that in the set of odd $j$ lines, there exists a $\mathbb{Z}_2$-order line\footnote{One can check that this fermionic line $f$ corresponds to the line $l(j,l):= (d,-\zeta')$ in the boundary language.} associated to the $\mathbb{Z}_2$ charge in the Lagrangian algebra:
\be
f := (k,1),\quad f^2=1 , \quad \theta(f)=-1,\quad \CB(f,l\in\CL_\rho')=1\,,
\ee
that is invisible to braiding phase, but it is dressed by the transparent fermionic line $S$ in the bulk in the Lagrangian algebra $\CL_\rho'$. 
Because of the $S$-dressing, the endpoint of a line $l(j,l)\in\CL_\rho'$ is fermion-odd exactly when $j$ is odd. So $(-1)^F$, which is living on the boundary acts on the group of the boundary charges $\mathbb{Z}^2/\la(2d,-2\zeta')  \ra$ as the character $(-1)^j$. This character is trivial on endpoints of $(2d,-2\zeta')$, but it equals $-1$ on $f$, whose boundary label is $(d,-\zeta')$. The SymTFT braiding, on the other hand, is trivial on $f$, because $f$ is transparent. Thus, we briefly conclude here that despite similarity in various derivations of bosonic and fermionic theories, the fermionic theory always has an additional $\mathbb{Z}_2^F$ hidden in its Lagrangian algebra which is blind to the braiding phase. This also extends our understanding of gauging in the fermionic case from \cite{Tachikawa:2017gyf}.

Now, we are going to study the anomalies of the gauged theory. The gauged group structure is similar to the bosonic case except this additional $\mathbb{Z}_2^F$ that comes with the spin structure.\footnote{We would like to remark that the complete fermionic anomaly of the gauged theory is actually captured by the bordism, i.e., $\Omega^\text{Spin}_3 (B(U(1)\times \mathbb{Z}_g))$ with $g$ odd. At the time being, we do not know the value of this bordism.} Therefore, we primarily extract the anomaly information from the topological spin and braiding phase of symmetry lines \eqref{eq:gauged-symmetry-lines}. 
In this fermionic case, we use the same notations with the bosonic case, i.e., $u_\sigma$ and $w_r$ for $U(1)'$ and $\mathbb{Z}_{g_f}$, which explicit forms are shown in the following:
\be
\begin{aligned}
    u_\sigma &:= \CW_{(0,\sigma/g_f)},\qquad \sigma\in [0,1),\\
    w_r&:= \CW_{(r,-rt/g_fd)},\qquad d':= d/g_f\,,
\end{aligned}
\ee
where $t\in\mathbb{Z}$ satisfying the condition $d'v+(\zeta'/g_f)t=1$ for some integer $v\in\mathbb{Z}$. The three levels of anomalies are then given by
\begin{itemize}
    \item[(i)] Pure $U(1)'$ anomaly is again captured by topological spin of $u_\sigma$ lines, i.e.,
    \be
    \theta(u_\sigma) = \exp(-\frac{2\pi i k'\sigma^2}{2})\,,\qquad k':= k/g_f^2 \in\mathbb{Z}_\text{odd}
    \ee
    \item[(ii)] Mixed $U(1)'\times\mathbb{Z}_{g_f}$ anomaly is given by the braiding phase between two symmetry lines, namely
    \be
    \CB(u_\sigma,w_r) = \exp(\frac{2\pi i\sigma}{g_f}r\nu_f')\,\qquad \nu_f' :=1 + \frac{kt}{g_fd}\in\mathbb{Z}\,.
    \ee
    As in the bosonic case, the shift $t\to t+d's$, where $s$ is some integer, changes $\nu'_f\to \nu'_f + k's$, and $\nu'_f$ is a unit modulo $\gcd(g_f,k')$. Hence the mixed anomaly can be removed by a change of basis if and only if $\gcd(g_f,k/g_f^2) = 1$, i.e., the anomaly is necessarily mixed if and only if $\gcd(g_f,k/g_f^2) > 1$.
    \item[(iii)] Pure $\mathbb{Z}_{g_f}$ anomaly is detected by
    \be
    \theta(w_r) = \exp(-\frac{2\pi i r^2}{2g_f^2}\mu_f'),\qquad \mu_f' := \frac{t}{d'}\left( 2+\frac{kt}{g_fd} \right) \in\mathbb{Z}\,,
    \ee
    where the pure $\mathbb{Z}_{g_f}$ coefficient is defined modulo $g_f$, i.e., $\mu_f'\in\mathbb{Z}_{g_f}$.
\end{itemize}
Similar to the bosonic case, one can derive the boundary $F$--move to identify the anomaly, where $x_1,x_2,x_3\in\CG_f'$, namely
\be
\bar{\alpha}(x_1,x_2,x_3) = p(\sigma_1,\sigma_2)\left[ -\frac{k'}{2}\sigma_3 + \frac{k't}{2d}r_3 \right] + p_g(r_1,r_2)\left(1+\frac{kt}{2g_f d} \right)\left[ \sigma_3 - \frac{t}{d}r_3 \right] \pmod{1/2}
\ee
Each term is read off from the topological spin and braiding phase as in the bosonic case; the reduction modulo $1/2$ is explained in Appendix \ref{app:for-sec-v}. Since $k':= k/g_f^2$ is odd, the gauged $U(1)$ theory has odd level $k'$, and its $\mathbb{Z}_2$ subgroup again carries a Gu-Wen layer with $\nu =2k'\pmod{8}$.

\section{Discussions and Outlooks} \label{section:discussions-and-outlooks}



Let us summarise the results and discuss possible future directions. The lattice SymTFT constructed in this work can be regarded as a concrete realisation of flat $U(1)$ bundles, namely of $BU(1)^\delta$, on which the symmetry defects of
the boundary theory are defined. Since the modified Villain construction enforces flatness, the level
$k\in H^4(BU(1);\mathbb Z)$ and its restriction to finite subgroups, i.e. the bosonic class in
$H^3(B\mathbb Z_d;U(1))$ and, for odd $k$, the Gu--Wen layer, appear on an equal footing as the
$F$-symbols of the same lattice operators. This complements the categorical and operator-algebraic
descriptions of continuous symmetries in \cite{Jia:2025vrj,Jia:2025uun,Jia:2026vcr}, in which the
defect network and its anomaly are characterised abstractly. On the lattice, the fusion, braiding and
junctions of the symmetry defects, including the Majorana fermions at the junctions, are realised by
explicit operators acting on a Hilbert space. It would be interesting to make the dictionary between
the two descriptions precise, and to understand whether the remaining Arf layer of
$\Omega_3^{\rm Spin}(B\mathbb Z_2)$ can be reached by embedding $\mathbb Z_2$ into a different
continuous group.

A natural next step is to apply the same strategy to other global anomalies whose evaluation proceeds through an embedding of symmetry into a larger group. One example is the non-perturbative $\mathbb{Z}_4$ anomaly of a four-dimensional theory with $U(1)\times \mathbb{Z}_2$ symmetry \cite{Davighi:2023mzg}. Its perturbative anomalies vanish, and the $\eta$--invariant that detects the anomaly was computed by embedding $\mathbb{Z}_2\subset U(1)'$, where the anomaly becomes a perturbative one in the $U(1)\times U(1)'$ theory.
Another is Witten's $SU(2)$ anomaly \cite{Witten:1982fp}, which becomes a local anomaly once $SU(2)$ is embedded in
$U(2)$ \cite{Davighi:2020bvi} and leads to fractionally quantised transport \cite{Davighi:2024qmh}. The
new $SU(2)$ anomaly \cite{Wang:2018qoy}, whose five-dimensional anomaly theory is $(-1)^{\int w_2w_3}$,
is of the same type. Extracting the symmetry algebra of these theories along the lines of this work
requires a lattice SymTFT in five dimensions and, in the latter cases, for a non-abelian group. While
SymTFTs for continuous non-abelian symmetries have been proposed in the continuum
\cite{Bonetti:2024cjk,Jia:2025vrj}, their lattice counterparts are less developed. Nevertheless, there are
indications of what such a SymTFT should capture. It was shown in \cite{Okada:2025kie} that, in
four-dimensional theories with the Witten anomaly, position-dependent $SU(2)$ transformations of winding
number one supported on disjoint regions anticommute. This was derived by evaluating the invertible
phase of the anomaly, namely the $\eta$-invariant of a doublet fermion, on $M_3\times T^2$, and is the
four-dimensional counterpart of the statement that, in two dimensions with odd $k$, position-dependent
$U(1)$ transformations of odd winding number are fermionic. Upon restricting $\mathbb Z_n\subset U(1)$
to an interval, the latter implies that the fusion operators at the endpoints are fermionic precisely
when the carry $p$ is non-zero \cite{Okada:2025kie}, which is what the Majorana fermions at the
junctions realise in our lattice SymTFT. It would therefore be interesting to derive the $SU(2)$
counterpart from an $SU(2)$ SymTFT. Encouragingly, a lattice instanton density, with an integer total
instanton number, and a lattice Chern--Simons--Yang--Mills term for $SU(N)$ gauge theory have recently
been proposed via a higher-categorical generalisation of the Villain construction \cite{Chen:2024ddr},
with an explicit construction for $SU(2)$ given in \cite{Zhang:2024sgm}. These may serve as building
blocks for such a SymTFT. Along similar lines, a discrete formula for the second Stiefel--Whitney class
has been obtained in band theory and interpreted in terms of gauged lattice sigma models and
Villainisation \cite{Shiozaki:2024xmq}. Since $b$ is $\mathbb{R}$--valued, the same lattice SymTFT also admits flat gauging of continuous subgroups. In the continuum non-compact SymTFT this produces non-invertible $\mathrm{T}$-duality defects at any radius \cite{Argurio:2024ewp}, and its boundary lattice realization was found in \cite{Argurio:2026txf}. It would be interesting to construct the bulk lattice counterpart, including the case of odd $k$.

Finally, it would be interesting to connect the lattice SymTFT to anomaly-induced transport. In the
continuum, the inflow action of the anomalous $U(1)$, $\frac{k}{4\pi}\int A\,dA$, is that of a Hall
response proportional to $k$, from which the anomaly-induced transport of the boundary theory follows.
Generalised Hall conductivities have recently been realised in local commuting projector models, whose
protected gapless boundaries are constructed with the modified Villain formulation
\cite{Hsin:2025jot}. It would be natural to compute the same responses directly from the lattice SymTFT
of this work. This is particularly interesting for global anomalies, which in four dimensions lead to
fractionally quantised transport coefficients \cite{Davighi:2023mzg,Davighi:2024qmh}. Effective field
theories derived from the SymTFT would be useful in this direction. For instance, the SymTFT with a
suitable Lagrangian algebra reproduces the effective action of the Goldstone mode in the
symmetry-broken phase, including the coupling to the background that matches the anomaly
\cite{Antinucci:2024bcm}; see also \cite{Bonetti:2025dvm} for non-linear realisations of continuous
symmetries in the SymTFT. A lattice version of this construction, including the fermionic boundary
conditions discussed here, would allow one to see how the junction data of discrete anomalies are
reflected in low-energy transport. We leave these questions for future work.
\section*{Acknowledgment}

We would like to thank Jing-Yuan Chen, Joe Davighi, Oleg Evnin, Yoshimasa Hidaka, Kantaro Ohmori, Nat Tantivasadakarn and Yuya Tanizaki for discussions. We would also like to thank Riccardo Argurio for their comments on the manuscript.
The authors are grateful for the opportunity to engage in discussions at the YITP conference  ``Emergent Phenomena in Quantum Matter: From Symmetry to Dynamics" (YITP-T-26-04). Discussions during the long-term workshop "New Frontiers in QCD 2026" at Yukawa Institute for Theoretical Physics (YITP-T-26-01) were useful as we finalize this work. NP would like to thank Yukawa Institute for Theoretical Physics, International Center for Interdisciplinary Science and Education (ICISE) and University of Cape Town for their hospitality. KS was supported in part by the Japanese Government (MEXT) Monbukakusho Scholarship. The work of NP received funding support
from the NSRF via PMU-B, grant number B39G690007.
\begin{appendix}

\section{Some useful identities} \label{app:useful}
This section summarises useful identities we use for the topological theory on the square or (hyper)cubic lattice. Formulas here are not new, see e.g. \cite{Chen:2021ppt,Sulejmanpasic:2019ytl,Jacobson:2023cmr} and \cite{hirani:2203} for a more complete exposition, and are placed here to ensure a uniform notation.

Here we work with the square lattice as a $2-$dimensional cubical cell complex; thus, we will use cell. On the lattice, there are three types of cell associated to different geometric objects: $0-$cell, $1-$cell, and $2-$cell associated to point, line, and face on the lattice. Having this in mind, one can define a chain $c_p\in C_p(\rld;R)$, where $R$ is ring of a group $G$, using cells as
\be \label{eq:p-chain}
c_p :=\sum_{i} c_i \sigma_{p,i}, \sigma_i \in R
\ee
where $c_i$ is a coefficient valued in $R$ and $\sigma_{p,i}$ is the $i-$th oriented $p-$cell. Then one can define a $p-$cochain $\alpha\in C^p(\rld;R)$ as a map $\alpha:C_p(\rld;R)\to R$.
\begin{itemize}
\item \textbf{Cochain Identification and Support:} Given arbitrary $p-$chain \eqref{eq:p-chain} $c_p = \sum_i c_i\sigma_{p,i}$ defined on the oriented $p-$simplices $\{ \sigma_{p,i}\}$, there exists a set of dual basis corresponding to cochain $\{ \sigma^{p,i}\}$ such that
\be
\sigma^{p,i}(\sigma_{p,j}) = \delta^{i}_j
\ee
Thus, we can define a cochain \emph{identification} $\hat{c}$ to represent a cochain spanned by the same dual basis, namely
\be
\hat{c} := \sum_{i}c_i\sigma^{p,i}
\ee
where $c_i$ are coefficients of the given chain
This can be understood as a delta function representing cochain.
In this present work, we are interested in the oriented simplices,; therefore, the associated cochain $\hat{c}$ is defined as
\be
\hat{c}(\sigma) = \begin{cases}
                    +1,\;\sigma \in C\;\text{with matching orientation} \\
                    -1,\;\sigma \in C\;\text{with opposite orientation} \\
                    0,\;\sigma\neq C
                    \end{cases}
\ee
It is necessary to state that $\hat{c}$ is just an identification of cochain not its support. For a given cochain, for example, $\alpha\in C^p(\rld;\bullet)$, the support of this cochain is a set of $p-$simplices, which are geometric objects living on the lattice, defined as
\be
\text{supp}(\alpha) := \{ \sigma_p\subset \rld \vert \alpha(\sigma_p)\neq 0\}
\ee
From the definition, the support is basically a collection of simplices representing geometric realization of cochain.
\item \textbf{Differential operators on lattice:}
It should be noted that there are multiple ways to define derivative and differential operators on the lattice. Let us start with the simplest one namely the forward derivative $\Delta_\mu$ and backward derivative $\nabla_\mu$ (see e.g. Appendix A of \cite{Xu:2024hyo})
\be 
(\Delta_\mu X)_{\bx,\bullet} := X_{\bx +\hat \mu,\bullet} - X_{\bx,\bullet} \,,\qquad (\nabla_\mu X)_{\bx,\bullet} := X_{\bx,\bullet} - X_{\bx -\hat\mu,\bullet}
\ee 
Summation by parts exchanges the two derivatives namely 
\be \label{eq:forward-backward-relation}
\begin{aligned}
\sum_\bx (\Delta_\mu X)_\bx Y_\bx &= \sum_\bx \left(X_{\bx+\hat \mu} Y_\bx - X_{\bx} Y_{\bx}\right)\,,\\
&= \sum_{\bx} X_\bx Y_{\bx-\hat \mu} - \sum_\bx X_{\bx} Y_{\bx} = -\sum_\bx X_\bx (\nabla_\mu Y)_\bx 
\end{aligned}
\ee 
There are two different differential operators on the lattice, the boundary operator $\partial$ and the coboundary/exterior derivative operator $d$ given by
\be
\begin{aligned}
    \partial: C_{n+1}(X;R) \to C_{n}(X;R),\qquad d: C^{n}(X;R) \to C^{n+1}(X;R)
\end{aligned}
\ee
where $R$ is a coefficient ring. Both are nilpotent i.e. $\d^2 = d^2 =0$ but behave differently. The boundary operator acting on 0- and 1-form can be written as
\be 
\begin{aligned} \label{eq: boundary derivative property}
(\d a^{(0)}) &= 0\,,\\
(\d a^{(1)})_\bx &= (a^{(1)}_{\bx-\hat 1,1} - a^{(1)}_{\bx,1}) + (a^{(1)}_{\bx-\hat 2,2} - a^{(1)}_{\bx,2}) = (\nabla_1 a^{(1)})_{\bx,1} + (\nabla_2 a^{(1)})_{\bx,2}\,,\\
(\partial a^{(2)})_{\bx,i}&=\epsilon_{ij}(a^{(2)}_{\bx,12}-a^{(2)}_{\bx -\hat j,12}) = \epsilon_{ij} (\nabla_j a^{(2)})_{\bx,12}
\end{aligned}
\ee 
and the coboundary operator is behave as 
\be 
\begin{aligned} \label{eq: exterior derivative property}
(da^{(0)})_{\bx,i} &= a^{(0)}_{\bx+\hat i} -a^{(0)}_\bx = (\Delta_i a^{(0)})_\bx\,,\\
(d a^{(1)})_{\bx,12} &= a^{(1)}_{\bx,1} + a^{(1)}_{\bx+\hat 1,2} - a^{(1)}_{\bx+\hat 2,1} - a^{(1)}_{\bx,2} = (\Delta_1 a^{(1)})_{\bx,2} - (\Delta_2 a^{(1)})_{\bx,1}\,,\\
(da^{(2)}) &=0
\end{aligned}
\ee 
\item \textbf{Cup product on lattice:}
The cochain can be combined with the Steenrod cup product $\cup_i$ which takes $a^{(p)}\in C^p(X;R)$ and $b^{(q)}\in C^q(X;R)$ on some ring $R$ via 
\be 
\cup_i \quad: C^p \cup C^q \to C^{p+q-i}
\ee 
The ordinary cup product $\cup_0 :=\cup$ satisfies the Leibnitz rule upon the coboundary operator 
\begin{equation}
\begin{aligned}
    d (a^{(p)} \cup b^{(q)}) &= d a^{(p)} \cup b^{(q)} + (-1)^p a^{(p)}\cup d b^{(q)} \\
    \int d a^{(p)} \cup b^{(q)} &= -(-1)^p \int a^{(p)}\cup d b^{(q)} + \text{boundary term}
\end{aligned}
\end{equation}
The conventional cup product is also not super-commutative as it involves terms with higher cup product, namely 
\be \label{eq:cup-commutator}
a^{(p)}\cup b^{(q)} - (-1)^{pq} b^{(q)}\cup a^{(p)} = (-1)^{p+q+1} \left[ d(a^{(p)}\cup_1 b^{(q)}) - d a^{(p)}\cup_1 b^{(q)} - (-1)^p a^{(p)}\cup_1 d b^{(q)} \right]
\ee 

The cup product of cochain on lattice can be completely expressed in terms of lattice components in two dimensions (see Appendix A of \cite{Jacobson:2023cmr} for graphical representations of the cup product), namely
\begin{equation} \label{eq:cup_prod}
\begin{aligned}
\left(a^{(0)}\cup b^{(1)}\right)_{\bx,1} &= a_{\bx} b_{\bx,1},\; \left( b^{(1)} \cup a^{(0)} \right)_{\bx,1} = b_{\bx,1} a_{\bx+\hat 1}
\,,\\
\left( a^{(1)} \cup b^{(1)} \right)_{\bx,12} &= a_{\bx,1}b_{\bx+\hat 1,2} - a_{\bx,2} b_{\bx+\hat 2,1},\;\left(a^{(0)}\cup b^{(2)}\right)_{\bx,12} = a_{\bx} b_{\bx,12}\,,\\
\left(b^{(2)}\cup a^{(0)}\right)_{\bx,12} &= b_{\bx,12} a_{\bx + \hat s}\,,\quad \hat s = \hat 1+\hat 2 \\
\left( a^{(1)}\cup_1 b^{(1)} \right)_{\bx,1} &= a_{\bx,1}b_{\bx,1},\;\left( a^{(1)}\cup_1 b^{(2)} \right)_{\bx,12} = -(a_{\bx,2} + a_{\bx+\hat 2,1})b_{\bx,12}\,,\\
\left( b^{(2)}\cup_1 a^{(1)} \right)_{\bx,12} &= b_{\bx,12}(a_{\bx,1}+a_{\bx+\hat 1,2}),\;\left( a^{(2)}\cup_2 b^{(2)} \right)_{\bx,12}= a_{\bx,12}b_{\bx,12}
\end{aligned}
\end{equation}
where $\bx$ is an arbitrary site of the physical lattice $\rld$, and $\hat{1}$ and $\hat{2}$ are the horizontal and vertical unit vectors, respectively. Point-like objects, such as $a_\bx$, are localized on lattice sites. A link object starting at $\bx$ and pointing in the $\hat{i}$ direction is denoted by $a_{\bx,i}$, while $b_{\bx,12}$ resides on the plaquette whose lower-left corner is located at $\bx$.
\item \textbf{Lattice Hodge star dual:} Another useful notation is the relation between the physical lattice $\rld$ and its dual lattice $\tilde{\rld}$. The dual lattice $\tilde\rld$ can be obtained by performing a diagonal translation by $\hat{s}/2$, where $\hat{s}=\hat{1}+\hat{2}$. Given a $p-$cell on $\rld$, there is a naturally associated $(d-p)-$simplex of the dual lattice $\tilde\rld$ which it intersects/pierces, and conversely. These intersecting objects are crucial for defining the geometric support of the $\star-$dual on the lattice. From the diagonal translation map between lattice and the definition of cochain identification, we can define the dual map $\star$, which can be understood as a lattice Hodge dual mapping between cochains.
\be
\star : C^{p}(\rld;\bullet)\rightarrow C^{d-p}(\tilde\rld;\bullet)
\ee
Then for a chain $c_p\in C_p(\rld;R)$ with basis $\{ \sigma_i \}$, there is an associated cochain defined by $\hat{c}_p(\sigma_p) = c_\sigma$. The corresponding $\star-$dual is then given by
\be
\star\hat{c}_p:C^p(\rld;R)\to C^{d-p}(\tilde\rld;R)
\ee
whose supports are $(d-p)-$cell living on the dual lattice, namely
\be
\text{supp}(\star\hat{c}_p) = \{ \tilde{\sigma}_{d-p}\in\tilde{\rld} \vert (\star\hat{c}_p)(\tilde{\sigma}_{d-p})\neq 0\}
\ee
Similarly, one can construct a $\star-$dual of chain $\tilde c_p\in C_p(\tilde\rld;R)$ living on the dual lattice $\tilde\rld$. Its $\star-$dual's supports will be living on the physical lattice $\rld$, but we are not going to use this throughout the paper. Here $\sigma_p$ denotes a generic oriented $p-$cell of $\rld$; explicitly $\sigma_{0} = \sigma_{\bx},\; \sigma_1 =\sigma_{\bx,i},\; \sigma_2 = \sigma_{\bx,12}$
We denote by $\hat{\sigma}_p$ the cochain associated with the cells $\sigma_p$. The supports of their $\star-$duals are
\be
\begin{aligned}
    \text{supp}(\star \hat{\sigma}_\bx) &= \{ \tilde{\sigma}_{\bx-\hat{s}/2,12} \}, \\
    \text{supp}(\star\hat{\sigma}_{\bx,1}) &= \{ \tilde{\sigma}_{\bx+\hat{1}/2-\hat{2}/2,2 } \} \\
    \text{supp}(\star\hat{\sigma}_{\bx,2}) &= \{ \tilde{\sigma}_{\bx-\hat{1}/2+\hat{2}/2,1} \} \\
    \text{supp}(\star\hat{\sigma}_{\bx,12} ) &= \{\tilde{\sigma}_{\bx+\hat{s}/2} \}
\end{aligned}
\ee
For $d=2,p=1$, the support of the $\star-$dual corresponds to a $90-$degree clockwise rotation of the original support link.
The corresponding $\star-$dual cochains carry orientation coefficients determined by the fixed branching structure. In particular,
\be
(\star \hat{\sigma}_{\bx,1}) (\tilde{\sigma}_{\bx+\hat{1}/2-\hat{2}/2,2}) = -1
\ee
because the clockwise rotation maps the $+\hat 1$ orientation to $-\hat 2$, which is opposite to the fixed branching structure.
We denote the geometric collection of dual cells specified by these supports collectively by $\widetilde{C}$.

\item \textbf{The translation dual} As discussed above, given a cochain on the dual lattice, its geometric
support is a collection of cells on $\tilde\rld$. We then define a cochain map from the dual lattice to the physical lattice by performing a diagonal translation on the cochain's supports, namely
\be
\begin{aligned} \label{eq: iota-map}
    \iota: C^p(\tilde\rld;R) &\to C^p(\rld;R) \\
    (\iota{\tilde{\alpha}})_{\tilde\bx,\bullet} &= {\tilde{\alpha}}_{\tilde\bx+\hat{s}/2.\bullet},\quad \tilde{\alpha}\in C^p(\tilde\rld;R)
\end{aligned}
\ee
where $\tilde\bx \in \tilde\rld$ are sites on the dual lattice. This map also induces a map between supports from one cochain to another, namely
\be
\text{supp}(\iota\tilde{\alpha}) = \iota(\text{supp}(\tilde{\alpha})),\quad \tilde{\alpha}\in C^p(\tilde\rld;R)
\ee
This means that the map $\iota$ translates every $p-$cell in the support of a cochain $\tilde\alpha\in C^p(\tilde\rld;R)$ on the dual lattice to the corresponding $p-$cell on the physical lattice. This map will be very useful when it is applied to the $\star$-dual of cochain $\hat{c}_p$, namely the $\hat{m}:=\iota(\star \hat{c})$ is given by
\be \label{eq:support-of-m-cochain}
\begin{aligned}
    \text{supp}(\iota(\star\hat{c}_\bx) &= \{ \sigma_{\bx,12} \} \\
    \text{supp}(\iota(\star\hat{\sigma}_{\bx,1})) &= \{\sigma_{\bx+\hat 1,2} \} \to \text{supp}(\iota(\hat{\sigma}_{\bx-\hat{1},1})) = \{\sigma_{\bx,2} \}\\
    \text{supp}(\iota(\star\hat{\sigma}_{\bx,2})) &= \{\sigma_{\bx+\hat 2, 1} \}\to \text{supp}(\iota(\hat{\sigma}_{\bx-\hat{2},2}) = \{ \sigma_{\bx,1} \}\ \\
    \text{supp}(\iota(\star\hat{\sigma}_{\bx,12})) &= \{ \sigma_{\bx+\hat{s}} \}
\end{aligned}
\ee
where the positions in the second and the third equations are relabeled for simplicity. This map is explicitly shown in Fig. \ref{fig:iota-map-explain}. The results of this map are supports living on the physical lattice; thus, we use the notation for physical lattice's cell $\sigma$ here. The second and the third equations imply that in the level of the associated cochain, the explicit component expression of $\hat{m}$ are
\be\label{eq:m-in-term-of-c}
\hat{m}_{\bx,1} = \hat{c}_{\bx-\hat{2},2},\;\qquad \hat{m}_{\bx,2} = -\hat{c}_{\bx-\hat{1},1}
\ee
In particular, one can check that using the property of boundary and coboundary operator, one can shown that
\be 
(d\hat{m})_{\bx,12} = \widehat{(\partial c)}_{\bx}
\ee
This equation geometrically implies that the base of $d\hat{m}$ coincides with the boundary of the chain $c$. Another important relation is that upon commuting $\d$ and $d$ on $\rld$ with $\star-$dual it turns into the dual operator on the dual lattice $\tilde\rld$ 
\be 
d\star = \star \tilde \d \,,\qquad \d \star = \star \tilde d
\ee 
Then by applying the pullback map $\iota$, the derivatives on the dual lattice are mapped back to the original lattice $\rld$, namely
\be 
d(\iota \star) = \iota(\star \d)
\ee 
\item \textbf{Intersection number on lattice:}
Note that there is a combination that will play a crucial role in defining the (framed) Wilson loop. For any two 1-cochains $\hat C$ and $a$, we have the following identities.
\be \label{eq:property-of-c-cup-a}
\begin{aligned}
\int_\rld \left(\hat C\cup a - a\cup \hat C\right)&= \sum_{\bx} \left[ \left(\hat C_{\bx,1} a_{\bx +\hat 1,2} - \hat C_{\bx,2} a_{\bx + \hat 2,1}\right) -\left(a_{\bx,1} \hat C_{\bx +\hat 1,2}- a_{\bx,2}\hat C_{\bx+\hat 2,1} \right)\right]
\,,\\
&=\epsilon^{ij}\sum_{(\bx,i)}\hat{C}_{\bx,i}(a_{\bx+\hat{i},j}+a_{\bx-\hat{j},j}) \,,\\
\end{aligned}
\ee 
Since \eqref{eq:property-of-c-cup-a} holds for any 1--cochain, setting $\hat{C}\to\hat{m}=\iota(\star\hat{C})$ and using \eqref{eq:m-in-term-of-c} gives
\be
\begin{aligned}
    \int_\rld\big(\hat m\cup a-a\cup\hat m\big)
    &=\sum_{\bx}\left(\hat C_{\bx-\hat2,2}\,a_{\bx+\hat1,2}+\hat C_{\bx-\hat1,1}\,a_{\bx+\hat2,1}
    +\hat C_{\bx,1}a_{\bx,1}+\hat C_{\bx,2}a_{\bx,2}\right)\\
    &=\sum_{(\bx,i)}\hat C_{\bx,i}\left(a_{\bx,i}+a_{\bx+\hat s,i}\right)
\end{aligned}
\ee 
where the last equality is obtained by relabeling $\bx\to\bx+\hat2$ and $\bx\to\bx+\hat1$ in the first and the second term, respectively.
\be \label{eq:def-intersection-number}
\begin{aligned}
    \text{Int}(C_1,C_2) &:= \frac{1}{2} \int (\hat C_1\cup \hat C_2 - \hat C_2\cup \hat C_1) = \int \hat C_1\cup \hat C_2 \\
    \int_\rld \hat{m}_1\cup\hat{m_2} &= \int_\rld \hat{C_1}\cup \hat{C}_2,\;\text{for}\;\hat{m}=\iota(\star\hat{C})
\end{aligned}
\ee 
where the last equality of the first line only holds in case $\partial C_1 = \partial C_2 = 0$. The lattice intersection number has a built-in framing (point-splitting) since the cup product evaluates its two factors at relatively shifted position.
\end{itemize}
\section{Detailed Derivations of Non-compact theory} \label{app: quantize-noncompact}
For non-compact theory, we decompose  the action into lattice and temporal components as follows.
\begin{align}
    a = a +dt a_0,\quad b = b + dt b_0. \quad d = d+dt\partial_0
\end{align}
where variables with subscript $0$ denote temporal components. Under these decompositions, the action becomes
\be 
\begin{aligned}
    -\CS &=\frac{i}{2\pi}\int_{\mathbb{R}}dt\int_{\rld} b\cup ( \dot a -da_0) - b_0\cup da + \frac{ik}{4\pi} \int_{\mathbb{R}}dt\int_{\rld} a\cup ( \dot a - da_0 ) - a_0\cup da \\
     &=\frac{i}{2\pi}\int_{\mathbb{R}}dt\sum_{p\in\rld} [b\cup (\dot{a} -da_0)]_p - [b_0\cup {da}]_p +\frac{ik}{4\pi}\int_{\mathbb{R}}dt\sum_{p\in\rld} [a\cup ( {\dot{a}} - da_0 )]_p - [a_0\cup{da}]_p
\end{aligned}
\ee
where the last equation is expressed in terms of the plaquette $p\in\rld$ sum on the lattice. We then translate this into a complete lattice expression.
\begin{align}
    -\CS &= \frac{i}{2\pi} \int_{\mathbb{R}_t}dt\sum_{\bx\in\rld}\epsilon^{ij}b_{\bx,i}\dot{a}_{\bx+\hat{i},j} - (d b)_{\bx-\hat{s},12}(a_0)_\bx -(b_0)_\bx(da)_{\bx,12} \nn\\
    &\qquad+\frac{ik}{4\pi}\int_{\mathbb{R}_t}dt\sum_{\bx\in\rld}\epsilon^{ij}a_{\bx,i}\dot{a}_{\bx+\hat{i},j}-(a_0)_\bx[ (da)_{\bx-\hat{s},12} + (da)_{\bx,12} ] \nn \\
    &= \frac{i}{2\pi}\int_{\mathbb{R}_t}dt\sum_{\bx\in\rld}\Big(\epsilon^{ij}[ b_{\bx,i}\dot{a}_{\bx+\hat{i},j} + \frac{k}{2}a_{\bx,i}\dot{a}_{\bx+\hat{i},j} ] - (b_0)_\bx(da)_{\bx,12} \nn \\
    &\qquad - (a_0)_\bx[(db)_{\bx-\hat{s},12}+\frac{k}{2}(da)_{\bx-\hat{s},12} + \frac{k}{2}(da)_{\bx,12}]\Big)
\end{align}
From the second equality, if we integrate over $b_0$ and $a_0$, we read the following constraints.
\be \label{eq: double R Gauss laws}
\begin{aligned} 
    (\CG_B)_{\bx,12} &= (da)_{\bx,12} = 0, \\
    (\CG_A)_{\bx,12} &= (db)_{\bx-\hat{s},12} + \frac{k}{2}(da)_{\bx-\hat{s},12} + \frac{k}{2}(da)_{\bx,12} = 0
\end{aligned}
\ee 
These are known as (modified) Gauss's laws on the lattice \cite{Chen:2017fvr}. Next, we can compute the canonical commutation relations by imposing a symplectic structure, obtaining the commutation relations \eqref{eq:BF-CS-commutator}.
To see how gauge transformations manifest in lattice computations, we must look for the commutators that generate them. Firstly, using lattice identity \eqref{eq: exterior derivative property}, one obtain that
\begin{align}
    (da)_{\bx,12} &= a_{\bx,1} + a_{\bx+1,2} - a_{\bx+\hat{2},1} - a_{\bx,2}, \nn\\
    (db)_{\bx,12} &= b_{\bx,1} + b_{\bx+\hat{1},2}-b_{\bx+\hat{2},1} - b_{\bx,2}
\end{align}
Then, by using commutators in \eqref{eq:BF-CS-commutator}, one can show that
\be
\begin{aligned}
[(da)_{\bx,12},b_{\by,i}] &= 2\pi i (\delta_{\bx,\by+\hat i} - \delta_{\bx,\by})\,,\\
[(db)_{\bx,12},a_{\by,i}] &= 2\pi i (\delta_{\bx+\hat i_\perp,\by} - \delta_{\bx+\hat s,\by})\,,\\
[(db)_{\bx,12},b_{\by,i}] &= -i\pi k  (\delta_{\bx,\by+\hat i} - \delta_{\bx,\by} + \delta_{\bx+\hat i_\perp,\by} - \delta_{\bx+\hat s,\by})\,,\\
\end{aligned}
\ee 
These commutators are more convenient to work with. We can now check how the Gauss's law operators act on these commutators; that is,
\be \nonumber
\begin{aligned}
\sum_\bx (\lambda_B)_\bx [(\CG_B)_{\bx,12}, a_{\by,i}] &= 0\\
\sum_\bx (\lambda_B)_\bx [(\CG_B)_{\bx,12}, b_{\by,i}]&=
\sum_\bx (\lambda_B)_\bx [(da)_{\bx,12}, b_{\bx,i}] = -2\pi i (d\lambda_B)_{\by,i}\,,\\
\sum_\bx (\lambda_A)_\bx [(\CG_A)_{\bx,12}, a_{\by,i}]&=\sum_\bx (\lambda_A)_\bx[ (db)_{\bx,12},a_{\bx,i}] = \sum_\bx (\lambda_A)_\bx( \delta_{\bx -\hat i,\by}- \delta_{\bx,\by})= -(d\lambda_A)_{\by,i}\,,\\
\sum_\bx (\lambda_A)_\bx [(\CG_A)_{\bx,12}, b_{\by,i}]&= \sum_\bx(\lambda_A)_\bx \Big([(db)_{\bx -\hat s,12},b_{\by,i}] + [(da)_{\bx -\hat s,12},b_{\by,i}] + [(da)_{\bx,12},b_{\by,i}]\Big)= 0
\end{aligned}
\ee 
where $\lambda_A$ and $\lambda_B$ are real-valued $0$--cochain, i.e. $\lambda_A,\lambda_B\in C^{0}(X;\mathbb{R})$
These imply that operators generating gauge transformations are generated by Gauss's law operators, namely gauge transformation operators are
\be
G[\lambda_A, \lambda_B] = \exp\left( -\frac{i}{2\pi}\sum_\bx (\lambda_A)_\bx (\CG_A)_{\bx,12}-\frac{i}{2\pi}\sum_\bx (\lambda_B)_\bx (\CG_B)_{\bx,12} \right)
\ee 
Moreover, it indicates that the only invariant operators are $\int_C a$ and $\int_C b$ with a closed $C$, i.e,. $\d C = 0$. Both of these are automatically topological due to the Gauss law which implies that both $da = db = 0$. Therefore, one concludes that corresponding topological operators of this theory are 
\be 
\begin{aligned}
W_q[C] &= \exp\left(i q\int_C a\right) = \exp\left(i q \sum_{\bx,i} \hat C_{\bx,i} a_{\bx,i} \right)\,,\\
V_\phi[C] &= \exp\left(i \phi\int_C b\right) = \exp\left(i \theta \sum_{\bx,i} \hat C_{\bx,i} b_{\bx,i} \right)
\end{aligned}
\ee 
where $q,\phi\in\mathbb{R}$. Both coefficients are $\mathbb{R}-$valued as there is no any condition to further constraints them.

\section{Detailed Derivations of Compact theory} \label{app: quantize-compact}
The SymTFT for a $U(1)-\mathbb{R}$ theory was already discussed in the continuum limit in \cite{Antinucci:2024zjp, Antinucci:2024bcm, Brennan:2024fgj} and its lattice version in \cite{Chen:2019mjw}.
In the previous section, we introduced a double $\mathbb{R}$ BF+CS theory using the modified Villain construction. To turn a non-compact theory into a compact, one can use a gauging mechanism; that is,
\be
\mathbb{R} \xrightarrow{\text{Gauge } 2\pi\mathbb{Z}} U(1) \cong \mathbb{R}/(2\pi\mathbb{Z})
\ee
The Villain construction in \cite{Jacobson:2023cmr,Jacobson:2024hov} dictates that there is also $ n \in C^2(\mathrm{\rld};\mathbb{Z})$, representing the local (integer-valued) magnetic flux. This integer flux $ n$, indeed, represents the first homotopy group of the $U(1)$ group. In particular, introducing large gauge transformations corresponds to gauging the $2\pi \mathbb{Z}$ subgroup. 
Note that $\lambda_A$ becomes a $U(1)$--valued cochain after one introduced the large gauge transformation. The modified Villain action for the $U(1)-\mathbb{R}$ theory that is invariant under the gauge transformations \eqref{eq:complete-gauge-transformation-U1} is given in equation \eqref{eq:action-BF-CS}, which must be invariant under both large and small gauge transformations. To perform canonical quantisation, one immediately finds that the commutators involving the 1-cochains $a$ and $ b$ are exactly identical to those in \eqref{eq:BF-CS-commutator}.
However, the compact theory is richer, and we have additional commutators from the extra fields $\tilde\varphi$ and $ n$, namely
\be \label{eq:commutator-CS-U1-n-lattice}
\left[\tilde\varphi_\bx,  n_{\by,12}\right] = i\delta_{\bx,\by},\; \tilde\varphi_\bx := \varphi_\bx -\frac{k}{2}(a_{\bx,2}+a_{\bx+\hat 2,1})
\ee 
One can then verify that the other commutators are all trivial, such that
\be 
[a,a] =[a, n] = [a,\tilde\varphi] = [b,\tilde\varphi] =[b, n] = [ n, n] = 0
\ee 
Note that, due to $[a,  n] = 0$, both the dressed scalar $\tilde{\varphi}$ and the bare scalar $\varphi$ commute with $ n$. This implies that both scalars $\tilde{\varphi}$ and $\varphi$ are conjugates of $ n$. However, the bare scalar $\varphi$ does not commute with $b$, namely one can check that from the symplectic structure
\be
[\varphi_\bx,b_{\by,i}] = \frac{2\pi i k}{2}\left( \delta_{i1}\delta_{\bx,\by+\hat{1}}-\delta_{i2}\delta_{\bx,\by} \right)
\ee
Similar to the previous case, the modified Gauss's laws generate small gauge transformations on the fields. First, we consider gauge transformations generated by $\CG_A$, one finds that
\begin{equation} \label{eq:action-of-gauss-A-op-on-fields}
    \begin{aligned}
        \sum_\bx(\lambda_A)_\bx [ (\CG_A)_{\bx,12},a_{\by,i} ] &= -2\pi i (d\lambda_A)_{\by,i},\;\quad \sum_\bx (\lambda_A)_\bx[(\CG_A)_{\bx,12}, n_{\bx,12}] =0\,,\\
        \sum_{\bx}(\lambda_A)_\bx[ (\CG_A)_{\bx,12},b_{\by,i} ] &= 0,\;\quad \sum_\bx(\lambda_A)_\bx[(\CG_A)_{\bx,12},\tilde\varphi_\by] = \frac{2\pi ik}{2}\left( (\lambda_A)_\by + (\lambda_A)_{\by+\hat{s}} \right)
    \end{aligned}
\end{equation}
Here the only tricky derivation is for the third equality, namely
\be 
\sum_\bx (\lambda_A)_\bx [(\CG_A)_{\bx,12},b_{\by,i}] = O_1+O_2 +O_3 = 0
\ee
where each term is given by
\be  
\begin{aligned}
O_1 &= \sum_\bx (\lambda_A)_\bx[ (db)_{\bx-\hat{s},12}, b_{\by,i}] = +i\pi k\Big( (d\lambda_A)_{\by,i}+ (d\lambda_A)_{\by+\hat{s},i} \Big)\,,\\
O_2 &= \frac{k}{2}\sum_\bx (\lambda_A)_\bx[(da)_{\bx,12},b_{\by,i}] =- i k \pi (d\lambda_A)_{\by,i} \,,\\
O_3&= \frac{k}{2}\sum_\bx(\lambda_A)_\bx[(da)_{\bx-\hat{s},12},b_{\by,i}] = -ik\pi (d\lambda_A)_{\by+\hat{s},i}
\end{aligned}
\ee 
One can see that $O_1 + O_2 + O_3 = 0$; therefore, $\CG_A$ does not induce any transformation on the field $b$. Second, we consider gauge transformations generated by $\CG_B$
\begin{equation} \label{eq:action-of-gauss-B-op-on-fields}
    \begin{aligned}
        \sum_{\bx}(\lambda_B)_\bx[(\CG_B)_{\bx,12},b_{\by,i}] &=-2\pi i (d\lambda_B)_{\by,i},\;\quad \sum_\bx (\lambda_B)_\bx [(\CG_B)_{\bx,12}, \tilde\varphi_{\by}] = 2\pi i (\lambda_B)_\by \,,\\
        \sum_\bx(\lambda_B)_{\bx}[(\CG_B)_{\bx,12},a_{\by,i}] &=  \sum_\bx(\lambda_B)_{\bx}[(\CG_B)_{\bx,12}, n_{\by,i}] = 0,\\
    \end{aligned}
\end{equation}
Then we define Gauss's law operator $G[\lambda_A,\lambda_B]$ \eqref{eq:gauss-law-operator-G} which acts as $G^\dagger[\lambda_A,\lambda_B]\mathcal{C}G[\lambda_A,\lambda_B]$ summarized in \eqref{eq:action-of-gauss-law-operato}

On the lattice, using the canonical commutation relations \eqref{eq:BF-CS-commutator} for the non-compact theory and \eqref{eq: complete-commutators-compact} for the compact theory, can give rise to an integer known as the intersection number.
\begin{subequations}\label{eq:BF-CS-commutator-integral}
\be 
\begin{aligned}
[\int_C a ,\int_{C'}b] &= \sum_{\bx,i}\sum_{\by,j} (-2\pi i \epsilon_{ij} \delta_{\bx,\by+\hat j}) = -2\pi i \text{Int}(C,C') \in 2\pi \mathbb{Z}
\end{aligned}
\ee 
where the term in the sum is nothing but the intersection number when $C,C'$ are two closed 1-chain. Similar game can be played for the other nontrivial commutator namely 
\be \label{eq:algebra-of-b}
[\int_C b,\int_{C'}b] = 2\pi k
 i\epsilon^{ij} \sum_{\bx} \left(\frac{ \hat C_{\bx,i}\hat {C'}_{\bx -\hat j,j} + \hat C_{\bx,i}\hat {C'}_{\bx+\hat i,j}     }{2}  \right) = 2\pi i k \mathrm{Int}(C,C')
\ee 
\end{subequations}
%
Terms in the sum becomes an ordinary intersection number, $\text{Int}(C,C')$, for closed $C,C'$.
Another neat thing is that for $C'\to C$, the r.h.s. shift the curve from $\bx \to \bx + \hat s$ and act as a built-in framing/regulator of the self-intersection on the lattice. From the main text, we constructed the following operators
\be \nn
\begin{aligned}
    W^\text{fr}_q[m]
&= \exp\left(\frac{iq}{2} \int_\rld \left[ m\cup  a - a \cup m + 2\pi m\cup_1 n\right]\right)\\
V_\phi[C] &:= \exp(i\phi\sum_{(\bx,i)}\hat{C}_{\bx,i}b_{\bx,i})
\end{aligned}
\ee
Note that the framed Wilson line is defined from summing over plaquettes, but here we use integration notation for convenience.
It is straightforward to see that $V_\phi$ is a topological operator, namely we consider deforming the line $C_{\bx,i}\to C_{\bx,i}+(\partial D)_{\bx,i}$, where $(\partial D)_{\bx,i}=\epsilon_{ij}(D_{\bx,12}-D_{\bx-\hat{j},12})=\epsilon_{ij}(\nabla_j D)_{\bx,12}$. To see the variation from this deformation, one can consider the following ratio
\be
\begin{aligned}
    \frac{V_\phi[C+\partial D]}{V_\phi[C]} &= \exp\left( i\phi\sum_{(\bx,i)}\epsilon_{ij}(\nabla_j D)_{\bx,12}b_{\bx,i} \right) =\exp(i\phi\sum_{(\bx,i)}\epsilon_{ij}D_{\bx,12}(b_{\bx,i}-b_{\bx+\hat{j},i}))\\
    &= \exp(i\phi\sum_{(\bx,i)}D_{\bx,12}(db)_{\bx,12}) = G[\lambda_A,\lambda_B]
\end{aligned}
\ee
To obtain the last equality, we can express $db$ in terms of Gauss laws, i.e., $(db)_{\bx,12}= (\CG_A)_{\bx+\hat{s},12} - \frac{k}{2}((\CG_B)_{\bx,12})+(\CG_B)_{\bx+\hat{s},12})$, then one can identify $(\lambda_A)_\bx := -2\pi \phi D_{\bx-\hat{s},12}$ and $(\lambda_B)_\bx:= \pi k \phi(D_{\bx-\hat{s},12}+D_{\bx,12})$. Thus, the variation of 't Hooft line yields a Gauss law operator, implying that the operator is topological without any constraint on $\phi$, i.e., $\phi\in\mathbb{R}$. 

Next, we can check that the framed Wilson line is also a topological operator. Recall that the cochain $\hat m:=\iota(\star\hat{C})\in C^1(\rld;\mathbb{Z})$, so that its variation is $m\to m+ d\sigma$, where $\sigma \in C^0(\rld;\mathbb{Z})$ is some 0--cochain. Again, let's compute the following ratio:
\be
\begin{aligned}
    \frac{W^\text{fr}_q[m+d\sigma]}{W^\text{fr}_q[m]} &= \exp(\frac{iq}{2}\sum_{\bx}d\sigma\cup a - a\cup d\sigma + 2\pi d\sigma\cup_1 n) \\
    &= \exp(\frac{iq}{2}\int_\rld -\sigma\cup da - da \cup\sigma + 2\pi \sigma\cup n - 2\pi n \cup\sigma) \\
    &= \exp(\frac{iq}{2}\int_\rld - (\sigma\cup F + F\cup \sigma) - 4\pi n\cup \sigma)
\end{aligned}
\ee
where $F_p = (\CG_B)_p$ is the Gauss law constraint on each plaquette.
To obtain the second equality, we applied by-part integration and the fact that $\sigma\cup n - n\cup \sigma = d\sigma\cup_1 n$. The final equality is obtained from grouping terms in the variation to reveal the Gauss law terms from the variation. Therefore, by Gauss law constraint, the framed Wilson line is topological under a requirement that $q\in \mathbb{Z}$, which is a constraint from the term $4\pi n\cup\sigma$.

One can also check that, using the commutation relation \eqref{eq:commutator-CS-U1-n-lattice}, $\mathcal{D}_\bx$ has a nontrivial action on the integer-valued magnetic flux $ n_{\bx,12}$.
\be
[\mathcal{D}_\bx, n_{\by,12}] = -\mathcal{D}_\bx \delta_{\bx,\by}
\ee
Therefore, $\mathcal{D}_\bx$ shifts the magnetic flux $ n_{\by,12}$ by one unit since
\begin{align}
    \mathcal{D}_\bx  n_{\by,12}\mathcal{D}_\bx^\dagger &=  n_{\by,12} - \delta_{\bx,\by} \\
    \mathcal{D}^\dagger_\bx  n_{\by,12} \mathcal{D}_\bx &=  n_{\by,12} + \delta_{\bx,\by}
\end{align}
%

There are important details that one ought to point out. The first is that for $k \in \mathbb{Z}_{\text{odd}}$, one can show that $U[C]$ and $U[C']$ for any 1-chains $C, C'$ act projectively on the Hilbert space, namely:
\be 
U[C]U[C'] = U[C']U[C] e^{i\Phi}
\ee 
where, for $X[C] = \sum_{\bx,i\in\rld} \Big(\hat C_{\bx,i}b_{\bx,i} +\tilde\varphi_\bx \widehat{(\d C)}_{\bx} + \frac{k}{2}\hat C_{\bx,i}(a_{\bx,i} + a_{\bx+\hat s,i})  \Big)$, we have 
\be 
\begin{aligned}
\Phi&= i [X[C],X[C']]\,,\\
&= \underbrace{i[ \la C,b\ra, \la C',b\ra]}_{-2\pi k \text{Int}(C,C')} + \underbrace{\frac{ik}{2}\left([\la C,b\ra, \la C'+ C'_{-\hat s},a\ra] + (C\leftrightarrow C') \right)}_{4\pi k \text{Int}(C,C')}\\
&= 2\pi k \text{Int}(C,C') = \pi k \int_\rld (\hat C\cup \hat C' - \hat C'\cup \hat C)
\end{aligned}
\ee 
This is exactly the same as that in Eq. (61) of \cite{Jacobson:2024hov}. It also implies that one cannot set $U[C] = 1$; thus, one cannot consistently define the physical Hilbert space. Using properties of the cup product, we can further check that this phase is non-trivial for $k \in \mathbb{Z}_{\text{odd}}$ if $\partial C,\partial C'\neq 0$ (equivalently $d\hat{m}\neq 0$), namely:
\be 
\pi k \int_\rld \left(\hat C \cup \hat C' - \hat C' \cup \hat C\right) = 2\pi k \int_\rld \hat C\cup \hat C' +  \pi k \overbrace{\int_\rld\left( d(\hat C\cup_1\hat C') -(d\hat C)\cup_1 \hat C' + \hat C\cup_1 (d\hat C')\right)}^{=0 \text{ if } d\hat C =d \hat C' = 0}
\ee 
the first term vanishes modulo $2\pi$, the bracket vanishes for $\partial C =\partial C'=0$. Therefore, we conclude that one is unable to set $U[C] = 1$ for an arbitrary chain $C$ whenever $k$ is odd. This requires a construction of the transparent fermionic operator $S$ as mentioned in the main text.

\section{Derivation of $R-$ and $F-$symbols for dyonic line operators $\CW_{(q,\phi)}$} \label{app:r-and-f-detailed}
In this section, we demonstrate explicit details on the derivation of the $R-$ and $F-$symbol for a generic dyonic line operators $\CW_{(q,\phi)}$ in Section \ref{section:u(1)-junction-computation}. This derivation is followed by explanation on how to project them on the boundary state described by the Lagrangian algebra \eqref{eq:even-lagrangian-state}-\eqref{eq:odd-lagrangian-state}.
For this derivation it is useful to denote identities of the carry $p(\phi_i,\phi_j)$ 
\be \label{eq:carry-prop-1}
p(\phi_1,\phi_2) + p([\phi_1+\phi_2],\phi_3) = p(\phi_2,\phi_3) + p(\phi_1,[\phi_2+\phi_3]) = \lfloor\phi_1+\phi_2+\phi_3\rfloor
\ee 
and the flooring functions 
\begin{equation} \label{eq:floor-prop-1}
\begin{aligned}
\lfloor [\phi_2+\phi_3] +\phi_4\rfloor = \lfloor \phi_2+\phi_3 +\phi_4 - \lfloor\phi_2+\phi_3\rfloor \rfloor = \lfloor \phi_2+\phi_3+\phi_4\rfloor - \lfloor \phi_2+\phi_3\rfloor\\
\lfloor \phi_2 +[\phi_3+\phi_4]\rfloor =  \lfloor \phi_2+\phi_3 +\phi_4 - \lfloor\phi_3+\phi_4\rfloor \rfloor = \lfloor \phi_2+\phi_3+\phi_4\rfloor - \lfloor \phi_3+\phi_4\rfloor
\end{aligned}
\end{equation}
%
The detailed derivations here are of the even $k$ case. For the odd $k$ case, one sees that the Majorana operator $\gamma_p$ cancel each other out upon taking the inner product and results in the same symbols. Recall the shorthand notation $\CW_a$ for the dyonic line where $a$ represent the pair $(q,\phi)$ and the composite line $ab$ carries charge $(q_{ab} := q_a +q_b - kp(\phi_a,\phi_b), \phi_{ab}:= [\phi_a+\phi_b])$. Phases obtained by moving the link $M^a_\ell$ across $\CD_\bx$ and one other can be read off from \eqref{eq:general-move-dynon-algebra}. However, specifically for the configuration in Figure \ref{fig:r-move} and \ref{fig:f-move-lattice}, the R-and F-symbol can be extracted from three simple rules 
\begin{itemize}
\item  Moves $M^a_\ell$ and $M^a_{\ell'}$ with parallel links $\ell\parallel \ell'$ commute
\item Moving $\CD^{-p}_\bx$ pass the link $M^a_\ell$ gain a phase $(-1)^{q_a p s(\ell,\bx)}$ where only nontrivial entries for $s(\ell,\bx)$ is when $\ell$ is a link between $\bx$ and $\bx-\hat 1$.  
\item The vertical move $M^a_{\bx_0+\hat 2,\bx_0}$ commutes with $M^b_{\bx_1,\bx_0}$ but not with $M^b_{\bx_0,\bx_{-1}}$.
At $\bx_0$ the two moves meet in three half-crossings: the Wilson piece of $a$ on $(\bx_0,2)$ with the flux of $b$ on $(\bx_{-1},1)$,
the flux of $a$ with the framing copy $a_{\bx_0+\hat 2,1}$ of $b$, and the two fluxes. This results in the phase
\be
M^b_{\bx_0,\bx_{-1}} M^a_{\bx_0+\hat 2,\bx_0} = e^{i\pi\left(q_a\phi_b+q_b\phi_a-k\phi_a\phi_b\right)}\, M^a_{\bx_0+\hat 2,\bx_0} M^b_{\bx_0,\bx_{-1}}
\ee
\end{itemize}

Let us start with the $R-$symbol. The left $\ket{R_L}$ and right $\ket{R_R}$ states in the $R-$move computation, as shown in Fig. \ref{fig:r-move} are explicitly given by series of moves and splits as in \eqref{def:state-F}
 \begin{subequations}\label{eq:detailed-r-move-even-k}
 \begin{equation}
\begin{aligned}
|R_L\ra &= M^b_{\bx_1,\bx_{-1}} M^a_{\bx_0+\hat 2,\bx_0} \left(N^{(b),(a)}_{ab}\right)_{\bx_{-1}} |(ab)_{\bx_{-1}}\ra \\
&= M^b_{\bx_1,\bx_{-1}} M^a_{\bx_0+\hat 2,\bx_0} \CD_{\bx_{-1}}^{-p(\phi_a,\phi_b)} M^a_{\bx_0,\bx_{-1}}  |(ab)_{\bx_{-1}}\ra \\
&= (-1)^{q_{ab}} M^b_{\bx_1,\bx_0} M^a_{\bx_0+\hat 2,\bx_0} \CD_{\bx_{-1}}^{-p(\phi_a,\phi_b)} M^b_{\bx_0,\bx_{-1}} M^a_{\bx_0,\bx_{-1}}|ab_{\bx_{-1}}\ra
\end{aligned}
\end{equation}
while for $|R_R\ra$, we have 
\be 
\begin{aligned}
|R_R\ra &= M^a_{\bx_0+\hat 2,\bx_0} M^b_{\bx_1,\bx_0} M^a_{\bx_0,\bx_{-1}} \left( N^{(a),(b)}_{ab} \right)_{\bx_{-1}}|ab_{\bx_{-1}}\ra \\
&=  M^a_{\bx_0+\hat 2,\bx_0} M^b_{\bx_1,\bx_0} M^a_{\bx_0,\bx_{-1}} \CD_{\bx_{-1}}^{-p(\phi_a,\phi_b)} M^b_{\bx_0,\bx_{-1}}|ab_{\bx_{-1}}\ra \\
&= M^b_{\bx_1,\bx_0} M^a_{\bx_0+\hat 2,\bx_0} \CD_{\bx_{-1}}^{-p(\phi_a,\phi_b)} M^b_{\bx_0,\bx_{-1}} M^a_{\bx_0,\bx_{-1}}|ab_{\bx_{-1}}\ra
\end{aligned}
\ee 
\end{subequations}
The $R-$symbol is then obtained via 
\be \label{eq:R-symbol-bulk}
R(a,b) = \la R_R|R_L\ra = \exp\left(i\pi (q_a\phi_b+ q_b\phi_a - k\phi_a\phi_b\right)
\ee 

For $F-$move, the left and right configurations in Fig. \ref{fig:f-move-lattice} can be written as
\begin{equation} \label{eq:detailed-f-move-even-k}
\begin{aligned}
|F_L\rangle&=M^{b}_{\mathbf x_1,\mathbf x_2}M^{a}_{\mathbf x_0,\mathbf x_1}(N^{(a),(b)}_{ab})_{\mathbf x_1}M^{c}_{\mathbf x_3,\mathbf x_2}(N^{(ab),(c)}_{abc})_{\mathbf x_1}|abc_{\mathbf x_1}\rangle\\
&=M^{b}_{\mathbf x_1,\mathbf x_2}M^{a}_{\mathbf x_0,\mathbf x_1}\mathcal D^{-p(\phi_a,\phi_b)}_{\mathbf x_1}M^{b}_{\mathbf x_2,\mathbf x_1}M^{c}_{\mathbf x_3,\mathbf x_2}\mathcal D^{-p(\phi_{ab},\phi_c)}_{\mathbf x_1}M^{c}_{\mathbf x_2,\mathbf x_1}|abc_{\mathbf x_1}\rangle\\
&=M^{a}_{\mathbf x_0,\mathbf x_1}\mathcal D^{-p(\phi_a,\phi_b)}_{\mathbf x_1}M^{c}_{\mathbf x_3,\mathbf x_2}\mathcal D^{-p(\phi_{ab},\phi_c)}_{\mathbf x_1}M^{c}_{\mathbf x_2,\mathbf x_1}|abc_{\mathbf x_1}\rangle\\
&=M^{a}_{\mathbf x_0,\mathbf x_1}M^{c}_{\mathbf x_3,\mathbf x_1}\mathcal D^{-\lfloor\phi_a+\phi_b+\phi_c\rfloor}_{\mathbf x_1}|abc_{\mathbf x_1}\rangle ,\\[4pt]
|F_R\rangle&=M^{c}_{\mathbf x_3,\mathbf x_2}(N^{(b),(c)}_{bc})_{\mathbf x_1}M^{bc}_{\mathbf x_1,\mathbf x_2}M^{a}_{\mathbf x_0,\mathbf x_1}(N^{(a),(bc)}_{abc})_{\mathbf x_1}|abc_{\mathbf x_1}\rangle\\
&=M^{c}_{\mathbf x_3,\mathbf x_2}\mathcal D^{-p(\phi_b,\phi_c)}_{\mathbf x_1}M^{c}_{\mathbf x_2,\mathbf x_1}M^{bc}_{\mathbf x_1,\mathbf x_2}M^{a}_{\mathbf x_0,\mathbf x_1}\mathcal D^{-p(\phi_a,\phi_{bc})}_{\mathbf x_1}M^{bc}_{\mathbf x_2,\mathbf x_1}|abc_{\mathbf x_1}\rangle\\
&=M^{c}_{\mathbf x_3,\mathbf x_2}\mathcal D^{-p(\phi_b,\phi_c)}_{\mathbf x_1}M^{c}_{\mathbf x_2,\mathbf x_1}M^{a}_{\mathbf x_0,\mathbf x_1}\mathcal D^{-p(\phi_a,\phi_{bc})}_{\mathbf x_1}|abc_{\mathbf x_1}\rangle\\
&=(-1)^{q_a\,p(\phi_b,\phi_c)}\,M^{a}_{\mathbf x_0,\mathbf x_1}M^{c}_{\mathbf x_3,\mathbf x_1}\mathcal D^{-\lfloor\phi_a+\phi_b+\phi_c\rfloor}_{\mathbf x_1}|abc_{\mathbf x_1}\rangle .
\end{aligned}
\end{equation}
Here, we using the property \eqref{eq:carry-prop-1} to combine the exponent of the local dyonic operator $\CD_{\bx_1}$ and the fact that only one move that does not commute with local dyonic operator $\CD_{\bx_1}$ is $M_{\bx_0,\bx_1}$ i.e.
\be
\begin{aligned}
    s(\bx_0\to\bx_1,\bx_1) &= -1 \\
    s(\bx_1\to\bx_0,\bx_1) &= +1
\end{aligned}
\ee
 Combining this together, one can evaluate the value of $F-$move as 
\be \label{eq:result-Fsymbol-bulk}
F(a,b,c) := \la F_R|F_L\ra = (-1)^{q_a p(\phi_b,\phi_c)}\,\Longrightarrow \qquad \alpha(a,b,c) = \frac{1}{2} q_a p(\phi_b,\phi_c)
\ee 
 where $\alpha$ is defined through $F = e^{2\pi i\alpha}$.
To understand the group cohomological nature of the symbol $\alpha(a,b,c)$, one can check that upon writing $F(a,b,c) = \exp(2\pi i \alpha(a,b,c))$ and with help of properties \eqref{eq:floor-prop-1}, the pentagon equation can be rewritten as
%
\begin{equation}
\begin{aligned}
\delta\alpha&=\alpha(b,c,d)-\alpha(ab,c,d)+\alpha(a,bc,d)-\alpha(a,b,cd)+\alpha(a,b,c)\\
&=\tfrac12\Big[q_b\,p(\phi_c,\phi_d)-q_{ab}\,p(\phi_c,\phi_d)+q_a\big(p([\phi_b+\phi_c],\phi_d)-p(\phi_b,[\phi_c+\phi_d])+p(\phi_b,\phi_c)\big)\Big]\\
&=\tfrac12\big[q_b-q_{ab}+q_a\big]\,p(\phi_c,\phi_d)\\
&=\frac k2\lfloor\phi_a+\phi_b\rfloor\lfloor\phi_c+\phi_d\rfloor ,
\end{aligned}
\end{equation}
indicating the r.h.s. is zero mod 1 for $k\in \mathbb{Z}_\text{even}$.

\section{Discrete Gauging and Anomalies} \label{app:for-sec-v}

In section \ref{section:gaugin-anomaly}, the boundary charges of the gauged theory are labeled by $(j,l)\in\mathbb{Z}^2$, subject to the identification induced by the bulk relation. For even $k$, the boundary identification reads $(j,l)\sim (j+d,l+\kappa)$, where $\kappa:=k/2d$, so the charge group is actually $\mathbb{Z}^2/\langle(d,\kappa)\rangle$. Quotients of this kind can be computed with the Smith normal form. Namely, for any integer matrix $A$, there exist unimodular matrices $P,Q$ such that $PAQ$ is diagonal. For a single vector this reduces to an elementary change of basis, which now we describe. One important identity used in this work is the Bézout's identity: If $a,b$ are non-zero integers, then there exist integers $u$ and $v$ such that
\be \label{eq:bezout-identity}
\gcd(a,b) = au+bv,
\ee
where $\gcd(a,b)$ is the greatest common divisor of $a$ and $b$. In the language of the Smith normal form, basis changing can be understood as turning the $1\times 2$ matrix $(a\; b)$ to $(g\;0)$ by column operation. Let $\mathbf{v}:= (a,b)\in\mathbb{Z}^2$ be non-zero, and let $g:=\text{gcd}(a,b)$. One can write $\mathbf{v}=g\mathbf{u}$ with
\be
\mathbf{u}:=\left(\frac{a}{g},\frac{b}{g} \right), \qquad \text{gcd}\left(\frac{a}{g},\frac{b}{g}\right) =1.
\ee
Then Bézout's identity \eqref{eq:bezout-identity} provides integers $x,y$ with
\be
x\frac{a}{g} + y\frac{b}{g} = 1.
\ee
Define a new $\mathbf{w}:=(-y,x)$ such that the matrix with columns $\mathbf{u}$ and $\mathbf{w}$ has 
\be
\text{det}\begin{pmatrix}
    a/g & -y \\
    b/g & x
\end{pmatrix} =1,
\ee
so it belongs to $SL(2,\mathbb{Z})$, and $\{\mathbf{u},\mathbf{w}\}$ is a new basis of $\mathbb{Z}^2$ such that $\mathbb{Z}^2=\mathbf{u}\mathbb{Z}\oplus \mathbf{w}\mathbb{Z}$. In this new basis, the relation involves only $\mathbf{u}$, namely $\langle \mathbf{v} \rangle = \langle g\mathbf{u} \rangle = g\mathbb{Z}\mathbf{u}$, where $g=g_b:= \text{gcd}(d,\kappa)$. Hence 
\be
\frac{\mathbb{Z}^2}{\langle (d,\kappa)\rangle} = \frac{\mathbf{u}\mathbb{Z}\oplus \mathbf{w}\mathbb{Z}}{g_b \mathbf{u}\mathbb{Z}} \cong \mathbb{Z}_{g_b}\oplus \mathbb{Z},
\ee
where the torsion factor $\mathbb{Z}_{g_b}$ is generated by $\mathbf{u}$ and the free factor by $\mathbf{w}$. For bosonic case, we apply $\mathbf{v}:=(d,\kappa)$, this gives
\be
\frac{\mathbb{Z}^2}{\langle (d,\kappa) \rangle}\cong \mathbb{Z}_{g_b}\oplus \mathbb{Z},\qquad g_b:=\text{gcd}(d,\kappa)
\ee

For fermionic case, the trivial charge is generated by $2(d,-\zeta')$ with $\zeta':=\frac{k}{d}\frac{d-1}{2}$ is an integer. The same construction gives
\be
\frac{\mathbb{Z}^2}{\langle  2(d,-\zeta') \rangle} \cong \mathbb{Z}_{2g_f}\oplus \mathbb{Z} \cong \mathbb{Z}_{g_f}\oplus \mathbb{Z}_2\oplus \mathbb{Z},\qquad g_f:=\text{gcd}(d,\zeta') = \text{gcd}(d,k/d),
\ee
where the second equality is obtained since $\text{gcd}(2,g_f)=1$. Note that $\text{gcd}(d,\zeta') = \text{gcd}(d,k/d)$ since $\gcd(d,\frac{d-1}{2})=1$. We also showed in the main text that this additional $\mathbb{Z}_2$ is generated by the transparent fermion $f$.

For $k$ even, the group structure reads $\CG'_b\cong U(1)'\times \mathbb{Z}_g$. Here we give a detailed derivation of this isomorphism explicitly,\footnote{Alternatively, one can extract the group structure by counting the connected components formed by the lines $L_n:=\{(x,y) \in\mathbb{R}^2:xd+\kappa y = n\}$, with $n\in\mathbb{Z}$ in the torus $T^2$} similar in spirit to the Smith normal form implemented above. Note that the following derivations are based on the bosonic case, but it straightforwardly works with the fermionic case and we will drop the subscript $b$ in $g_b$ for convenience. Namely, we perform an $SL(2,\mathbb{Z})$ change of coordinate on $T^2$ that brings the defining condition of $\CG'_b$ to the form from which the group structure can be read off. Let $g:= \gcd(d,\kappa)$, and define $d':=d/g$ and $\kappa'=\kappa/g$, so that $\gcd(d',\kappa')=1$. By Bézout's identity, there are integers $t,v\in\mathbb{Z}$ with
\be
d'v-\kappa' t = 1
\ee
We define
\be \label{eq:sl(2,z)-from-bezout}
M:= \begin{pmatrix}
    d' & \kappa' \\
    t  & v
\end{pmatrix}, \quad \det(M) = d'v-\kappa' t =1.
\ee
Since $M$ has a unit determinant, its inverse is also an integer matrix, so that $M\mathbb{Z}^2=\mathbb{Z}^2$, i.e., $M\in SL(2,\mathbb{Z})$. Hence $M$ induces an automorphism $\mathrm{\Phi}$ of the torus, acting on the torus coordinates as
\be
\mathrm{\Phi}:\ \begin{pmatrix} x \\ y \end{pmatrix} = M\begin{pmatrix} x \\ y \end{pmatrix} = \begin{pmatrix}d'x+\kappa' y \\ tx + vy\end{pmatrix} := \begin{pmatrix}\widetilde{X},\widetilde{Y} \end{pmatrix},
\ee
where $(x,y)$ and $(\widetilde{X},\widetilde{Y})$ are the original and new bases in $\CG'_b$, respective. Because $xd+\kappa y = g\widetilde{X}$, it maps the defining condition to
\be
\mathrm{\Phi}(\CG'_b) = \left\{ (\widetilde{X},\widetilde{Y})\in T^2: g\widetilde{X}\in\mathbb{Z}  \right\}.
\ee
The condition constrains only $\widetilde{X}$ and $\Phi$ is a group automorphism, so the group product is direct. In terms of the probes, given $(x,y)=(\frac{n}{d}-\kappa\alpha,\alpha d)$ gives
\be
\begin{aligned}
    \widetilde{X} = n/g,\quad \widetilde{Y}= \frac{tn}{d} + \alpha g,\qquad n\in\mathbb{Z}.
\end{aligned}
\ee
where we have used $vd-t\kappa =g$. Thus the $\mathbb{Z}_g$ label is $n\text{ mod }g$ while the $U(1)$ factor is parameterized by $\widetilde{Y}$. We now construct the bulk lines representing these symmetry defects.

\emph{$U(1)'$ factor.} A pure flux lines $(0,\alpha)$ in the bulk label maps to $(\widetilde{X},\widetilde{Y}) = (0,g\alpha)$ in the new basis, so the circle closes at $\alpha = 1/g$. The generating class is
\be
u_\sigma := \CW_{(0,\sigma/g)},\qquad \sigma\in [0,1).
\ee

\emph{$\mathbb{Z}_g$ factor.} By definition, the $\mathbb{Z}_g$ factor is the subgroup $\widetilde{Y}=0$ of $\mathrm{\Phi}(\CG'_b)$. A Wilson line $(r,0)$ maps to $(r/g,tr/d)$, so its $\widetilde{X}$ component carries the $\mathbb{Z}_g$ label, but $\widetilde{Y}=tr/d\neq 0$. This means that it is not a subgroup, and we know from the previous computation that the group structure is in a product form. We can correct this by adding some flux so that $\widetilde{Y}=0$, namely
\be
w_r:= \CW_{(r,-\frac{tr}{gd})}\,,\quad r\in\mathbb{Z}_g\,,
\ee
where $t$ is an integer satisfying $vd'-\kappa' t=1$ for some $v\in\mathbb{Z}$, i.e., $-\kappa' t \equiv 1 \pmod{d'}$. This generates the desired $\mathbb{Z}_g$ alone. One can then check that $w_r \mapsto (\widetilde{X},\widetilde{Y})=(r/g,0)$. Moreover, one can verify that the label $r$ is guaranteed to be mod $g$ because $w_g = \CW_{(g,-t/d)}\in \CL'$, and we know that $\CG'_b = \CA/\CL'$. The fusion rules of these line labels are 
\be \label{eq:boundary-sym-line-fusion}
\begin{aligned}
    u_{\sigma_1} u_{\sigma_2} &= u_{\sigma_1+\sigma_2} = u_{[\sigma_1+\sigma_2]}u_1^{p(\sigma_1,\sigma_2)},\qquad u_1:= \CW_{(0,1/g)} \\
    w_{r_1}w_{r_2} &= w_{r_1+r_2} = w_{[r_1+r_2]_g}w_g^{p_g(r_1,r_2)},\qquad w_g := \CW_{(g,-t/d)}
\end{aligned}
\ee
where $u_1$ and $w_g$ are carries of the $u_\sigma$ and $w_r$, respectively. One can check that these carries lines are indeed living in the Lagrangian algebra, $\CL'$.
Here we only concern about fusion of line labels, so we do not include the local dyonic operator $\CD$. The local dyonic operators $\CD$ are already taken into account in our previous $F$--move derivations. The same strategy applies to the fermionic case, since the group structure is the same, with the additional $\mathbb{Z}_2^F$ factor that is blind to the conventional SymTFT as shown in the main text.

We show the derivation of the boundary $F$--move of the symmetry defect lines $x\in\CG'_b$ used in Section \ref{section:gaugin-anomaly}. We discuss the bosonic theory in detail, while the generalization to the fermionic case is given at the end. In this appendix, a product of bulk lines such as $l_al_b$ is the plain product of labels, i.e., $\CW_{(q,\phi)}\CW_{(q',\phi')}=\CW_{(q+q',\phi+\phi')}$. The composite line $ab$ of Section \ref{section:u(1)-junction-computation} has label $(q_a+q_b - kp(\phi_a,\phi_b),[\phi_a+\phi_b])$. So in this Appendix's notation, it equals the plain product $ab$ times $\nu^{-p(\phi_a,\phi_b)}$, where we define $\nu:= \CW_{(k,1)}\in\CL'$ in this appendix. From section \ref{section:u(1)-junction-computation}, the boundary $F$--move is the phase \eqref{eq:M6jpentagon},
\be
\bar\alpha(x_a,x_b,x_c) = \alpha(a,b,c) + \delta\bar\beta(a,b,c):= \Omega(a,b,c)\,,\nn
\ee 
where $\alpha(a,b,c)=\frac{1}{2}q_ap(\phi_b,\phi_c)$ is the bulk $F$--move. It must satisfy the constraint \eqref{eq:condensate-Omega},
\be \nn
\Omega(l,a,b) = \Omega(a,l,b) = \Omega(a,b,l) = 0\,,
\ee
for any $l\in\CL'$ and any bulk line $a$. This ensures that any condensed lines drop out and do not contribute to the boundary $F$--move. Therefore, we have to find $\bar\beta$ such that all bulk information is removed from $\Omega$. For this we need to separate the boundary part of a bulk line from its condensed part. 
For convenience we rewrite the lift lines corresponding to the symmetry defects $x=(\sigma,r)\in\CG'_b$ as
\be
s(\sigma,r) := u_\sigma w_r = \CW_{(r,\frac{\sigma}{g}-\frac{rt}{gd})}\,.
\ee
Since $\CG'_b=\CA/\CL'$, every bulk line $a$ determines a symmetry line $x_a\in\CG'_b$, and the class $as(x_a)^{-1}$ lies in $\CL'$. Hence every bulk line can be decomposed as
\be \label{eq:decompose-bulk-line-label}
a = s(x_a)l_a,\qquad x_a := (\sigma_a,r_a) \in \CG'\,, l_a\in\CL'\,.
\ee
The first one is the boundary label, the second is a line in the Lagrangian algebra. By the boundary fusion rules \eqref{eq:boundary-sym-line-fusion}, one finds that the lifts fuse as
%
\be \label{eq:lift-line-labels}
s(x_a)s(x_b) = s(x_ax_b)c_{ab},\qquad c_{ab} := u_1^{p(\sigma_a,\sigma_b)}w_g^{p_g(r_a,r_b)}\in \CL'\,,
\ee
%
%
where $c_{ab}$ contains all the carry information on the boundary. On the other hand, the fusion of the bulk lines has $\nu:= \CW_{(k,1)}$ as a carry. Decomposing both sides with \eqref{eq:decompose-bulk-line-label} and use the above boundary fusion rules, one read the following relation
%
\be \label{eq:label-constriant}
l_{ab} = c_{ab}l_al_b\nu^{-p(\phi_a,\phi_b)}
\ee
This is a constraint between the symmetry lines and the Lagrangian algebra lines. This ensures that the bulk carry $\nu^{p(\phi_a,\phi_b)}$ equals the boundary carry $c_{ab}$ times a mismatch of the condensed lines.
%
In this appendix, we use the bilinear pairing $\la a,b \ra$, an asymmetric half of the braiding, i.e., $\exp(2\pi i (\la a,b \ra + \la b,a \ra))=\CB(a,b)$, such that
\be
\langle a,b \rangle := q_a\phi_b - \frac{k}{2}\phi_a\phi_b\,,
\ee
On the bosonic $\CL'$, this pairing is integer-valued on condensed lines, i.e., $\langle l(j,l),l(j',l') \rangle =lj' \in\mathbb{Z}$ so there is no phase contribution from the condensed lines. (On the fermionic $\CL_\rho'$, it is half-integral; see the discussion after \eqref{eq:fermionic-boundary-f-move}.) To remove the bulk dependence in the bulk $F$--move, we can move the carry of $\alpha$ into the first two legs, where the constraint \eqref{eq:label-constriant} applies. With $\alpha_R := \frac{k}{2}p(\phi_a,\phi_b)\phi_c$, one can verify that
\be
\alpha(a,b,c) = \frac{1}{2}q_a p(\phi_b,\phi_c) = \alpha_R-\delta\left( \frac{1}{2}q_a\phi_b \right)\,,
\ee
and $\alpha_R=\langle \nu^{p(\phi_a,\phi_b)},c \rangle$. Inserting the bulk-boundary constraint \eqref{eq:label-constriant}, and $c = s(x_c)l_c$, and dropping $\langle c_{ab},l_c \rangle\in\mathbb{Z}$, $\alpha_R$ in the bilinear pairing form reads
\be\label{eq:alpha_R-binlinear}
\alpha_R = \langle c_{ab},s(x_c) \rangle + \langle l_al_bl_{ab}^{-1},c \rangle \pmod1\,.
\ee
The first term depends only on the boundary labels, while the second term contains all condensed lines and must be canceled, i.e., we need some $\bar{\beta}$ such that $\delta(\bar{\beta} - \frac{1}{2}q_a\phi_b)=-\langle l_al_bl_{ab}^{-1},c \rangle$. Then one can check that 
\be
\delta\langle l_a,b \rangle = \langle l_al_bl_{ab}^{-1},c \ra \pmod1\,.
\ee
Hence, the choice
\be
\bar\beta (a,b) = \frac{1}{2}q_a\phi_b - \langle  l_a,b\rangle\,,
\ee
cancels the condensed part leaving the boundary $F$--move with only boundary information as
\be
\Omega(a,b,c) = \bar{\alpha}(x_a,x_b,x_c) := \la c_{ab},s(x_c) \ra
\ee
%
This expression is completely described by the boundary labels as we desired. The condition \eqref{eq:condensate-Omega} holds true since if $a$ or $b$ is a condensed line, its boundary label is trivial. On the other hand, if $c$ is condensed, then its boundary label is trivial, i.e., $s(x_c)=1$, and pairing with the trivial line vanishes. Next, we substitute $c_{ab}$ and $s(x_c)$ and obtain that
\be\label{eq:alphabar-finalform-evenk}
\bar{\alpha}(x_1,x_2,x_3) = p(\sigma_1,\sigma_2)\left[-\frac{k'}{2}\sigma_3 + \frac{\kappa't}{g}r_3 \right] + p_g(r_1,r_2)(1+\kappa' t) \left[ \sigma_3 - \frac{t}{d}r_3\right]\,,
\ee
Note that this is not a unique way to determine the boundary $F$--move. We can also move the bulk-dependent term into the last two slots. The two representations give the two orientations of the boundary, with opposite classes, so that all anomaly coefficients change sign. This does not affect the conclusion of anomaly.

For odd $k$, we can also compute the boundary $F$--move. Namely, the above derivation goes through with a $1/2$ factor and now the Lagrangian algebra is $\CL'_\rho = \{l(j,l)=(\zeta j+ld,j/d)\}$. The additional term in the $F$--move is basically $\frac{k/d}{2}jj'$, since now we only have the condition that $k/d\in\mathbb{Z}$. Then, the boundary $F$--move for the fermionic case is given by
\be \label{eq:fermionic-boundary-f-move}
\bar{\alpha}(x_1,x_2,x_3) = p(\sigma_1,\sigma_2)\left[ -\frac{k'}{2}\sigma_3 + \frac{k't}{2d}r_3 \right] + p_g(r_1,r_2)\left(1+\frac{kt}{2g_f d} \right)\left[ \sigma_3 - \frac{t}{d}r_3 \right] \pmod{1/2}\,,
\ee
The pairing $\la \bullet,\bullet\ra$ is only half-integer on $\CL'_\rho$ so the anomaly cancellation condition in \eqref{eq:alpha_R-binlinear}-\eqref{eq:alphabar-finalform-evenk} holds mod 1/2. Namely, the bilinear pairing reads $\la l(j,l),l(j',l') \ra = (k/d)jj'/2 + lj'$, and the remaining sign is the $S$-dressing sign \eqref{eq:modified-f-move-odd-k}.

Next, we can show that $F$--move can be removed using the fact that $\bar{p}$ is a trivial class. First, from the definitions $\bar{p}(a,b)$ and $h(a)$ in section \ref{section:gaugin-anomaly}, one can check that the following identities hold
\be
\begin{aligned}
    &\bar{p}(a,b) = h(a) + h(b) + h(ab), \quad \bar{p}(ab,c) = h(ab)+h(c)+h(abc), \\
    &\bar{p}(b,c) = h(b)+h(c) + h(bc),\quad \bar{p}(a,bc)=h(a) + h(bc)+h(abc)
\end{aligned}
\ee
Then the modified $F$--move can be computed using the left and right configurations of Fig. \ref{fig:f-move-lattice}, namely
\be
\begin{aligned}
    |\widetilde{F}_L\rangle&=\widetilde{M}^{b}_{\mathbf x_1,\mathbf x_2}\widetilde{M}^{a}_{\mathbf x_0,\mathbf x_1}(\widetilde{N}^{(a),(b)}_{ab})_{\mathbf x_1}\widetilde{M}^{c}_{\mathbf x_3,\mathbf x_2}(\widetilde{N}^{(ab),(c)}_{abc})_{\mathbf x_1}|abc_{\bx_1}\rangle \\
    |\widetilde{F}_R\rangle&=\widetilde{M}^{c}_{\mathbf x_3,\mathbf x_2}(\widetilde{N}^{(b),(c)}_{bc})_{\mathbf x_1}\widetilde{M}^{bc}_{\mathbf x_1,\mathbf x_2}\widetilde{M}^{a}_{\mathbf x_0,\mathbf x_1}(\widetilde{N}^{(a),(bc)}_{abc})_{\mathbf x_1}|abc_{\mathbf x_1}\rangle
\end{aligned}
\ee
Generally, for the fermionic case, the fusion vertices always carry Majorana fermions $\gamma^{h(abc)}_{\bx_1}$ which will play a crucial role in this derivation. For the both configuration, one can use the fact that $h\in\mathbb{Z}_2$ and the fact that $\gamma_{\bx}^2=1$ and $\{ \gamma_{\bx},\gamma_{\bx'} \}=2\delta_{\bx,\bx'}$ to reduce some phases from moving Majorana fermions around.
Since the bosonic $M_a$, and $\CD$ parts do not change upon dressing, we can only consider the fermionic contributions from Majorana fermions, namely
\be
\begin{aligned}
    |\widetilde{F}_L\rangle &\propto T^{h(b)}_{\bx_1,\bx_2}T^{h(a)}_{\bx_0,\bx_1}T^{h(b)}_{\bx_2,\bx_1}\gamma^{\bar{p}(a,b)}_{\bx_1}T^{h(c)}_{\bx_3,\bx_2}T^{h(c)}_{\bx_2,\bx_1}\gamma_{\bx_1}^{\bar{p}(ab,c)}\gamma^{h(abc)}_{\bx_1}|0\rangle \\
    &\propto \gamma^{h(b)}_{\bx_1}\gamma^{h(b)}_{\bx_2}\gamma^{h(a)}_{\bx_0}\gamma^{h(a)}_{\bx_1}\gamma^{h(b)}_{\bx_2}\gamma^{h(b)}_{\bx_1}\gamma^{\bar{p}(a,b)}_{\bx_1}\gamma^{h(c)}_{\bx_3}\gamma^{h(c)}_{\bx_2}\gamma^{h(c)}_{\bx_2}\gamma^{h(c)}_{\bx_1}\gamma^{\bar{p}(ab,c)}_{\bx_1}\gamma^{h(abc)}_{\bx_1}|0\rangle\\
    &\propto (-1)^{\bar{p}(a,b)h(c)}\gamma^{h(c)}_{\bx_3}\gamma^{h(b)}_{\bx_1}\gamma^{h(a)}_{\bx_0}|0\rangle \\
    |\widetilde{F}_R\rangle &\propto T^{h(c)}_{\bx_3,\bx_2}T^{h(c)}_{\bx_2,\bx_1}\gamma^{\bar{p}(b,c)}_{\bx_1} T^{h(bc)}_{\bx_1,\bx_2}T^{h(a)}_{\bx_0,\bx_1}T^{h(bc)}_{\bx_2,\bx_1}\gamma^{\bar{p}(a,bc)}_{\bx_1}\gamma^{h(abc)}_{\bx_1}|0\rangle \\
    &\propto \gamma^{h(c)}_{\bx_3}\gamma^{h(c)}_{\bx_2}\gamma^{h(c)}_{\bx_2}\gamma^{h(c)}_{\bx_1}\gamma^{\bar{p}(b,c)}_{\bx_1}\gamma^{h(bc)}_{\bx_1}\gamma^{h(bc)}_{\bx_2}\gamma^{h(a)}_{\bx_0}\gamma^{h(a)}_{\bx_1}\gamma^{h(bc)}_{\bx_2}\gamma^{h(bc)}_{\bx_1}\gamma^{\bar{p}(a,bc)}_{\bx_1}\gamma^{h(abc)}_{\bx_1}|0\rangle \\
    &\propto \gamma^{h(c)}_{\bx_3}\gamma^{h(b)}_{\bx_1}\gamma^{h(a)}_{\bx_0}|0\rangle
\end{aligned}
\ee
where we use $\bar{p}(ab,c)+h(abc)=h(ab)+h(c)$ and $\bar{p}(a,bc)+h(abc)=h(a)+h(bc)$ to merge the original Majorana fermion sitting at the fusion junction with the rest for the left and the right configurations, respectively. Since both left and right configurations contain the same Majorana configurations, one can directly compute the inner product to obtain the redefined $F$--move, namely
\be
\frac{ \langle \widetilde{F}_L| \widetilde{F}_R \rangle }{\langle F_L|F_R \rangle} = (-1)^{\bar{p}(a,b)h(c)}
\ee
The sign has a simple origin. In $\ket{\widetilde{F}_L}$, the fusion of $(a,b)$ takes place after $c$ has been split off together with its fermion, so the junction Majorana of $(a,b)$ must be exchanged with the fermion of $c$. While in $\ket{\widetilde{F}_R}$ no such exchange occurs.
\end{appendix}
\bibliographystyle{JHEP.bst}

\bibliography{Reference}

\end{document}